\documentclass[11pt,          
               ms,           
               doublespacing, 
               ]{ncsuthesis}

\usepackage{booktabs}  
\usepackage{textcomp}  
\usepackage{subfig}    
\usepackage[toc,page]{appendix}
\usepackage{soul}
\usepackage{relsize}
\usepackage[utf8]{inputenc}
\usepackage{csquotes}

\RequirePackage[
			bibstyle=ieee,
			citestyle=numeric-comp,
			sorting=none,					
			hyperref=false,
			firstinits=true,
			backend=bibtex,
			url=false,
			isbn=false,
			sortcites=true,
			doi=false]{biblatex}

\makeatletter

\newrobustcmd*{\parentexttrack}[1]{%
  \begingroup
  \blx@blxinit
  \blx@setsfcodes
  \blx@bibopenparen#1\blx@bibcloseparen
  \endgroup}

\AtEveryCite{%
  }

\makeatother
\renewcommand{\cite}[1]{\parencite{#1}}

\renewbibmacro{in:}{%
  \ifentrytype{article}{}{%
  \printtext{\bibstring{in}\intitlepunct}}}
  
\AtEveryBibitem{\clearfield{language}}

\defbibheading{myheading}[BIBLIOGRAPHY]{
 \chapter*{#1}
 \markboth{#1}{#1}}

\usepackage{dcolumn}
\usepackage{bm}
\usepackage{cancel}
\usepackage{verbatim}
\usepackage{ifthen}
\usepackage{url}
\usepackage{sectsty}
\usepackage{balance} 
\usepackage{graphicx} 
\usepackage{lastpage}
\usepackage[format=plain,justification=justified,singlelinecheck=false,font=small,labelfont=bf,labelsep=space]{caption} 
\usepackage{fancyhdr}
\usepackage{epstopdf}

\usepackage{abbrevs}
\usepackage{etoolbox}
\robustify{\DateMark} 

\usepackage{units} 

\usepackage[sharp]{easylist} 

\usepackage{array,booktabs}
\usepackage{colortbl}
\newcolumntype{L}{@{}>{\kern\tabcolsep}l<{\kern\tabcolsep}}

\usepackage{titlesec}

\titleformat{\paragraph}
{\normalfont\normalsize\bfseries}{\theparagraph}{1em}{}
\titlespacing*{\paragraph}
{0pt}{3.25ex plus 1ex minus .2ex}{1.5ex plus .2ex}

\usepackage{placeins}

\dispositionformat{\bfseries}            

\headingformat{\large\MakeUppercase}   

\usepackage[Glenn]{fncychap}

\usepackage{microtype}

\usepackage[T1]{fontenc}
\usepackage[sc]{mathpazo}

\committeesize{3}

\chair{Dror Baron}
\memberI{David Lalush}
\memberII{Michael Daniele}

\student{Ryan Zachary}{Pilgrim} 

\program{Electrical Engineering}

\thesistitle{Source Coding Optimization for Distributed Average Consensus}

\usepackage{color}

\usepackage{calc}
\newlength{\chaptercapitalheight}
\newlength{\chapterfootskip}
\renewcommand{\bibname}{BIBLIOGRAPHY}

\newlength\graphht

\usepackage{pdflscape}
\usepackage{tikz}
\fancypagestyle{lscapedplain}{%
  \fancyhf{}
  \fancyfoot{%
    \tikz[remember picture,overlay]
      \node[outer sep=1cm,above,rotate=90] at (current page.east) {\thepage};}

}
\usepackage{amsmath,amssymb,amsfonts,cleveref}
\usepackage{mathabx}
\crefrangeformat{equation}{(#3#1#4)--(#5#2#6)}
\crefmultiformat{equation}{(#2#1#3)}%
{ and~(#2#1#3)}{, (#2#1#3)}{ and~(#2#1#3)}

\newcommand{\norm}[1]{\lVert#1\rVert}

\newcommand{\reals}{{\mathbb R}}

\newcommand{\tp}{^{\top}}

\newcommand{\fourspace}{\,\,}
\newcommand{\tindex}[1]{(#1)}
\newcommand{\vindex}[1]{_{#1}}
\newcommand{\eightspace}{\fourspace\fourspace}

\newcommand{\A}{{\mathbf{A}}}
\newcommand{\z}[1]{{{\bf z}\tindex{#1}}}
\newcommand{\zt}[1]{{{\bf z}\tp \tindex{#1}}}
\newcommand{\qe}[1]{{{\boldsymbol{\delta}}\tindex{#1}}}
\newcommand{\qet}[1]{{\boldsymbol{\delta}\tp\tindex{#1}}}
\newcommand{\q}[1]{{Q(#1)}}
\newcommand{\W}[1]{{{\bf W}^{#1}}}
\newcommand{\D}[1]{{D\tindex{#1}}}

\newcommand{\I}{{\bf I}}

\newcommand{\MSE}[1]{{\mathrm{MSE}}\tindex{#1}}
\newcommand{\MSELL}[1]{{\mathrm{MSE}_{\mathrm{lossless}}\tindex{#1}}}

\newcommand{\E}[1]{{\mathbb{E}\left[#1\right]}}
\newcommand{\tr}[1]{{\mathrm{tr}}\left(#1\right)}

\newcommand{\ones}{{\mathbf{1}}}
\newcommand{\avg}{{\frac{1}{m}{\ones \ones \tp}}}

\renewcommand{\th}{{\text{th}}}
\newcommand{\meanz}[1]{{{\boldsymbol{\mu}}}}

\newcommand{\var}[1]{{\mathrm{var}\left(#1\right)}}

\DeclareMathOperator*{\argmin}{arg\,min}
\newcommand{\diff}[1]{{{{\mathbf{e}}\tindex{#1}}}}
\newcommand{\difft}[1]{{{\mathbf{e}\tp\tindex{#1}}}}

\newcommand{\M}{{\left( \I - \avg \right)}}

\newcommand{\zmarg}[2]{{z\vindex{#1}\tindex{#2}}}

\newcommand{\margvar}[2]{{\nu_{#1}\tindex{#2}}}

\newcommand{\ze}[1]{{\boldsymbol{\zeta}\tindex{#1}}}

\newcommand{\zet}[1]{{\boldsymbol{\zeta} \tp \tindex{#1}}}
\newcommand{\et}[1]{{\boldsymbol{\eta}\tindex{#1}}}
\newcommand{\ett}[1]{{\boldsymbol{\eta} \tp \tindex{#1}}}
\newcommand{\Om}{{\boldsymbol{\Omega}}}

\newcommand{\OsubI}{{\left( \Om - \I \right) }}

\newcommand{\WsubI}{{\left( \W{} - \I \right)}}

\newcommand{\Dv}{\mathbf{D}}
\newcommand{\w}{\mathbf{w}}

\newcommand{\x}{{\mathbf x}}
\newcommand{\Dmarg}[2]{{D_{#1}\tindex{#2}}}
\newcommand{\What}[1]{{\widehat{\mathbf{W}}\tindex{#1}}}
\newcommand{\Whatt}[1]{{\widehat{\mathbf{W}}\tp\tindex{#1}}}
\newcommand{\what}[2]{{\widehat{\mathbf{w}}_{#1}\tindex{#2}}}
\newcommand{\whatt}[2]{{\widehat{\mathbf{w}}_{#1}\tp\tindex{#2}}}

\newcommand{\Wcar}[1]{{\widecheck{\mathbf{W}}\tindex{#1}}}
\newcommand{\Wcart}[1]{{\widecheck{\mathbf{W}}\tp\tindex{#1}}}
\newcommand{\wcar}[2]{{\widecheck{\mathbf{w}}_{#1}\tindex{#2}}}
\newcommand{\wcart}[2]{{\widecheck{\mathbf{w}}_{#1}\tp\tindex{#2}}}
\newcommand{\Rhat}{{\mathbf{R}}}

\newcommand{\etal}{{\em{et al.}}}

\newcommand{\mean}[2]{\boldsymbol{\mu}_{#1}\tindex{#2}}
\newcommand{\meant}[2]{\boldsymbol{\mu}_{#1}\tp\tindex{#2}}
\newcommand{\etav}{{\boldsymbol{\eta}}}
\newcommand{\zetav}{{\boldsymbol{\zeta}}}

\newcommand{\Ragg}{R_{\text{agg}}}
\newcommand{\MM}{{\mathbf{M}}}
\newcommand{\zv}[2]{{\boldsymbol{\zeta}_{#1}(#2)}}
\newcommand{\zvmarg}[3]{ {\zeta}^{#2}_{#1}(#3) }
\newcommand{\zebar}{\widebar{\boldsymbol{\zeta}}}
\newcommand{\zestar}{\boldsymbol{\zeta}^*}
\newcommand{\qes}[1]{\boldsymbol{\epsilon}(#1)}
\newcommand{\qesu}{\boldsymbol{\epsilon}}
\newcommand{\qevmarg}[3]{ {\delta}^{#2}_{#1}(#3) }
\newcommand{\deltav}{{\boldsymbol{\delta}}}
\newcommand{\Sig}[2]{{\boldsymbol{\Sigma}_{#1}\tindex{#2}}}
\newcommand{\spacedvert}{{\,\,\Big\vert\,\,}}

\newcommand{\Rggp}[1]{R_{\text{GGP}}(#1)}
\newcommand{\RhatGGP}{\Rhat_{\text{GGP}}}
\newcommand{\bll}{\boldsymbol{\ell}}

\newcommand{\figpctg}{0.405}
\newcommand{\figpctgtwo}{0.435}

\newcommand*{\QED}{\hfill{\Huge\ensuremath{\square}}}
\newcommand{\seeqs}{eq:se-data-mean-simple,eq:se-error-mean-simple,eq:se-data-simple,eq:se-error-simple}

\newtheorem{lemma}{Lemma}
\newtheorem{assumption}{Assumption}

\begin{document}
\pagestyle{plain}
\frontmatter

\thispagestyle{empty}
{\bf Changelog:}
\begin{itemize}
\item December 2, 2021
\begin{itemize}
\item Reworded Assumption 2 (p. 37).
\item Removed ``Lemma 1,'' which was actually an informal argument not referenced anywhere else in the thesis, and replaced with an intuitive discussion
\item Substituted ``Multivariate'' for ``Multidimensional''
\end{itemize}
\end{itemize}

\pagebreak

\begin{abstract}

Consensus is a common method for computing a function of the data distributed among the nodes of a network. Of particular interest is distributed average consensus, whereby the nodes iteratively compute the sample average of the data stored at all the nodes of the network using only near-neighbor communications. In real-world scenarios, these communications must undergo quantization, which introduces distortion to the internode messages. In this thesis, a model for the evolution of the network state statistics at each iteration is developed under the assumptions of Gaussian data and additive quantization error. It is shown that minimization of the communication load in terms of aggregate source coding rate can be posed as a generalized geometric program, for which an equivalent convex optimization can efficiently solve for the global minimum. Optimization procedures are developed for rate-distortion-optimal vector quantization, uniform entropy-coded scalar quantization, and fixed-rate uniform quantization. Numerical results demonstrate the performance of these approaches. For small numbers of iterations, the fixed-rate optimizations are verified using exhaustive search. Comparison to the prior art suggests competitive performance under certain circumstances but strongly motivates the incorporation of more sophisticated coding strategies, such as differential, predictive, or Wyner-Ziv coding.

\end{abstract}

\makecopyrightpage

\maketitlepage

\begin{dedication}
 \centering To everyone who has stood by me in tough times: my parents, my brothers, and my girlfriend, Kristin. My success rests on a foundation of love and support from you all, and I am grateful.
\end{dedication}

\begin{biography}
The author was born in Boulder, Colorado in 1991. His childhood was spent in Colorado, Virginia, Mississippi, and North Carolina. After graduating from Hickory High school, he attended the University of North Carolina at Asheville. He then transferred to North Carolina State University to study Biomedical Engineering, graduating \textit{magna cum laude} with a Bachelor of Science in 2014.

His research interests span statistical signal processing, compressive sensing, and machine learning. He also enjoys programming and embedded systems development. In his spare time, he plays disc golf and the guitar.

The author completed an internship at Southwest Research Institute in 2015, where he helped create a GPS spoofing detection system. He is currently a Project Engineer at Vadum, Inc. in Raleigh, North Carolina.
\end{biography}

\begin{acknowledgements}
I would like to thank my advisor, Dror Baron, for his patient mentorship and excellent advice. Thanks also to my collaborators Waheed Bajwa and Junan Zhu for their many helpful inputs regarding consensus and the research process. I would furthermore like to acknowledge Yanting Ma for helping to expand the main theorem of the thesis to the more general, variable distortion setting.

Special thanks to Mehmet Ercan Yildiz and Anna Scaglione for graciously providing their code for comparison, and to Yaoqing Yang and Pulkit Grover for helpful discussions about their information theoretic work with Soummya Kar on the consensus problem.
\end{acknowledgements}

\thesistableofcontents

\thesislistoftables

\thesislistoffigures

\mainmatter

\pagestyle{plain}
\newgeometry{margin=1in,lmargin=1.25in,footskip=\chapterfootskip, includefoot}

\chapter{INTRODUCTION}
\label{chap-one}
The wide availability of wireless sensors and large data sets in recent years has provided significant motivation for the development of distributed computing methods. Large sensor networks and big, distributed data sets pose a number of unique optimization opportunities, including energy management in battery-powered sensor networks~\cite{pottie2000,estrin2002} and run time reduction in cloud computing applications~\cite{EC2}.

In many distributed computing problems, it is necessary to compute a function of data that may be dispersed among a number of computing nodes, for instance in sensor networks, where each agent observes a different measurement of a physical process, or large-scale server farms, where the size of the data set requires distributed storage~\cite{NoorshamsWainwright2011}. The class of distributed algorithms considered in this thesis computes these functions with only local, near-neighbor interactions. This distributed approach offers several potential advantages, including robustness to link and node failure and maintenance of privacy~\cite{Dimakis2010}. One popular approach to distributed function computation is called consensus~\cite{Olfati-Saber2007,Dimakis2010}, which has many variants. Consensus protocols have found application in a wide variety of settings, including oscillator synchronization, robot swarm control, load balancing in computer networks, distributed sensor fusion, belief propagation~\cite{Olfati-Saber2007}, detection, estimation, target tracking~\cite{Dimakis2010}, distributed optimization, filtering, and environmental monitoring~\cite{Nokleby2013}. Consensus algorithms have also found applications in computer vision~\cite{Tron2011}, where they can be used for null-space and least-squares estimation, distributed singular value decomposition, principal components and generalized principal components analyses, point triangulation, and linear pose estimation. Furthermore, consensus algorithms have recently found application in distributed eigenvector estimation and dictionary learning~\cite{Raja2016}.

Although it has a wide variety of engineering applications, consensus originated as a model of social interaction in the management science and statistics communities~\cite{DeGroot74} during the 1960s, according to Olfati-Saber \etal~\cite{Olfati-Saber2007}. Its application to the problem of distributed networked computation was first studied by Borkar and Varaiya~\cite{BorkarVaraiya82} and Tsitsiklis~\cite{Tsitsiklis84,TsitsiklisBertsekasAthans86}, where it was applied to distributed estimation, optimization, and decision making. Much of the early work on consensus assumed that the nodes could communicate real-valued data to one another~\cite{Frasca2008}. In realistic scenarios, the nodes must communicate within bandwidth and energy constraints, which can have a significant impact on the convergence of distributed averaging algorithms. Although many papers have been published on quantized consensus in recent years~\cite{CarliBulloZampieri2009,Chamie2014,CarliFagnaniFrascaZampieri2009,FangLi2009,AysalCoatesRabbat2008,Thanou2013,YildizScaglione2008,YildizScaglione2008b,Frasca2008,MosqueraLopezValcarceJayaweera2010,RajagopalWainwright2011,NoorshamsWainwright2011,Nedic2009,Zhu2015,KarMoura2010,Nokleby2013,HuangHua2011}, a large portion consider trade-offs among run time, communication load, and final accuracy in ways that do not take advantage of the tools of constrained optimization. This work attempts to take a more structured approach to the reduction of communication in consensus than past efforts by considering rate-distortion (RD) theory~\cite{Berger71,Cover06} and convex optimization techniques~\cite{BoydVandenberghe2004}.

\section{Summary of prior art}
\label{sec:priorart}

\subsection{Overview}
Many previous works have addressed consensus with limited coding rate. Due to the multidisciplinary nature of the subject, several approaches have been taken to address the issue of transmitting limited-precision information. These include deterministic and dithered quantization strategies using both static and dynamic quantization schemes. Because the classic linear-update consensus algorithms suffer from divergence in the presence of quantized or noisy exchanges~\cite{XiaoBoydKim2007}, many of these works focus on proposing new algorithms and proving their asymptotic convergence properties. A variety of methods has been employed to sidestep the many issues quantization poses, including adapting the quantization range, tuning the weight link sequence, using dithering or randomized quantization, and filtering the past and present state values. This section provides a survey of these works before introducing and distinguishing our approach.

\subsection{Static, dynamic, and dithered quantization schemes}
Early publications on the topic, such as the work of Xiao \etal~\cite{XiaoBoydKim2007}, show that introducing perturbations of constant variance (such as quantization error) into the traditional consensus state update prevents convergence due to the limited precision of the quantizer. Similar convergence issues preclude convergence when the quantization range is held constant~\cite{Frasca2008}. Chamie \etal~\cite{Chamie2014}, like Frasca \etal~\cite{Frasca2008}, considered the case of nonadaptive, deterministically quantized consensus, and they showed that in finite time, consensus is achieved in the sense that the network state converges to one of the quantization levels. 

To surmount the convergence issue associated with quantization, Aysal \etal~\cite{AysalCoatesRabbat2008} proposed a randomized quantization scheme, and they showed it was equivalent to dithering. With this probabilistic quantization scheme, Aysal \etal~proved that consensus can be achieved almost surely at one of the neighboring quantization values and analyzed its performance relative to unquantized consensus. In the case of networks with quantization and random link failures (i.e., time-varying topology), Kar and Moura~\cite{KarMoura2010} showed that by subjecting the communication links to a persistence condition, dithered quantized consensus can converge. They also derived probability bounds on consensus within a certain mean square error (MSE) range. Under similar topological assumptions, Nedi\'{c} \etal~\cite{Nedic2009} assessed the performance degradation resulting from quantization and provided tight polynomial bounds on the convergence time for a wide class of distributed averaging algorithms.

Due to the difficulties associated with quantization error, many works address the incorporation of dynamic encoding/decoding strategies into consensus protocols. However, many of these schemes do not explicitly consider the RD trade-off and offer certain heuristics to optimize communication performance within their proposed frameworks.

Carli \etal~\cite{CarliFagnaniFrascaZampieri2009,CarliBulloZampieri2009} assessed the performance of a ``zoom-in, zoom-out'' strategy originally studied in the context of quantized feedback control system stabilization. In this scheme, the quantization range grows in the case of saturation and shrinks otherwise. The authors demonstrated convergence for certain topologies, numbers of quantizer levels, and initial quantizer range values. Interestingly, they showed that under certain circumstances, convergence speed can be faster than in the case of ideal, unquantized transmission. 

Similarly to Carli \etal~\cite{CarliFagnaniFrascaZampieri2009,CarliBulloZampieri2009}, Rego \etal~\cite{Rego2015} proposed an algorithm with progressively shrinking quantization range and derived conditions on the design parameters to guarantee bounded steady-state error. To demonstrate the efficacy of their approach, they simulated a vehicle formation control problem.

Li \etal~\cite{Li2011} proposed a control scheme capable of guaranteeing asymptotic consensus with as little as one-bit data exchange per iteration. The authors quantified the convergence rate in terms of network properties and coding rate. To achieve convergence while transmitting a single bit at each node per iteration, the authors relied on differential encoding/decoding~\autocite[Sec.~7.2]{GershoGray1993}. Fang and Li~\cite{FangLi2009} proposed a scheme that adjusts its quantizer parameters by learning from previous runs, and they showed convergence under the assumption that the quantization error converges to zero. Similarly, Thanou \etal~\cite{Thanou2013} proposed a differential encoding strategy with exponentially decaying quantization range. Their work is based on an ``average-case'' analysis that does not provide the same strict performance guarantees as Li \etal~\cite{Li2011}, but allows for better typical performance. The numerical results demonstrate lower MSE than a number of previous works (including Li \etal~\cite{Li2011}, Fang and Li~\cite{FangLi2009}, and Carli \etal~\cite{CarliFagnaniFrascaZampieri2009,CarliBulloZampieri2009})  with equal communication load, and an error decay that matches the rate of the unquantized algorithm in the limit of many iterations. Both of these works assume a constant coding rate throughout all iterations and rely on differential encoding/decoding to achieve convergence.

Yildiz and Scaglione~\cite{YildizScaglione2008b,YildizScaglione2008,YildizScaglione2007}, unlike other authors, explicitly considered the RD trade-off to achieve an asymptotic MSE value in consensus with Gaussian states. They proposed schemes based on differential~\cite{YildizScaglione2008b}, predictive, and Wyner-Ziv coding~\cite{YildizScaglione2008,YildizScaglione2007}. In Yildiz and Scaglione~\cite{YildizScaglione2008}, the authors exploited the correlation of the current network state with that of previous iterations, and showed that bounded steady-state error is possible under their schemes using shrinking coding rates. Modeling the quantization error as an additive noise, they also provided a necessary and sufficient condition on the variance of the quantization error to ensure convergence. The main focus of the work~\cite{YildizScaglione2008} is to show that convergence can be achieved using asymptotically decreasing coding rates under predictive and Wyner-Ziv coding schemes.  In~\cite{YildizScaglione2008b}, Yildiz and Scaglione considered a simpler case of differential coding and showed similar results for this approach. Yildiz and Scaglione~\cite{YildizScaglione2008b} also imposed a parametric form on the
distortion sequence and examined the effect of varying the convergence rate on the aggregate
coding rate required to achieve a target asymptotic MSE.

\subsection{Topology and weight tuning}
Another approach to combat the effect of quantization and channel noise, and to improve energy consumption, is to tune the network topology or estimation update rule. In many of the most popular consensus algorithms, the update rule consists of a convex combination of the current node state with those of its neighbors. By adapting which nodes are considered neighbors or by altering the weights associated with each node in the neighborhood, it is possible to improve convergence properties in the presence of quantization. In the absence of quantization, Xiao and Boyd~\cite{XiaoBoyd2004} showed that the fastest linear iterative update can be found by semidefinite programming. They also listed a number of heuristics for weight design in the case of incomplete knowledge of the network topology. In their later work, Xiao \etal~\cite{XiaoBoydKim2007} considered the problem of noisy exchanges and showed that the weights corresponding to the smallest steady-state error can be found via convex optimization. 

Mosquera \etal~\cite{MosqueraLopezValcarceJayaweera2010} considered a greedy approach to updating the weight sequence by minimizing the minimum MSE (MMSE) of the estimates at each node during each iteration and proposed a modified scheme that only requires statistical knowledge about the topology. The modified scheme approximates a random geometric network by a regular graph, for which each node has the same number of neighbors. 

\subsection{Sequence filtering}
To suppress the perturbations resulting from quantization, some authors have considered the possibility of using the history of past state values. Zhu \etal~\cite{Zhu2015,ZhuSohXie2015} considered the problem of distributed parameter estimation with quantization error, proposed a scheme to reduce randomness using a moving average, and bounded its almost sure performance. Similarly, Thanou \etal~\cite{Thanou2010} considered the problem of quantized distributed averaging and demonstrated a technique to find the optimal polynomial filter coefficients to minimize the effect of quantization error. Fang and Li~\cite{FangLi2010} developed a sequence averaging approach with convergence properties that improved over Frasca \etal~\cite{Frasca2008}. In contrast to Zhu \etal~and Thanou \etal, Fang and Li only computed the average in the final iteration. These approaches both reduce the randomness introduced by quantization and accelerate the convergence of consensus algorithms.

\subsection{Wireless considerations}
In cases where the network is wireless, certain aspects of the communication medium can be exploited. Nokleby \etal~\cite{Nokleby2013} assessed the resource consumption of consensus under a wireless path-loss model. They considered total transmit energy, elapsed time, and time-bandwidth product and showed that by recursively forming geographic clusters, consensus can be achieved with nearly order-optimal performance relative to all three metrics. For comparison, they also assessed the performance of some popular gossip algorithms. Similarly, Huang and Hua~\cite{HuangHua2011} designed an energy planning algorithm for progressive estimation and consensus in multihop wireless sensor networks (WSNs). They formulated energy models based on assumptions on the wireless channels, and proposed energy optimization approaches for both types of estimation in the presence of quantization. However, Huang and Hua~\cite{HuangHua2011} only considered the simple fixed-rate uniform quantizer and did not allow the coding rate to vary over the iterations or nodes in their analysis of consensus. 

\subsection{Information theoretic approaches}
Although much of the literature considers the design of specific protocols for consensus averaging, a handful of works have explored the fundamental RD limits in the distributed computation problem. Two of these derive bounds on computation time using information theoretic inequalities~\cite{AyasoShahDahleh2010,XuRaginsky2014}, but they do not consider RD theory~\cite{Berger71,Cover06} in their analyses. Recently, Su and El Gamal~\cite{SuGamal2010} considered the problem of computing the RD function for distributed average consensus using peer-to-peer communication protocols. The authors derived a closed form for the RD function for a two-node network, and derived upper and lower bounds on the RD function for weighted-sum and gossip-based protocols in arbitrary networks. However, the authors restricted their attention to a class of protocols that uses communication between node pairs and assumes time-invariant normalized distortion. The analysis of this thesis, by contrast, assumes communication among more than one agent at each time step (corresponding to broadcast, rather than peer-to-peer, protocols), greater flexibility in the selection of edge weights, and time-varying distortions. Furthermore, Su and El Gamal did not present results on the tightness of their weighted-sum or gossip bounds, nor did they demonstrate the achievability of these bounds except in the case of a large star network with a centralized protocol.

In a similar vein, Yang \etal~\cite{YangGroverKar2016} considered RD bounds for two lossy in-network function computation scenarios. The first of these is data aggregation, in which data is routed through a tree network to a fusion center. Along the way, each node computes a partial result before communicating its result to the next node in the path. The second scenario is that of consensus, in which each node forms an estimate of the desired quantity. Although Yang \etal~\cite{YangGroverKar2016} provided bounds on the RD relationship for consensus in tree networks and proved the achievability of the derived bounds, their analysis is limited to the setting where the network is tree-structured. Often it is beneficial to consider more flexible topologies, such as random geometric graphs~\cite{Penrose2003}, which have been used to model WSNs~\cite{BoydMixingTimes}. In general, random geometric graphs and their real-world WSN counterparts have loops.

\section{Motivation and contributions}
\label{sec:contributions}
This thesis presents a framework for attaining an estimate of the network sample mean at each node, within a desired average level of accuracy, with finite run time and minimal total communication cost using either deterministic or dithered quantization. Our analysis is informed by the results of RD theory~\cite{Berger71,Cover06} and convex optimization~\cite{BoydVandenberghe2004}, which allow a more structured approach than some of the prior studies in the literature that neglect constrained optimization. The proposed cost function, which originally appeared in the work of Zhu and coauthors~\cite{ZhuBaronMPAMP2016ArXiv,ZhuBeiramiBaron2016ISIT,ZhuDissertation2017,HanZhuNiuBaron2016ICASSP,ZhuPilgrimBaron2017} is simple but capable of modeling diverse networks, from large-scale server farms and cloud services~\cite{EC2} to battery-powered WSNs~\cite{pottie2000,estrin2002}. Unlike much of the prior art, which is focused on showing asymptotic convergence for particular protocols, this thesis takes a constrained optimization approach similar to Huang and Hua~\cite{HuangHua2011} to optimizing the communication scheme for finite run time, subject to an accuracy constraint. Unlike Huang and Hua~\cite{HuangHua2011}, however, we allow time- and node-varying coding rates. We explicitly assess the trade-offs among run time, total communication load as measured by aggregate source coding rate, and quantizer complexity to attain a final MSE across the network. The focus of this work is on a Gaussian-distributed initial network state, but the results can also be used to design fixed-rate uniform quantization schemes for other distributions. Additionally, for the case of fixed-rate coding, we present a heuristic that reduces optimization complexity but scales well compared to the exact problem. The accuracy of this heuristic is verified using numerical experiments.

The advantages of our approach are ({\em i}) ignorance about the parametric form of the distortions, which allows greater flexibility in the selection of the optimal rates, ({\em ii}) support for different rates at each node and iteration of the algorithm, and ({\em iii}) optimization with respect to exact MSE quantities for finite iteration count.
  
Point ({\em i}) distinguishes our approach from Yildiz and Scaglione~\cite{YildizScaglione2008b}, who restricted their study to optimal distortion sequences that formed convergent series. Point ({\em ii}) contrasts this thesis with the work of Huang and Hua~\cite{HuangHua2011} and Thanou \etal~\cite{Thanou2013}, who required the use of a fixed-rate uniform quantizer with a rate that was constant over both nodes of the network and iterations of the algorithm. Point ({\em iii}) differentiates this thesis from both Huang and Hua~\cite{HuangHua2011} and Yildiz and Scaglione~\cite{YildizScaglione2008,YildizScaglione2008b}; Huang and Hua~\cite{HuangHua2011} optimized with respect to a bound on the MSE, and Yildiz and Scaglione~\cite{YildizScaglione2008,YildizScaglione2008b} considered the {\em asymptotic} MSE as the number of iterations goes to infinity.

\section{Organization}
The remainder of this thesis is organized as follows. Chapter 2 presents the previous results in the literature, including the basics of consensus, quantization, rate-distortion theory, the additive quantization noise model, and dithering. Therefore, Chapter 2 does not contain any original results. In Chapter 3, we present the main contributions of this thesis, namely the state-evolution model for the modified consensus iteration of Frasca \etal~\cite{Frasca2008} and the application of generalized geometric programming (GGP) to optimize the source coding in consensus. Chapter 4 presents numerical results demonstrating the performance of the GGP approach and compares its performance to some of the prior art. Finally, Chapter 5 includes some concluding remarks on the contributions of this thesis, the merits and drawbacks of the presented approach, and possible topics for future research.

\section{Notation and acronyms}
\label{sec:notation}
\subsection{Notation}
In this thesis, uppercase bold letters (e.g., $\A$) will be used to denote matrices, and lowercase bold letters (e.g., $\x$) will be used to denote vectors. Vectors that vary with time are assigned a time index, so that a time-varying $\x$ becomes $\x(t)$. Scalars that vary with time are denoted similarly (e.g., $x\tindex{t}$). A superscript on a square matrix (e.g., $\A^k, \fourspace \A \in \reals^{n \times n},$ $k$ a positive integer) denotes raising that matrix to the $k^{\text{th}}$ power (i.e, multiplying that matrix by itself $k-1$ times). Random variables (RVs) are not distinguished by notation to avoid confusing vectors and matrices.

The following list enumerates frequently used notation and variables and their associated meanings. The meaning of the following notation may not be clear until later, and it is provided here for the reader's reference.
\begin{itemize}
  \item $\{\cdot\}\tp$: transpose
  \item $Q(\cdot)$: quantization function
  \item $\I$: identity matrix
  \item $\boldsymbol{0}$: the matrix or vector of all zeros
  \item $\ones$: vector of all ones
  \item $\reals$: real numbers
  \item $\mathbb{Z}$: integers
  \item $\mathbb{Z}_{\geq 0}$: nonnegative integers
  \item $\mathbb{Z}_{> 0}$: positive integers
  \item $\E{\cdot}$: statistical expectation
  \item $\mathcal{N}\left(\boldsymbol{\mu}, \boldsymbol{\Sigma}\right)$: multivariate Gaussian distribution with mean $\boldsymbol{\mu}$ and covariance $\boldsymbol{\Sigma}$
  \item $\norm{\cdot}_p$: $\ell_p$ norm
  \item $\mathcal{G}$: undirected graph
  \item $\mathcal{V}$: vertex (node) set of a graph
  \item $\mathcal{E}$: edge set of a graph
  \item $\mathcal{N}_i$: neighborhood of node $i$
  \item $\mathcal{X}, \mathcal{Y}$: finite subsets of $\reals$
  \item $H(x)$: entropy of the random variable $x$
  \item $I(x;y)$: mutual information between the random variables $x$ and $y$
  \item $p(x)$: probability mass function (PMF) of the discrete random variable $x$
  \item $f(x)$: probability distribution function (pdf) of the continuous random variable $x$
  \item $\Phi_x(\omega)$: characteristic function of the random variable $x$
  \item $\mathcal{U}(a,b)$: uniform distribution with minimum $a$ and maximum $b$
  \item $[ \cdot ]_i$: $i^\th$ component of a vector
  \item $[ \cdot ]_{ij}$: $(i,j)^\th$ component of a matrix
  \item $\mean{\x}{t}$: mean of the random vector $\x(t)$
  \item $\Sig{\x}{t}$: covariance matrix of the random vector $\x(t)$
  \item $\tr{\cdot}$: the trace of a matrix
  \item $\lvert \cdot \rvert$: the cardinality of a set
  \item $\deg i$: the degree of node $i$ (i.e., $\lvert \mathcal{N}_i \rvert$)
\end{itemize}

\subsection{Acronyms}
Here we define a number of acronyms that will be used throughout the work.
\begin{itemize}
  \item AWGN: additive white Gaussian noise
  \item CLT: central limit theorem
  \item ECSQ: entropy-coded scalar (uniform) quantization/quantizer
  \item GP: geometric program/programming
  \item GGP: generalized geometric program/programming
  \item i.i.d.: independent and identically distributed
  \item LMMSE: linear minimum mean square error
  \item LSE: log-sum-exponential or log-sum-exp
  \item MMSE: minimum mean square error
  \item MSE: mean square error
  \item PMF: probability mass function
  \item pdf: probability density function
  \item RD: rate-distortion
  \item RV: random variable
  \item RVec: random vector
  \item SDR: signal-to-distortion ratio
  \item SNR: signal-to-noise ratio
  \item VQ: vector quantization/quantizer
  \item WSN: wireless sensor network
\end{itemize}

\chapter{QUANTIZED DISTRIBUTED AVERAGE CONSENSUS}
\label{chap-two}

This chapter presents much of the background required to understand consensus and source coding. The results presented here are not original, and they are presented to make the thesis self-contained. Our original results appear in Chapters 3 and 4.

\section{Overview}
One of the many attractive features of the variety of distributed average consensus algorithms explored in this thesis is its linearity. At each iteration, every node takes a weighted average of incoming messages from its neighbors, and in the absence of quantization errors, the performance is elegantly described using concepts from spectral graph theory~\cite{Olfati-Saber2007,XiaoBoyd2004}.

Unfortunately, quantization introduces nonlinearity into the state update, which complicates analysis. Using the additive quantization noise model~\cite{WidrowKollar2008} and dithering~\cite{Lipshitz1992}, however, it is possible to linearize the impact of quantization on the performance. The use of dithering has been found in previous work~\cite{AysalCoatesRabbat2008} to greatly simplify the design of distributed averaging algorithms, because the quantization error becomes uncorrelated with the quantizer input signal.

Although dithering and high-resolution assumptions greatly simplify the analysis of distributed average consensus, close attention must be paid to the design of the state update. If quantization errors are introduced into the lossless algorithm, then the network state will not converge to the true sample average~\cite{YildizScaglione2008,XiaoBoydKim2007}. However, by making a simple modification to the state update, Frasca \etal~\cite{Frasca2008} guaranteed convergence to within one quantization bin. In contrast, prior efforts~\cite{XiaoBoydKim2007,YildizScaglione2008,YildizScaglione2008b} could only guarantee bounded asymptotic mean square error (MSE) due to the drift from the sample average in the presence of quantization error.

In this chapter, we explore the fundamentals of distributed average consensus, algorithmic modifications to account for quantization effects, rate-distortion (RD) theory, and the basics of the additive noise model for quantization error.

\section{Mathematical preliminaries}
Before introducing the consensus problem, it is first necessary to introduce a number of mathematical concepts related to linear algebra and graph theory.

\subsection{Matrix and graph theory concepts}
To represent the network of interest, it is necessary to model the nodes, which represent the computing elements or agents. These can be wireless sensors~\cite{Nokleby2013,HuangHua2011}, servers~\cite{Raja2016}, cameras~\cite{Tron2011}, or robots~\cite{Olfati-Saber2007}. We assume that each node can only communicate with a subset of the other nodes of the network, so it is also necessary to model the presence or absence of communication links between them, which can be wireless channels or wired connections. These relationships are modeled by a graph~\cite{XiaoBoyd2004}. 

In this thesis, the communication links are bidirectional connections, so we model the network as an undirected graph. This graph can be represented in a number of ways. One of the simplest representations of the graph is a pair of sets, 
\begin{align}
\label{eq:G-definition}
\mathcal{G} = \{\mathcal{V}, \mathcal{E}\},
\end{align}
where the graph $\mathcal{G}$ is comprised of a set of vertices (nodes) $\mathcal{V}$ and a set of edges $\mathcal{E}$ between pairs of vertices~\autocite[Sec. 1.2]{GraphsOptimizationAndAlgos}. Because the communication links are bidirectional, each edge $\{i,j\} \in \mathcal{E}$ is represented as an unordered pair of vertices $i$ and $j$~\cite{XiaoBoyd2004}.

\section{The consensus problem}
In the simplest case of the consensus problem, each node $i \in \{1,\ldots,m\}$ has an initial scalar quantity 
\begin{align}
\label{eq:scalar-state-def}
z_i(0) \in \reals,
\end{align} and the goal is to have all nodes of the network agree upon the sample mean of these quantities by iteratively exchanging messages with their neighbors~\cite{XiaoBoyd2004}. The quantities $z_i(t)$ will be referred to as ``states,'' which in this thesis are assumed to be real-valued scalar random variables (RVs) with known distribution. More formally, let the (discrete) iteration index be a nonnegative integer,\footnote{We denote the positive subset of a set $\mathcal{S}$ by $\mathcal{S}_{>0}$. The nonnegative subset is similarly denoted $\mathcal{S}_{\geq 0}$. The integers are denoted by $\mathbb{Z}$, and the real numbers by $\reals$.} $t \in \mathbb{Z}_{\geq 0}$. At $t=0$, the states $\{z_i(t)\}_{i=1}^m$ are the initial values to be averaged by the consensus algorithm. For $t \geq 1$, the state $z_i(t)$ represents the estimate of the sample average 
\begin{align}
\widebar{z} := \frac{1}{m} \sum_{i=1}^m z_i(0)
\end{align}
at node $i$. The objective of consensus is for the state $z_i(t)$ to eventually equal the sample mean of the initial states~\cite{XiaoBoyd2004}. Mathematically, this is expressed as $\lim_{t\rightarrow\infty} z_i(t) = \widebar{z}$, $\forall i \in \{1,\ldots,m\}$~\cite{XiaoBoyd2004}. We leave the discussion of vector-valued states for the following chapter. In this thesis, we restrict our attention to deterministic, synchronous-update consensus algorithms. We assume the following: ({\em i}) the communication link topology of the network is fixed and does not change with time, ({\em ii}) at each iteration, every node of the network exchanges messages with only its neighbors, and ({\em iii}) the communication channels between nodes are noiseless.

Given the above assumptions on communication, one of the most popular algorithms for consensus relies on linear updates~\cite{XiaoBoyd2004,Olfati-Saber2007}. Each node is assigned an index $i \in \{1,\ldots,m\}$. Let $z_i(t)$ denote the state of the $i^{\text{th}}$ node at iteration $t$. Each node updates its state by taking a weighted sum of its own state with those of its neighbors~\cite{XiaoBoyd2004},
\begin{align}
\label{eq:node-update}
z_i(t+1) &= w_{ii} z_i(t) +  \sum_{j \in \mathcal{N}_i} w_{ij} z_j(t) ,
\end{align}
where $w_{ij} > 0$ $\forall i,j$, $\sum_{k=1}^m w_{ik} = 1$, and $\mathcal{N}_i$ denotes the neighborhood of node $i$,
\begin{align}
\mathcal{N}_i := \{ j \fourspace | \fourspace \{i,j\} \in \mathcal{E} \}.
\end{align} 

The degree of node $i$ is defined as
\begin{align}
\deg i := | \mathcal{N}_i |,
\end{align}
where $|\cdot|$ represents set cardinality. By this definition, $\deg i$ is the number of neighbors of node $i$.
The weights $w_{ij}$ are designed such that~\cite{XiaoBoyd2004}
\begin{align}
\lim_{t\rightarrow \infty} z_i(t) = \widebar{z}.
\end{align}
If the state of each node of the network is collected in a vector 
\begin{align}
\label{eq:z-definition}
\z{t} := [z_1(t), \ldots, z_m(t)]\tp,
\end{align} 
and the averaging weights $w_{ij}$ are collected in a matrix,
\begin{align}
\label{eq:W-definition}
\left[\W{}\right]_{ij} := w_{ij},
\end{align} 
then the above update equation~\eqref{eq:node-update} can be written in matrix-vector form as~\cite{XiaoBoyd2004}
\begin{align}
\label{eq:update-mv}
\z{t+1} = \W{} \z{t} .
\end{align}
The design of the weight matrix $\W{}$ that yields the fastest convergence is a well studied problem; the interested reader is referred to Xiao and Boyd~\cite{XiaoBoyd2004}. To converge asymptotically, that is,
\begin{align}
\lim_{t \rightarrow \infty} \z{t} = \frac{1}{m} \ones \ones \tp \z{0} = \widebar{z}\,\ones,
\end{align}
the weight matrix must be doubly stochastic, $\sum_i w_{ij} = \sum_j w_{ij} = 1$, and the modulus of its largest eigenvalue must be less than unity~\cite{XiaoBoyd2004}.

\subsection{Lossy consensus}
By introducing quantization error into the internode messages, the simple linear iteration above~\eqref{eq:update-mv} is not guaranteed to converge. Instead, we use the modified iteration proposed by Frasca \etal~\cite{Frasca2008}, which allows the sample average to be preserved in the presence of quantization error. 

Let $Q: \reals^m \to \mathcal{X}^m$ represent quantization to a finite set of representation levels $\mathcal{X}^m \subset \reals^m$ (i.e., $\q{\z{t}} = [Q_1(\zmarg{1}{t}), \ldots, Q_m(\zmarg{m}{t})]\tp$). The subscripts on $Q$ indicate that each node can use a different quantizer in general. State updates for the case of consensus with quantized messages have been studied by a number of authors, as discussed in Chapter~\ref{chap-one}. In this thesis, we use the update proposed by Frasca \etal~\cite{Frasca2008}, which is
\begin{align}
\label{eq:frasca-node-update}
z_i(t+1) = z_i(t) + \sum_{j=1}^m w_{ij} \left(Q_j(z_j(t))-Q_i(z_i(t)) \right),
\end{align}
or in matrix-vector form,
\begin{align}
\label{eq:lossy-iteration}
\z{t+1} &= \z{t} + \left(\W{} - \I\right)\q{\z{t}} .
\end{align} 
The key advantage of this update is that the average $\frac{1}{m} \sum_{i=1}^m z_i(t)$ of the states $z_i(t)$ is preserved at each step $t$, despite the presence of quantization error~\cite{Frasca2008}. Defining the quantization error
\begin{align}
  \label{eq:qe-definition}
  \qes{t} := \q{\z{t}}-\z{t},
\end{align}
the preservation of the sample mean can be seen by rewriting~\eqref{eq:lossy-iteration} as
\begin{align}
\z{t+1} &= \z{t} + \left( \W{} - \I \right) (\z{t} + \qes{t}) \nonumber \\
&= \W{} \z{t} + \left( \W{} - \I \right) \qes{t} , \label{eq:lossy-alt}
\end{align}
and then taking the sample mean~\cite{Frasca2008},
\begin{align}
\frac{1}{m} \ones \tp \z{t+1} = \frac{1}{m} \ones \tp \W{} \z{t} + \frac{1}{m} \ones \tp \left( \W{} - \I \right) \qes{t}.
\end{align}
Because $\W{}$ is doubly stochastic, $\ones \tp \W{} = \ones \tp$, and the mean at iteration $t+1$ is equal to the mean at iteration $t$~\cite{Frasca2008}:
\begin{align}
\frac{1}{m} \sum_{i=1}^m \zmarg{i}{t+1} &= \frac{1}{m} \ones \tp \z{t+1} \\
&= \frac{1}{m} \ones \tp \W{} \z{t} + \frac{1}{m} \ones \tp \left( \W{} - \I \right) \qes{t} \\
&= \frac{1}{m} \ones \tp \z{t} + \frac{1}{m} \underbrace{\left( \ones \tp - \ones \tp \right)}_{= \mathbf{0}} \qes{t} \\
&= \frac{1}{m} \ones \tp \z{t} = \frac{1}{m} \sum_{i=1}^m \zmarg{i}{t} \label{eq:mean-pres-end}.
\end{align}
In this thesis, we model the quantization error $\qes{t}$ as additive noise.

Note that the target state of consensus, termed the {\em average consensus state}, can be written,
\begin{align}
\mathbf{z}^* := \widebar{z}\ones,
\end{align}
which means that $z_i(t) = \widebar{z}$, $\forall i \in \{1, \ldots, m\}$. We also define the {\em average consensus operator} $\avg$, so that~\cite{XiaoBoyd2004}
\begin{align}
\avg \z{0} = \mathbf{z}^*.
\end{align}

Defining the error from the true sample mean,
\begin{align}
\diff{t} &:= \z{t} - \mathbf{z}^* \nonumber \\
&= \z{t} - \avg \z{0},  \label{eq:e-definition}
\end{align} 
and noting that the average is preserved over the iterations, (i.e., $\frac{1}{m} \ones \tp \z{t} = \frac{1}{m} \ones \tp \z{t+1}$), the error $\diff{t}$ can be expressed as~\cite{Frasca2008}
\begin{align}
\diff{t} &= \M \z{t}.
\label{eq:error-from-iterate}
\end{align}

\section{Source coding, quantization, and rate-distortion theory}
Digital systems rely on the ability to transmit and store information. To perform these tasks reliably, the information takes the form of strings of symbols belonging to some finite alphabet~\cite{Cover06}. As a cornerstone of digital communication and storage, coding has been studied quite extensively. In this section, we present some of the fundamental coding and quantization concepts required to understand this thesis.

\subsection{Coding and information theory}
Understanding data compression requires a few concepts from information theory~\cite{Cover06}, initially developed by Shannon in the 1940s~\cite{Shannon48}. This theory offers insight into the fundamental limiting performance of communication and compression systems~\cite{Cover06}. In this subsection, we focus on the case where the source to be encoded, $\x \in \reals^n$, is mapped to a point $\hat{\mathbf{x}}$ in some finite set $\mathcal{X}^n$ by the quantization operator $\q{\cdot}$.

The {\em entropy} of a scalar RV $x \in \mathcal{X}$ with probability mass function (PMF) $p(x)$ is given by~\cite{Cover06} 
\begin{align}
H(x) := \fourspace - \hspace{-6pt} \sum_{x \in \mathcal{X}} p(x) \log_2{p(x)}.
\end{align}
Intuitively, the entropy of an RV is a measure of its uncertainty or information content~\cite{Cover06}. Let $\x \in \reals^n$ represent a long sequence $x$, the entries of which are independent and identically distributed (i.i.d.) according to the probability mass function (PMF) $p(x)$. Then $n H(x)$ is the minimum expected binary sequence length required to describe $\x$ without error~\cite{Cover06}. That is, if we wish to describe the sequence $\x$ by a string of binary digits $\tilde{\x} \in \{0,1\}^M$, then a code exists such that the original sequence $\x$ can be noiselessly reconstructed from $\tilde{\x}$ provided that $\E{M} \geq n H(x)$~\cite{Cover06}. To extend this concept to sequences of arbitrary length, we define the {\em coding rate} per symbol as $\frac{1}{n}\E{M}$, which is the average number of bits used to describe a single source symbol~\cite{Cover06}.


Another important quantity in information theory is the mutual information between two RVs. Let $x$ and $y$ be RVs with joint PMF $p(x,y)$ and marginal PMFs $p(x)$ and $p(y)$, respectively.\footnote{This is an abuse of notation: it would be more correct to write $p_x(x)$ and $p_y(y)$, indicating that the two functions are, in general, different. Here the subscripts are omitted for simplicity.} Then the {\em mutual information} between $x \in \mathcal{X}$ and $y \in \mathcal{Y}$ is given by~\cite{Cover06}
\begin{align}
I(x;y) := \sum_{x \in \mathcal{X}} \sum_{y \in \mathcal{Y}} p(x,y) \log_2 \frac{p(x,y)}{p(x)p(y)}. 
\end{align}
The utility of these information-theoretic quantities will become apparent during the discussion of RD theory.

\subsection{Rate-distortion theory}
\label{sec:RD-theory}
The previous discussion of coding considered only RVs that take on a finite set of values. If we wish to digitally communicate or store a continuous source, it must first be quantized~\autocite[Ch. 1]{GershoGray1993}. The two most basic elements of a quantizer are a set of {\em representation levels}, which are used to approximate the unquantized signal, and a set of {\em decision thresholds}, which determine the mapping from the input set to the output set~\autocite[\ppno~133--135]{GershoGray1993}. If we imagine the source data as an RV $\x \in \reals^n$, then the quantizer $Q: \reals^n \rightarrow \mathcal{X}^n$ maps $\x$ to one of finitely many representations $\hat{\x} \in \mathcal{X}^n$~\cite{GershoGray1993}. 
\begin{figure}
  \begin{center}
    \includegraphics[width=0.8\linewidth]{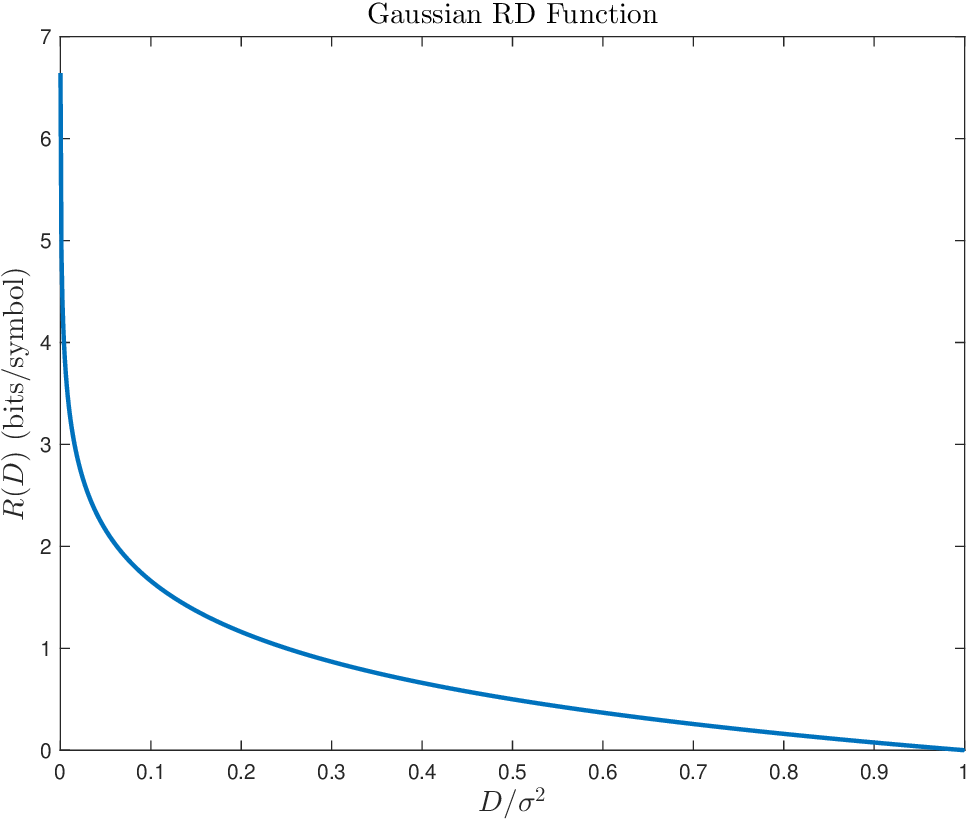}
    \caption{The RD function $R(D)$ of a Gaussian memoryless source. The function is a nonincreasing, convex function of the normalized distortion $\frac{D}{\sigma^2}$, where $\sigma^2$ is the variance of the Gaussian source. The RD pairs above the curve are achievable, while those below the curve are not.}
    \label{fig:gauss-RD}
  \end{center}
\end{figure}
For real-valued sources, quantizers necessarily introduce a certain expected distortion $D$ into their representation of the input signal~\autocite[\ppno~144--145]{GershoGray1993}. This distortion can be quantified using a number of metrics, but for the purpose of this thesis, we use the square error metric~\cite{GershoGray1993} 
\begin{align}
d(\x,\hat{\x}) = \norm{\x - \hat{\x}}_2^2,
\end{align} 
so that the expected distortion per entry or {\em per letter} is given by $D = \frac{1}{n} \E{\norm{\x-\hat{\x}}_2^2}$. In general, using a higher coding rate $R$ results in a lower distortion $D$, with the drawback of greater communication load. RD theory~\cite{Berger71,Cover06} quantifies the best possible trade-off between coding rate and distortion using the tools of information theory.

Assume that we wish to encode a memoryless source, which is represented by the random vector (RVec) $\x \in \reals^n$, $x_i \eightspace \text{i.i.d.} \sim f(x_i), \forall i \in \{1,\ldots,n\}$. For the remainder of this subsection, we drop the indices on $x_i$ due to the assumption that all entries of $\x$ are i.i.d. The minimum coding rate $R$ required to produce an expected distortion less than or equal to a particular value $D$ is given by the so-called RD function $R(D)$~\cite{Berger71,Cover06},
\begin{align}
R(D) := \min_{p(\hat{x}|x):\E{d(x,\hat{x})}\leq D} I(x;\hat{x}).
\end{align}
\begin{figure}
  \begin{center}
    \includegraphics[width=0.8\linewidth]{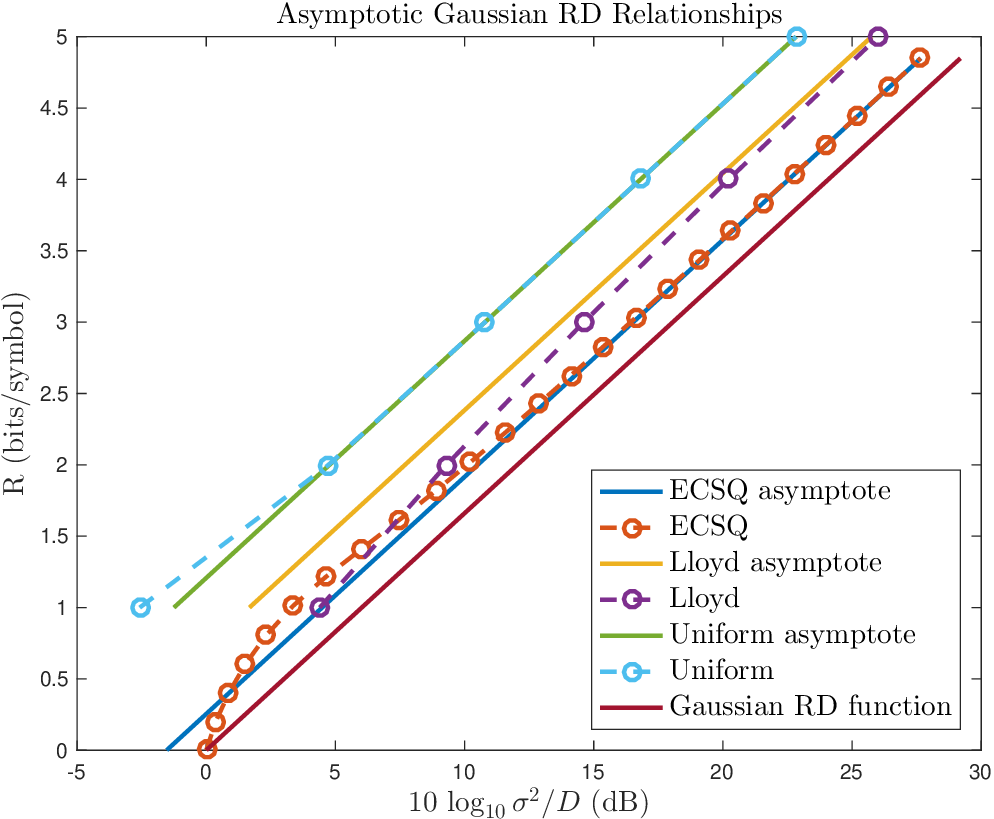}
    \caption{Operational RD relationships of practical quantizers: ECSQ, Lloyd-Max, and fixed-rate uniform quantization. Note that all of these approach an asymptote that is a constant offset from the Gaussian RD function as $R \rightarrow \infty$. The ECSQ performance corresponds to the uniform scalar quantizer with a level at zero. The representation level placement may be suboptimal, but it allows for coding rates below one bit per symbol. This figure was inspired by a similar plot in lecture notes from a TU Berlin source coding course~\cite{TU-Berlin-lecture}.}
    \label{fig:quantizer-asymptotes}
  \end{center}
\end{figure}
In words, the RD function is the minimum of the mutual information over all possible ``test channels'' $p(\hat{x}|x)$, subject to the constraint that the expected distortion per entry $\E{d(x,\hat{x})}$ is less than a specified value $D$~\cite{Cover06}. Operationally, the RD function is the minimum number of bits per symbol required to describe a long i.i.d. source within the prescribed distortion $D$~\cite{Cover06}. The computation of a closed form for $R(D)$ is difficult in general; however, $R(D)$ can be computed numerically~\cite{Arimoto72,Blahut72,Rose94}. When a particular quantizer is used, it will often have an RD trade-off curve that differs from $R(D)$, which is a bound on the best possible performance~\cite{GershoGray1993}. In this thesis, we term such a trade-off curve for a particular practical quantizer an {\em operational RD relationship} to avoid ambiguity. An example RD function for the Gaussian case is given in Figure~\ref{fig:gauss-RD}. Figure~\ref{fig:quantizer-asymptotes} shows the performance of practical quantizers compared to both the Gaussian RD function and their respective operational RD relationships. Note that the signal-to-distortion ratio (SDR),
\begin{align}
  \text{SDR} := \frac{\var{x}}{D},
\end{align} 
increases approximately exponentially in the high-rate limit for all practical quantizers in Figure~\ref{fig:quantizer-asymptotes}. This exponential growth is a common feature of Gaussian operational RD relationships that will be exploited in the following chapter.

\subsection{Uniform and nonuniform quantization}
From a conceptual standpoint, one of the simplest quantizers is the familiar {\em uniform scalar quantizer}, for which both the representation levels and the decision thresholds are uniformly spaced~\cite{GershoGray1993}. For a given input $\x$, a uniform scalar quantizer simply rounds each $x_i$ to the representation level $\hat{x}_i$ nearest to $x_i$~\cite{GershoGray1993}. This type of quantizer is frequently encountered in analog-to-digital conversion and various digital signal processing systems. Despite its simplicity, this quantizer performs surprisingly well in certain scenarios.

Consider a source $\x \in \reals^n$ that is not uniformly distributed. For a uniform scalar quantizer, each representation level $\hat{x}_i$ will occur with different probability. In this case, we can choose the representation levels based on the input statistics to obtain a low-distortion quantizer~\cite{GershoGray1993}. This idea was explored by Lloyd~\cite{Lloyd82} and Max~\cite{Max60}, who developed necessary and sufficient conditions for quantizer optimality, given a fixed number of representation levels and a certain source distribution. Lloyd and Max also developed algorithms for achieving an efficient quantizer by iteratively updating the thresholds and representation levels~\cite{GershoGray1993}. Because the quantizer is optimized for a particular number of representation levels, Lloyd-Max quantizers are useful for fixed-length codes, where the encoded sequence $\tilde{\x} \in \{0,1\}^{nR}$. However, more sophisticated coding techniques, such as Huffman coding~\cite{Huffman52}, can be used to approach the previously discussed entropy lower bound on code length if the length $M$ of the binary code sequence $\tilde{\x} \in \{0,1\}^M$ is allowed to vary with the input~\cite{GershoGray1993}.

In the case of variable-length codes, we encounter a surprising result. For certain source distributions, uniform scalar quantization followed by entropy coding, called entropy-coded scalar quantization (ECSQ), is the optimal scalar quantizer asymptotically as $R \rightarrow \infty$~\cite{GishPierce68,GershoGray1993}. In fact, for the memoryless Gaussian source, ECSQ has been proved to approach the RD limit within 0.255 bits as $R \rightarrow \infty$~\cite{GershoGray1993,GrayNeuhoff1998}. For many memoryless sources, the performance of ECSQ is similar~\cite{GershoGray1993,GrayNeuhoff1998}: Farvardin and Modestino~\cite{FarvardinModestino84} showed less than 0.3 bits redundancy over $R(D)$ in the worst case for all the memoryless distributions they studied. Similarly, the operational RD relationships for Lloyd-Max quantization and fixed-rate uniform quantization approach asymptotes that are a constant offset from the $R(D)$ function in the limit $R \rightarrow \infty$~\cite{GershoGray1993}. The asymptotic offset is illustrated in Figure~\ref{fig:quantizer-asymptotes}.

The curious reader might wonder how the optimal quantizer can differ from the bound if the bound is, in fact, achievable. Better performance can be achieved by quantizing long sequences of the source jointly. Interestingly, this is advantageous even if the source symbols are independent~\cite{Cover06}. This technique is termed {\em vector quantization (VQ)}~\cite{GershoGray1993}.

Because of the performance improvement associated with VQ, we also consider the multivariate extension of the scalar quantizer, which is the lattice quantizer~\cite{GershoGray1993}. These quantizers approach the RD lower bound for memoryless Gaussian sources~\cite{ZamirFeder96} as the block dimension approaches infinity, while also permitting the application of dithering, which we will discuss in the following sections.

\subsection{The quantization noise model and dithering}
One of the nicest properties of distributed averaging algorithms is their linearity~\cite{Frasca2008}. At each iteration, a weighted sum of the incoming messages is computed at each node~\cite{XiaoBoyd2004,Olfati-Saber2007}. This allows us to exploit the properties of linear operators, and, for Gaussian-distributed initial states, we can describe the entire network state statistics using linear algebra~\cite{YildizScaglione2008}. However, quantization is inherently nonlinear, which complicates the analysis~\cite{Frasca2008}. Luckily, under certain conditions, the quantization error can be modeled as an additive noise that is uncorrelated with the source~\cite{SripadSnyder77,Lipshitz1992,WidrowKollar2008}.

In undithered quantization systems, the conditions for i.i.d. uniform quantization error are rather restrictive~\cite{Lipshitz1992}. In most cases, these conditions are not met exactly, and the additive quantization noise model is used as an approximation to the true behavior that simplifies system design~\cite{WidrowKollar2008}. 
In general, the approximation resulting from assuming the additive quantization noise model improves with higher coding rates~\autocite[Sec.~9.5]{WidrowKollar2008}.

More precisely, let $\x \in \reals^n$ represent a vector to be quantized. The quantizer $\q{\cdot} : \reals \rightarrow \mathcal{X}^n$ maps the values of $\x$ to a set of representation levels $\hat{\x} \in \mathcal{X}^n$~\cite{GershoGray1993}. Defining the quantization error,
\begin{align}
\qesu := \hat{\x} - \x,
\end{align}
the quantization noise model can, for our purposes, be summarized by the following three relationships: ({\em i}) the quantization error is zero-mean,
\begin{align}
\label{eq:q-zero-mean}
\E{\qesu} = \mathbf{0},
\end{align} 
({\em ii}) the quantization error is uncorrelated with the source,
\begin{align}
\label{eq:q-uncorrelation}
\E{ \qesu \x\tp} = \mathbf{0},
\end{align}
and ({\em iii}) the quantization error entries $\left[ \qesu \right]_i$ and $\left[\qesu \right]_j$, $i \neq j$, are uncorrelated,
\begin{align}
\label{eq:q-independence}
\E{\qesu \qesu\tp} = \frac{b^2}{12} \I,
\end{align}
where $b$ is the quantizer bin size and $\I$ is the identity matrix~\cite{WidrowKollar2008}.

In situations that discourage or forbid the use of high rates, a technique called {\em dithering} can be used to cause the quantization error to be uncorrelated with the source~\cite{Lipshitz1992}. Dithering is the process of intentionally adding noise to the source signal $\x \in \reals^n$ to randomize the quantization error, which is otherwise a deterministic function of the quantizer input~\cite{Lipshitz1992}. If the i.i.d. noise $\w \in \reals^n$ satisfies certain technical conditions, and the receiver subtracts $\w$ from the reconstructed signal $\q{\x}$, then the previously discussed additive quantization noise model becomes exact in the sense that~\cite{Lipshitz1992}
\begin{align}
\E{(\q{\x+\w}-(\x+\w)) \fourspace \x \tp} = \mathbf{0}
\end{align}
and
\begin{align}
\E{(\q{\x+\w}-(\x+\w))(\q{\x+\w}-(\x+\w))\tp} = \frac{b^2}{12} \I.
\end{align} 
Therefore, the quantization error can be modeled as an additive white noise with variance $\frac{b^2}{12}$, which is uncorrelated with the input signal $\x$~\cite{Lipshitz1992}. The subtraction of the dither can be achieved in practice using a particular pseudorandom number generator with a seed that is known to both the transmitter and receiver.

Perhaps the simplest dither signal that causes the quantization error to be independent of the source and spectrally white is the uniform distribution $\mathcal{U}(-\frac{b}{2}, \frac{b}{2})$~\cite{Lipshitz1992}. Uniformly distributed dither $\mathbf{w} \sim \mathcal{U}(-\frac{b}{2},\frac{b}{2})$ is used throughout this thesis whenever dithering is mentioned.

\chapter{CODING RATE OPTIMIZATION IN CONSENSUS}
\label{chap-three}

Depending on the system of interest, communication may be more or less expensive relative to run time~\cite{ZhuBaronMPAMP2016ArXiv}. For instance, consider two extremes of distributed computing: wireless sensor networks (WSNs)~\cite{pottie2000,estrin2002} and multiserver networks~\cite{EC2}. WSNs have limited battery life and must communicate over wireless channels, which consumes a large amount of power. In many WSN applications, communication is expensive due to the low-power constraints at each node~\cite{ZhuBaronMPAMP2016ArXiv}. By contrast, multiserver networks, such as those used in cloud services~\cite{EC2} have less stringent communication requirements, and the run time is more significant to the user~\cite{ZhuBaronMPAMP2016ArXiv}. In these networks, the nodes communicate over wired connections and do not have battery power constraints. To produce an optimization that is useful in diverse distributed settings, then, we pursue a general and simple approach that models the relative costs of communication and computation. 

The contents of this chapter constitute the main analytical results of this thesis. The contributions of this chapter are twofold: ({\em i}) a set of state evolution equations for the network's sufficient statistics are derived under the assumption of initial Gaussian distribution and additive white quantization error, and ({\em ii}) it is proven that the problem of minimizing the aggregate rate required to achieve a certain mean square error (MSE) can be posed as a generalized geometric program (GGP). 
  
The first contribution is minor, since similar approaches to computing the network state statistics were derived in Yildiz and Scaglione~\cite{YildizScaglione2008} and Huang and Hua~\cite{HuangHua2011}; however, our derivation allows for the explicit computation of MSE expressions in closed form using the iteration~\eqref{eq:frasca-node-update} introduced in Frasca \etal~\cite{Frasca2008}. Yildiz and Scaglione~\cite{YildizScaglione2008} only derived a closed form for the asymptotic MSE as the iteration index $t \rightarrow \infty$, and Huang and Hua optimized with respect to an MSE bound. Our analysis allows an exact optimization to be performed in the finite-iteration (total iterations $T < \infty$) case.

The second contribution allows the problem to be formulated in a well-understood framework. The GGP model can be converted into convex form and efficiently solved using common numerical techniques~\cite{BoydVandenberghe2004}. Furthermore, due to the convexity of the equivalent transformed optimization problem, the solver is guaranteed to find the globally optimal solution~\cite{BoydVandenberghe2004}. In addition to the GGP model, we present a heuristic approach for optimization of fixed-rate source coding schemes. This model reduces the size of the optimization problem, which still provides the same solutions as exhaustive search.

\section{Problem formulation}

\label{sec:minimization}

At every iteration $t \in \{0,\ldots,T-1\}$ of the consensus process, each node $i \in \{1,\ldots,m\}$ uses a coding rate $R_i(t)$ to encode its state for transmission to the neighboring nodes.\footnote{The choice of zero as the initial iteration is arbitrary.} In general, $R_i(t)$ can vary across both nodes and iterations. In practice, we want to terminate the consensus process after a finite number of iterations $T$. We simplify notation by defining the {\em coding rate vector}
\begin{align}
\label{eq:r-vector}
\Rhat := \left[R_1(0), \ldots, R_m(0), \ldots, R_1(T-1), \ldots R_m(T-1) \right]\tp.
\end{align}
We denote the expected distortions per entry incurred by transmitting states using the coding rates $\Rhat$ by the {\em distortion vector}
\begin{align}
\label{eq:d-vector}
\Dv := \left[D_1(0), \ldots, D_m(0), \ldots, D_1(T-1), \ldots D_m(T-1) \right]\tp.
\end{align}
One key quantity we use to determine the cost of running the consensus process is the {\em aggregate coding rate}~\cite{ZhuBaronMPAMP2016ArXiv,ZhuPilgrimBaron2017}:
\begin{align}
\label{eq:ragg-def}
R_{\text{agg}} := \sum_{t=0}^{T-1} \sum_{i=1}^m R_i(t),
\end{align}
which represents the total coding rate used over the $T$ iterations of the consensus algorithm by all $m$ nodes of the network.
To address the trade-off between communication and computation, we borrow the following composite cost function from Zhu and coauthors~\cite{ZhuBaronMPAMP2016ArXiv,ZhuBeiramiBaron2016ISIT,ZhuDissertation2017,HanZhuNiuBaron2016ICASSP,ZhuPilgrimBaron2017} that models both communication and run-time costs:
\begin{align}
\label{eq:cost}
C_{\text{total}}(\Rhat, T) := K_1 R_{\rm agg} + K_2 T,
\end{align}
where $K_1$ and $K_2$ represent the costs of communication and computation, respectively. Since the iteration \eqref{eq:lossy-iteration} consists only of a quantization and a weighted sum at each node, the total run-time cost is modeled as linear in the number of iterations. This cost function can be used to model both battery-powered sensor networks and server networks by varying the relative values of $K_1$ and $K_2$~\cite{ZhuBaronMPAMP2016ArXiv,ZhuPilgrimBaron2017}.

In this chapter, we present minimization strategies of the subcost 
\begin{align}
\label{eq:comm-cost}
C_{\text{comm}}(\Rhat,T):= K_1 R_{\text{agg}}
\end{align}
for fixed- and variable-length codes for ({\em i}) Gaussian-distributed sources using a variety of quantizers, and ({\em ii}) arbitrarily distributed sources using a specific fixed-rate uniform quantizer. For both proposed schemes, the overall solution to the minimization of \eqref{eq:cost} can be found by fixing $T$ and minimizing the communication cost $C_{\text{comm}}(\Rhat,T)$, subject to the constraint that $\Rhat$ achieves a target mean square error (MSE) value. Repeating this procedure for $T\in\{T_{\text{min}},\ldots,T_{\text{max}}\}$ is equivalent to minimizing over $T$ and $\Rhat$ jointly.

To efficiently encode the data stored across the network, it is necessary to understand the evolution of its distribution over time. In general, this is not tractable due to the need to track dependences among source elements and transformations of source densities. Instead, we propose an efficient optimization scheme for entropy-coded uniform scalar quantization (ECSQ)~\cite{GershoGray1993} of stationary Gaussian states and rate-distortion-optimal (RD-optimal) vector quantization (VQ)~\cite{Berger71,Cover06} of memoryless Gaussian-distributed initial states.\footnote{For the scalar quantization schemes, we assume that the state of each node is a sample from a stationary, ergodic Gaussian random process. The scalar quantizer acts elementwise, so we expect the RD performance of correlated sources with scalar quantization and entropy coding to be the same as the memoryless RD performance.} We also use the resulting strategies to design a coding rate optimization heuristic for a particular dithered fixed-rate uniform quantizer~\cite{Lipshitz1992,WidrowKollar2008}, for which the expected distortion is the same regardless of source distribution~\cite{Lipshitz1992}.

In this thesis, we assume that the node-to-node communication is taking place in a {\em broadcast} manner~\cite{YildizScaglione2007}, which means that the total rate expended by a transmitting node is independent of the size of its neighborhood. This also simplifies the analysis of the introduced distortion, because each neighbor of a particular node will receive the same quantized message.

\section{Abstractions for vector-valued data}
\label{sec:abstractions}
The concepts presented in this section are included for clarity. We will later present simplifications for the case that each node has a stationary Gaussian source. Although not used directly in our optimization model, the mathematics presented in this section is useful for computing the source statistics across the network in a general setting.

To model the vector-valued data stored at each node during the averaging process, a useful abstraction is introduced, based on similar approaches by Huang and Hua~\cite{HuangHua2011}. Let each node $i \in \{1, \ldots, m\}$ at iteration $t \in \{0, \ldots, T-1\}$ of consensus have state $\zv{i}{t} \in \reals^L$. We represent the $j^\th$ element of the state $\zv{i}{t}$ as 
\begin{align}
\label{eq:zvmarg-definition}
\zvmarg{i}{j}{t} := \left[\zv{i}{t} \right]_j .
\end{align} 
These states are Gaussian random vectors (RVecs) that may be correlated, so that $\E{\zvmarg{i}{k}{t} \zvmarg{j}{l}{t}} \neq 0$ in general.

To model the state of the entire network, we define the {\em network state supervector} $\ze{t} \in \reals^{mL}$ as follows:
\begin{align}
\label{eq:network-state-def}
\ze{t} := \left[ \zvmarg{1}{1}{t} \cdots \zvmarg{m}{1}{t} \spacedvert \zvmarg{1}{2}{t} \cdots \zvmarg{m}{2}{t} \spacedvert \cdots \spacedvert \zvmarg{1}{L}{t} \cdots \zvmarg{m}{L}{t} \right]\tp .
\end{align}
During consensus, each partition of $\ze{t}$ in~\eqref{eq:network-state-def} is averaged independently (i.e., the vector states are averaged over the network nodes elementwise). To represent this elementwise action, we use the {\em direct sum} operator for matrices~\cite{DirectSum}:
\begin{align}
\bigoplus_{i=1}^n \mathbf{A}_i := \begin{bmatrix} 
\A_1 & \mathbf{0} & \cdots & \mathbf{0} \\ 
\mathbf{0} & \A_2 & \cdots & \mathbf{0}  \\ 
\vdots & \vdots & \ddots & \vdots \\
\mathbf{0} & \mathbf{0} & \cdots & \A_n
\end{bmatrix}.
\end{align}
Therefore, the operator $\bigoplus$ creates a block-diagonal matrix from the matrices $\{\A_i\}_{i=1}^n$. Note that $\bigoplus_{i=1}^n \A_i$ is shorthand for $\A_1 \oplus \A_2 \oplus \dots \oplus \A_n$. This is an alternative to the Kronecker product notation and the matrix structure introduced by Huang and Hua~\cite{HuangHua2011}.

Defining the weighted averaging matrix $\Om \in \reals^{mL\times mL}$:
\begin{align}
\label{eq:Omega-definition}
\Om := \bigoplus_{l=1}^L \W{}
\end{align}
as the operator equivalent to the weight matrix $\W{}$~\eqref{eq:W-definition} for vector consensus, the state update equivalent to~\eqref{eq:lossy-iteration} for vector consensus is
\begin{align}
\label{eq:vector-iteration}
\ze{t+1} &= \ze{t} + \OsubI \q{\ze{t}},
\end{align}
where 
\begin{align}
Q(\ze{t}) := \left[Q_1(\zvmarg{1}{1}{t}) \cdots Q_m(\zvmarg{m}{1}{t}) \spacedvert Q_1(\zvmarg{1}{2}{t}) \cdots Q_m(\zvmarg{m}{2}{t}) \spacedvert \cdots \spacedvert Q_1(\zvmarg{1}{L}{t}) \cdots Q_m(\zvmarg{m}{L}{t}) \right]\tp.
\end{align}
As in the scalar consensus case, we define the quantization error,
\begin{align}
\label{eq:quant-err-def}
\qe{t} := \q{\ze{t}} - \ze{t}.
\end{align}
This is the vector consensus analog to $\qes{t}$~\eqref{eq:qe-definition}.
Using the preceding definitions, the state update~\eqref{eq:vector-iteration} can be rewritten as
\begin{align}
\label{eq:direct-sum-quant-noise}
\ze{t+1} = \Om \ze{t} + \OsubI \qe{t}.
\end{align} 
The goal of vector consensus is to iteratively drive each $\zv{i}{t}$ to the sample mean of the initial states $\zv{i}{0}$. Define $\zebar \in \reals^L$ as the sample mean (over nodes) of the initial states,
\begin{align}
\label{eq:zebar-definition}
\zebar := \frac{1}{m} \sum_{i=1}^m \zv{i}{0}.
\end{align}
The goal of consensus is thus
\begin{align}
\label{eq:vec-consensus-state}
\lim_{t\rightarrow\infty} \zv{i}{t} = \zebar, \fourspace \forall i \in \{1, \ldots, m\}.
\end{align}
We term the state supervector $\zestar \in \reals^{mL}$ for which $\zv{i}{t} = \zebar$, $\forall i \in \{1, \ldots, m\}$, the {\em vector average consensus state.}  This state has the following form:
\begin{align}
\zestar = \left[ \vphantom{\sum_{i=1}^n} \left[ \zebar \,\right]_1 \cdots \left[ \zebar \,\right]_1 \spacedvert \left[ \zebar \,\right]_2 \cdots \left[ \zebar \,\right]_2 \spacedvert \cdots \spacedvert \left[ \zebar \,\right]_L \cdots \left[ \zebar \,\right]_L \right] \tp.
\end{align}
The {\em vector average consensus operator} is defined as
\begin{align}
\label{eq:M-def}
\MM := \bigoplus_{l=1}^L \avg,
\end{align}
so that 
\begin{align}
\label{eq:zestar-def}
\MM \ze{0} = \zestar.
\end{align}
The {\em state estimation error}, which is the error in the current estimate of the true average, is defined as
\begin{align}
\label{eq:err-def}
\et{t} := \ze{t} - \MM \ze{0}.
\end{align}
Because $\MM \ze{t+1} = \MM \ze{t} = \MM \ze{0}$ [this follows from~\eqref{eq:mean-pres-end}], the error~\eqref{eq:err-def} can also be expressed as
\begin{align}
\et{t} = \left(\I - \MM \right) \ze{t}.
\end{align}

\section{Assumptions and analytical results}

In this section, we present certain assumptions and results on the distribution of $\ze{t}$~\eqref{eq:network-state-def}, the network graph structure (i.e., $\mathcal{G}=\{\mathcal{V},\mathcal{E}\}$~\eqref{eq:G-definition}), and the quantization error $\qe{t}$~\eqref{eq:quant-err-def}. We also present relationships among the MSE, the variance of $\zvmarg{i}{j}{t}$~\eqref{eq:zvmarg-definition}, and the distortions introduced, $\Dv$~\eqref{eq:d-vector}.

\subsection{Additive noise model}

Following Frasca \etal~\cite{Frasca2008} and Yildiz and Scaglione~\cite{YildizScaglione2008,YildizScaglione2008b}, we adopt the additive model for quantization noise~\cite{WidrowKollar2008}. This model assumes that the errors introduced by the quantization process are uncorrelated with the quantizer input. A formal statement of the corresponding assumptions is provided.

We assume that the quantization error $\qe{t}$ is spectrally white:
\begin{assumption}
  \label{assumption:quantization-noise-properties}
  The quantization error element $\qevmarg{i}{k}{t}$ corresponding to $\zvmarg{i}{k}{t}$~\eqref{eq:zvmarg-definition} (i.e., quantization error at the $k^\th$ element of the $i^\th$ node at iteration $t$) is uncorrelated with all other quantization error elements: if $j \neq i$, $l \neq k$, or $\tau \neq t$, $\E{\qevmarg{i}{k}{t} \qevmarg{j}{l}{\tau}} = \E{\qevmarg{i}{k}{t}}\E{\qevmarg{j}{l}{\tau}} = 0$. 
\end{assumption}
We further assume that the quantization error is independent of the source to be quantized:
\begin{assumption}
  \label{assumption:quantization-noise-properties-2}
  The correlation between the quantization error $\qe{t}$ is and the state supervector $\ze{t}$ is zero (i.e., $\E{\ze{t}\qet{t}} = \mathbf{0}$).
\end{assumption}
These assumptions are well motivated for uniform quantizers with high coding rate~\cite{WidrowKollar2008}, or for uniform quantizers using dither at arbitrary rates~\cite{Lipshitz1992}. In general, probability-density-optimized (pdf-optimized) quantizers will have quantization error that is linearly correlated with the input signal (i.e., $\E{x(\q{x}-x)} = -D$, where $x$ is a scalar RV to be quantized, $\q{x}$ is the result of quantization, and $D$ is the expected distortion)~\autocites[\pno~182]{GershoGray1993}{Kabal2011}. Although dither for nonuniform quantization is not as well studied as for uniform quantization, the problem of correlated quantization noise may be alleviated by dithering using a compander as shown in a recent paper by Aykol and Rose~\cite{AykolRose2013}. For quantizers with nonuniform level spacing, Assumptions~\ref{assumption:quantization-noise-properties}~and~\ref{assumption:quantization-noise-properties-2} are not rigorously justified, but numerical results are included to assess performance.

\subsection{Preservation of Gaussian distribution}
In this work, we assume that the initial state supervector $\ze{t}$~\eqref{eq:network-state-def} has a multivariate Gaussian distribution. Intuitively, if we assume that the degree of each node can be lower-bounded by a reasonable constant (say 30), the distribution of the state supervector $\ze{t}$ will be approximately Gaussian for all $t \in \{0, \ldots, T-1\}$ by the Central Limit Theorem (CLT) and Assumptions~\ref{assumption:quantization-noise-properties} and~\ref{assumption:quantization-noise-properties-2}. For high-dimensional VQ, the quantization noise approaches a white Gaussian process~\cite{ZamirFeder96}, which further justifies the approximation.

\subsection{State evolution for vector consensus}
Because the network state supervector $\ze{t}$~\eqref{eq:network-state-def} is Gaussian for all $t$, the sufficient statistics of $\ze{t}$ are a mean vector and covariance matrix. Likewise, because the error $\et{t}$~\eqref{eq:err-def} is a linear transformation of $\ze{t}$, it has a multivariate Gaussian distribution. By deriving formulas for the evolution of these statistics, we can minimize the cost function \eqref{eq:comm-cost} to find the optimal source coding rates $\Rhat_{\text{opt}}$.

We define the mean of $\ze{t}$~\eqref{eq:network-state-def} as
\begin{align}
\label{eq:meanz-definition}
\mean{\zetav}{t} := \E{\ze{t}},
\end{align}
and the mean of $\et{t}$~\eqref{eq:err-def} as
\begin{align}
\label{eq:meane-definition}
\mean{\etav}{t} := \E{\et{t}}.
\end{align}
The covariance matrices of $\ze{t}$ and $\et{t}$ are defined as
\begin{align}
\label{eq:Z-definition}
\Sig{\zetav}{t} := \E{\ze{t} \zet{t}} - \mean{\zetav}{t} \meant{\zetav}{t} ,
\end{align}
and
\begin{align}
\label{eq:H-definition}
\Sig{\etav}{t} := \E{\et{t} \ett{t}} - \mean{\etav}{t} \meant{\etav}{t},
\end{align}
respectively.
\begin{lemma}
\label{lemma:se-supervector}
The quantities \cref{eq:meanz-definition,eq:meane-definition,eq:Z-definition,eq:H-definition} can be computed as follows,
\begin{align}
\label{eq:se-mean-data}
\mean{\zetav}{t} = \Om^t \mean{\zetav}{0},
\end{align}
\begin{align}
\label{eq:se-mean-error}
\mean{\etav}{t} = \left(\I - \MM \right) \mean{\zetav}{0}, 
\end{align}
\begin{align}
\label{eq:se-data}
\Sig{\zetav}{t} = \Om^t \Sig{\zetav}{0} \Om^t + \sum_{s=0}^{t-1} \Om^{t-s-1} \OsubI \Sig{\deltav}{s} \OsubI \Om^{t-s-1},
\end{align}
and 
\begin{align}
\label{eq:se-error}
\Sig{\etav}{t} = \left(\I - \MM \right) \Sig{\zetav}{t} \left(\I - \MM \right), 
\end{align}
where $\Sig{\deltav}{t} := \E{\qe{t} \qet{t}}$ is the quantization error~\eqref{eq:quant-err-def} covariance matrix.
\end{lemma}
{\em Proof:} The state update used in vector consensus is~\eqref{eq:vector-iteration}
\begin{align}
\ze{t+1} &= \ze{t} + \OsubI \q{\ze{t}},
\end{align}
where $\ze{t}$ is the state supervector~\eqref{eq:network-state-def}.
The mean $\mean{\zetav}{t}$ can be written in terms of the weight matrix $\Om$~\eqref{eq:Omega-definition} and the quantization error $\qe{t}$~\eqref{eq:quant-err-def} as
\begin{align}
\mean{\zetav}{t} = \E{\Om \ze{t-1}} + \E{\OsubI \qe{t-1}}.  
\end{align} 
Because $\Om$ and $\OsubI$ are constants, this expression is equivalent to
\begin{align}
\mean{\zetav}{t} = \Om \E{\ze{t-1}} + \OsubI \E{\qe{t-1}}.
\end{align}
Recalling that the mean of the quantization error is assumed to be zero, 
\begin{align}
\mean{\zetav}{t} = \Om \mean{\zetav}{t-1}.
\end{align}
By recursion on the above equation, 
\begin{align}
\mean{\zetav}{t} = \Om^t \mean{\zetav}{0},
\end{align}
which is~\eqref{eq:se-mean-data}.

The covariance $\Sig{\zetav}{t}$ can be written
\begin{align}
\Sig{\zetav}{t} = &\E{ \left(\Om \ze{t-1} + \OsubI \qe{t-1} \right) \left(\Om \ze{t-1} + \OsubI \qe{t-1} \right)\tp } \\
&- \E{\Om \ze{t-1}} \E{\left( \Om \ze{t-1} \right) \tp}.
\end{align}
Noting that $\Om$ is a constant, and that $\Om = \Om \tp$, expanding the above expression gives\footnote{The weight matrix $\W{}$ is symmetric for the optimal weights found by Xiao and coauthors~\cite{XiaoBoyd2004,XiaoBoydKim2007} and other simple weighting strategies, such as Metropolis-Hastings~\cite{XiaoBoydKim2007} and max-degree~\cite{XiaoBoyd2004}.}
\begin{align}
\Sig{\zetav}{t} = &\E{\Om \ze{t-1} \zet{t-1} \Om} + \E{\OsubI \qe{t-1} \zet{t-1} \Om} \\ &+ \E{\Om \ze{t-1} \qet{t-1} \OsubI} + \E{\OsubI \qe{t-1} \qet{t-1} \OsubI} \\
&- \Om \E{\ze{t-1}}\E{\zet{t-1}} \Om.
\end{align}
By Assumption~\ref{assumption:quantization-noise-properties-2}, the second and third terms above are zero, so that
\begin{align}
\Sig{\zetav}{t} = \Om \E{\ze{t-1} \zet{t-1}} \Om + \OsubI\Sig{\deltav}{t-1}\OsubI - \Om \mean{\zetav}{t-1}\meant{\zetav}{t-1}\Om.
\end{align}
Grouping terms,
\begin{align}
\Sig{\zetav}{t} = \Om \left( \E{\ze{t-1} \zet{t-1}} - \mean{\zetav}{t-1} \meant{\zetav}{t-1} \right) \Om + \OsubI\Sig{\deltav}{t-1}\OsubI.
\end{align}
Therefore,
\begin{align}
\Sig{\zetav}{t} = \Om \Sig{\zetav}{t-1} \Om + \OsubI\Sig{\deltav}{t-1}\OsubI.
\end{align}
Performing recursion on the above equation gives
\begin{align}
\Sig{\zetav}{t} = \Om^t \Sig{\zetav}{0} \Om^t + \sum_{s=0}^{t-1} \Om^{t-s-1} \OsubI \Sig{\deltav}{s} \OsubI \Om^{t-s-1},
\end{align}
which is \eqref{eq:se-data}. 

Because the error from the true sample mean is $\et{t} = \left(\I - \MM \right) \ze{t}$~(from \cref{eq:err-def,eq:M-def}) and $\left(\I - \MM \right) = \left(\I - \MM \right)\tp$, it follows that
\begin{align}
\mean{\etav}{t} = \left(\I - \MM \right) \mean{\zetav}{t},
\end{align}
which is~\eqref{eq:se-mean-error}, and
\begin{align}
\Sig{\etav}{t} = \left(\I - \MM \right) \Sig{\zetav}{t} \left(\I - \MM \right),
\end{align}
which is~\eqref{eq:se-error}. Similar steps are taken to obtain the simplified forms of these expressions for scalar consensus. \QED

The equations~\cref{eq:se-mean-data,eq:se-mean-error,eq:se-data,eq:se-error} can be simplified for optimization modeling in the case that the states $\zv{i}{0}$ are stationary Gaussian processes. In this scenario, the distortion incurred by uniform scalar quantization at node $i$ and iteration $t$ can be modeled as the distortion from quantization of an independent and identically distributed (i.i.d.) source under certain conditions~\autocite[Sec.~9.5]{WidrowKollar2008}. In particular, the RD performance can be described well by the marginal variance of the state $\zv{i}{t}$,
\begin{align}
\margvar{i}{\Dv, t} = \E{\left(\zvmarg{i}{1}{t} \right)^2} - \E{\zvmarg{i}{1}{t}}^2 = \cdots = \E{\left(\zvmarg{i}{L}{t} \right)^2} - \E{\zvmarg{i}{L}{t}}^2,
\end{align} 
provided that the correlations $\E{\zvmarg{i}{k}{t}\zvmarg{i}{l}{t}}, k \neq l,$ are not extreme and the quantizer bin size $b$ is not too coarse~\autocite[Sec.~9.5]{WidrowKollar2008}. 
For this scenario, we can consider the scalar states $z_i(t) \in \reals$~\eqref{eq:scalar-state-def}, $i \in \{1,\ldots,m\}$ and the state vector $\z{t} \in \reals^m$~\eqref{eq:z-definition} in place of $\zv{i}{t} \in \reals^L$, $i \in \{1, \ldots, m\}$ and $\ze{t} \in \reals^{mL}$, respectively, where $z_i(t)$ is distributed according to the marginal pdf $f(\zvmarg{i}{1}{t})$, $\forall i \in \{1,\ldots,m \}$.

We define the means of $\z{t}$~\eqref{eq:z-definition} and $\diff{t}$~\eqref{eq:e-definition},
\begin{align}
\label{eq:simple-mean-def}
\mean{\mathbf{z}}{t} := \E{\z{t}}
\end{align}
and
\begin{align}
\label{eq:simple-mean-error-def}
\mean{\mathbf{e}}{t} := \E{\diff{t}},
\end{align}
and the covariances of $\z{t}$ and $\diff{t}$,
\begin{align}
\label{eq:simple-covariance-def}
\Sig{\mathbf{z}}{t} := \E{\z{t}\zt{t}} - \mean{\mathbf{z}}{t} \meant{\mathbf{z}}{t}
\end{align}
and
\begin{align}
\label{eq:simple-covariance-error-def}
\Sig{\mathbf{e}}{t} := \E{\diff{t}\difft{t}} - \mean{\mathbf{e}}{t} \meant{\mathbf{e}}{t}.
\end{align}
The equations \cref{eq:se-mean-data,eq:se-mean-error,eq:se-data,eq:se-error} reduce to
\begin{align}
\label{eq:se-data-mean-simple}
\mean{\mathbf{z}}{t} = \W{t} \mean{\mathbf{z}}{0},
\end{align}
\begin{align}
\label{eq:se-error-mean-simple}
\mean{\mathbf{e}}{t} = \M \W{t} \mean{\mathbf{e}}{0}, 
\end{align}
\begin{align}
\label{eq:se-data-simple}
\Sig{\mathbf{z}}{t} = \W{t} \Sig{\mathbf{z}}{0} \W{t} + \sum_{s=0}^{t-1} \W{t-s-1} \WsubI \Sig{\qesu}{s} \WsubI \W{t-s-1},
\end{align}
and 
\begin{align}
\label{eq:se-error-simple}
\Sig{\mathbf{e}}{t} = \M \Sig{\mathbf{z}}{t} \M . 
\end{align}
Given the above state-evolution equations~\cref{eq:se-data-mean-simple,eq:se-error-mean-simple,eq:se-data-simple,eq:se-error-simple}, we can determine the variance $\margvar{i}{\Dv, t}$ and MSE, $\text{MSE}_i(\Dv,t)$, at each node $i \in \{1,\ldots,m\}$. The source variance at node $i$ and iteration $t$ is defined as
\begin{align}
\label{eq:margvar-def}
\margvar{i}{\Dv,t} := \E{z_i^2(t)} - \E{z_i(t)}^2.
\end{align}

Let us discuss our state estimation error~\eqref{eq:e-definition} metrics. We define the {\em node MSE}, which represents the MSE at node $i$ and iteration $t$, as
\begin{align}
\label{eq:node-mse-def}
\mathrm{MSE}_i(\Dv,t) &:= \E{e_i^2(t)},
\end{align} 
where $e_i(t)$ is the $i^\th$ element of $\diff{t}$~\eqref{eq:e-definition}.
We now turn to quantifying the estimation error over the entire network. For this purpose, we define the {\em network MSE}, which is the arithmetic mean of the node MSEs, $\{\mathrm{MSE}_i(\Dv,t)\}_{i=1}^m$,
\begin{align}
\label{eq:net-mse-def}
\MSE{\Dv,t} &:= \frac{1}{m} \sum_{i=1}^m \mathrm{MSE}_i(\Dv,t).
\end{align}
Using these definitions, we present the following mathematical relationships, which we term the {\em state evolution equations}. These equations allow us to perform the optimization of the coding rate vector $\Rhat$ using the cost function~\eqref{eq:comm-cost}.

\begin{figure}
  \begin{center}
    \includegraphics[width=\linewidth]{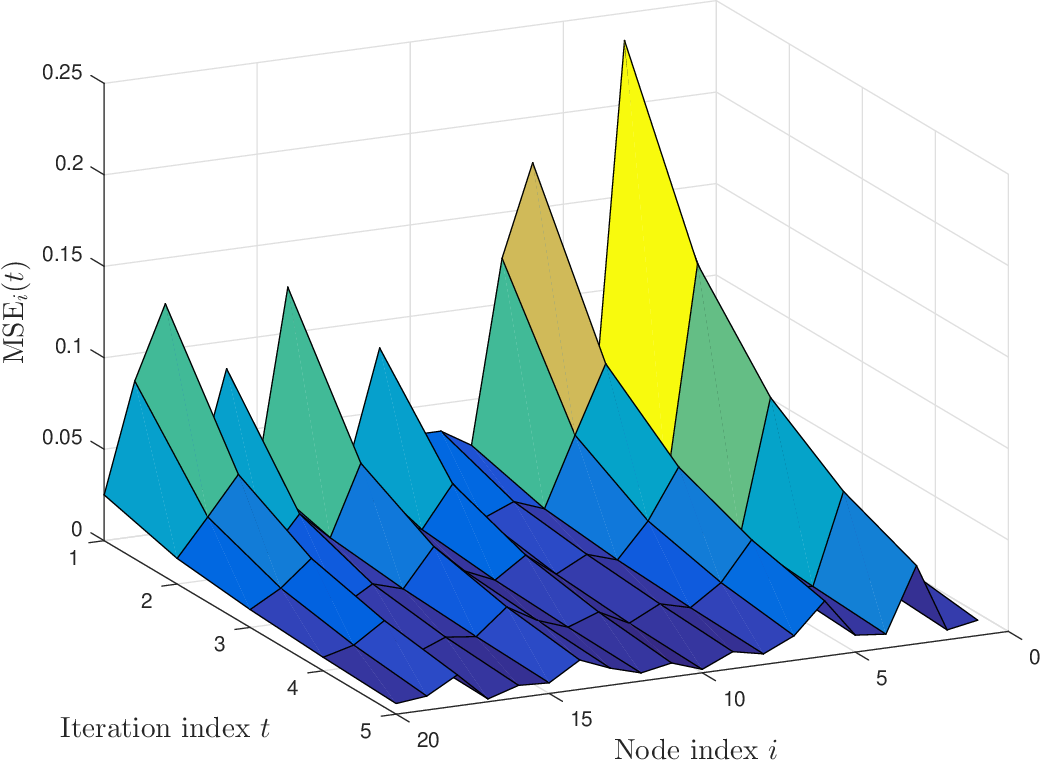}
    \caption{Surface showing the evolution of $\text{MSE}_i(t)$ for the case of lossless consensus. Note that the value of $\text{MSE}_i(t)$ is not equal across all nodes for a particular iteration, and the evolution of $\text{MSE}_i(t)$ is heavily dependent on network topology. Also observe that the MSE decreases monotonically at each node as $t \rightarrow \infty$. The variance $\margvar{i}{t}$ evolves in a similar fashion.}
    \label{fig:mseSurface}
  \end{center}
\end{figure}

\begin{lemma}
  Given the above equations~\eqref{eq:se-data-simple}~and~\eqref{eq:se-error-simple} for the state and error covariances, the variance at node $i$ at iteration $t$~\eqref{eq:margvar-def} is given by
  \begin{align}
  \label{eq:se-var}
  \margvar{i}{\Dv, t} &= \left[ \Sig{\mathbf{z}}{t} \right]_{ii},
  \end{align}
  the node MSE~\eqref{eq:node-mse-def} is given by
  \begin{align}
  \label{eq:se-mse}
  \mathrm{MSE}_i(\Dv, t) &= \left[  \Sig{\mathbf{e}}{t} + \mean{\mathbf{e}}{t}\meant{\mathbf{e}}{t} \right]_{ii} ,
  \end{align}
  and the network MSE~\eqref{eq:net-mse-def} is given by
  \begin{align}
  \label{eq:se-mse-net}
  \MSE{\Dv, t} &= \frac{1}{m} \tr{\Sig{\mathbf{e}}{t} + \mean{\mathbf{e}}{t}\meant{\mathbf{e}}{t} },
  \end{align}
  where $\tr{\cdot}$ denotes the trace of a matrix.
\end{lemma}
The derivations of equations~\cref{eq:se-data-mean-simple,eq:se-error-mean-simple,eq:se-data-simple,eq:se-error-simple} take almost identical steps to those in the proof of Lemma~\ref{lemma:se-supervector}. An example of the $\text{MSE}_i(t)$ predicted by~\eqref{eq:se-mse} is given in Figure~\ref{fig:mseSurface}.

\section{Optimization via generalized geometric programming}
\label{sec:cvx}
The key insight of this work is the ability to pose the optimization of the coding rate vector $\Rhat$ as a generalized geometric program (GGP), whose global optimum can be found efficiently~\cite{BoydKimVandenbergheHassibi2007}. Using variable-length codes~\cite{Huffman52}, the coding rates
\begin{align}
R_i(t) \in \reals_{>0}, \fourspace \forall i \in \{1,\ldots,m\}, \fourspace \forall t \in \{0, \ldots, T-1 \},
\end{align} 
in contrast to the fixed-length coding case where $R_i(t) \in \mathbb{Z}_{> 0}$, $\forall i \in \{1,\ldots,m\}$, $\forall t \in \{0, \ldots, T-1 \}$. For the case of variable-length codes, we propose a high-rate approximation to the operational RD relationship to find an efficient rate vector $\Rhat$ capable of achieving a target value of $\MSE{\Dv,T}$~\eqref{eq:se-mse-net}.\footnote{Note that the final network MSE corresponds to iteration index $T$, and not $T-1$. This is because $\MSE{\Dv,t}$ is the network MSE {\em after the end of} $t$ iterations. The quantity $\MSE{\Dv,t-1}$ would thus correspond to the end of the $(t-1)^\th$ iteration, or the beginning of the $t^\th$ iteration.} This modeling strategy will later be used to design an optimization procedure for fixed-length coding, where $\Rhat \in \mathbb{Z}_{> 0}^{mT}$.
%

\subsection{Cost function formulation}
For the quantizers of interest operating on Gaussian sources, the operational RD relationship (the reader is reminded that the this term was defined in Section~\ref{sec:RD-theory}) in the high-rate regime has the form
\begin{align}
\label{eq:pattern}
R(D) \approx \begin{cases}
\dfrac{1}{2} \log_2 \left(\dfrac{\sigma^2}{D} \right) + R_c, \fourspace &D \in (0, \sigma^2 D_{\text{max}}] \\
0, \fourspace &\text{otherwise}
\end{cases},
\end{align}
where $\sigma^2$ represents the variance of the data to be encoded and $D_{\text{max}}$ is some constant~\cite{GishPierce68}. In some cases, such as infinite-dimensional VQ with Gaussian memoryless sources and dithered scalar uniform quantization~\cite{ZamirFeder96}, the relationship~\eqref{eq:pattern} holds for all rates. Note that~\eqref{eq:pattern} is equal to the Gaussian RD function
\begin{align}
R_G(D) =
\begin{cases}
\dfrac{1}{2} \log_2 \left( \dfrac{\sigma^2}{D} \right), \fourspace &D \in (0, \sigma^2] \\
0, \fourspace &\text{otherwise}
\end{cases} ,
\end{align}
with a modified domain and offset. In the following discussion, it is easier to consider the inverse to the operational RD relationship, denoted $D(R)$.

For uniform scalar quantization of a memoryless Gaussian source followed by block entropy coding, the true operational RD relationship is approximated well in the high-rate regime by the Gaussian RD function plus 0.255 bits~\autocites[\pno~302]{GershoGray1993}{GrayNeuhoff1998}. More generally, the operational distortion-rate relationship for ECSQ at high rates is determined by
\begin{align}
\label{eq:high-rate-ecsq}
D(R) \approx \frac{1}{12} 2^{2 h(x)} 2^{-2 R},
\end{align} 
where $h(x)$ is the differential entropy of the source $x$~\cite{GrayNeuhoff1998}. Similar results hold for Lloyd-Max quantization~\cite{PanterDite51,GrayNeuhoff1998} and fixed-rate uniform quantization~\cite{GrayNeuhoff1998}. For Lloyd-Max quantization of memoryless Gaussian sources at high resolution, the operational RD relationship is approximated by
\begin{align}
\label{eq:high-rate-lloyd}
D(R) \approx \frac{1}{12} \left[ \int_{-\infty}^\infty f^{1/3}(x) dx \right]^{3} 2^{-2R},
\end{align}
where $f(x)$ is the probability distribution of the source $x$~\cite{PanterDite51,GrayNeuhoff1998}. For Gaussian densities, both~\cref{eq:high-rate-ecsq,eq:high-rate-lloyd} scale with the source variance $\sigma^2$. This scaling can be proven by straightforward computation.

As in Section~\ref{sec:RD-theory}, we denote the representation levels of the $N$-codeword quantizer as $\{\hat{x}_i\}_{i=1}^N$. The maximum representation level of the quantizer is denoted $\hat{x}_{\text{max}}$, and the minimum representation level is denoted $\hat{x}_{\text{min}}$. For subtractively dithered fixed-rate uniform quantization of any source at all rates with $\Delta := \frac{1}{\sigma}(\hat{x}_{\text{max}} - \hat{x}_{\text{min}})$ (neglecting overload distortion)~\cite{GrayNeuhoff1998},
\begin{align}
D(R) = \frac{\Delta^2 \sigma^2}{12} 2^{-2R}.
\end{align}

In our case, the source variance is $\margvar{i}{\Dv, t}$, defined in~\eqref{eq:se-var}, which is a function of the initial state vector covariance $\Sig{\mathbf{z}}{0}$~\eqref{eq:simple-covariance-def} and the distortion vector $\Dv$~\eqref{eq:d-vector}. It evolves as described by~\eqref{eq:se-var}. The operational RD relationship at all nodes $i \in \{1,\ldots,m\}$ and iterations $t\in\{0,\ldots,T-1\}$ can be expressed as follows. Any Gaussian scalar quantizer will have a maximum possible distortion for a particular source. Because the distortion is proportional to the source variance $\margvar{i}{\Dv,t}$ in the Gaussian case, we consider the maximum distortion when $\margvar{i}{\Dv,t}=1$. Letting $D_{\text{max}}$ denote this maximum possible distortion for a unit-variance Gaussian source,
\begin{align}
R_i(\Dv, t) = \begin{cases} 
\dfrac{1}{2} \log_2 \left( \dfrac{\margvar{i}{\Dv,t}}{\Dmarg{i}{t}} \right) + R_c, \fourspace &\Dmarg{i}{t} \in \left( 0, \margvar{i}{\Dv,t} D_{\text{max}}  \right] \nonumber \\
0, &\Dmarg{i}{t} > \margvar{i}{\Dv,t} D_{\text{max}} 
\end{cases} .
\end{align}
This operational RD relationship is equivalent to
\begin{align}
R_i(\Dv, t) &=\frac{1}{2} \max \left\{ \log_2 \left( \frac{\margvar{i}{\Dv,t}}{\Dmarg{i}{t}} \right), c \right\} + R_c \nonumber \\
&= \frac{1}{2}  \log_2 \left( \max \left\{ \frac{\margvar{i}{\Dv,t}}{\Dmarg{i}{t}}, k \right\} \right) + R_c , \fourspace D_i(t) > 0, \label{eq:rd}
\end{align}
where $c$ and $k$ are terms (constant in $\Dv$ and $t$) that depend on $D_\text{max}$. Concretely, $k = 2^{-2R_c} >0$.
 
Given a number of iterations $T$, the goal is to minimize the aggregate coding rate~\eqref{eq:ragg-def}, subject to a constraint on the final MSE, $\MSE{\Dv,T} \leq \text{MSE}^*$~\eqref{eq:se-mse-net}. For a particular number of iterations $T$ and target ${\rm{MSE}}^*$, the minimum number of iterations required to achieve that MSE, $T_{\text{min}}= \argmin_T \left \lbrace T \,\, \vert \,\, \MSELL{\Dv, T} < {\rm{MSE}}^* \right \rbrace$, can be readily obtained using the state-evolution equations~\cref{eq:se-data-mean-simple,eq:se-error-mean-simple,eq:se-data-simple,eq:se-error-simple}. More formally, using the operational RD relationship~\eqref{eq:rd}, the optimization problem is
\begin{equation}
\begin{aligned}
& \underset{\Dv}{\text{minimize}}
& & \sum_{t=0}^{T-1} \sum_{i=1}^m \frac{1}{2} \log_2\left( \max \left\{ \frac{\margvar{i}{\Dv,t}}{\Dmarg{i}{t}}, k \right\} \right) + R_c, \\
& \text{subject to}
& & \MSE{\Dv, T} \leq \text{MSE}^*, \\
& & &  \Dmarg{i}{t} > 0, \eightspace \forall i,t .
\end{aligned}
\end{equation}

Note that the above optimization can be rewritten as
\begin{equation}
\begin{aligned}
\label{eq:cfun}
&\underset{\Dv}{\text{minimize}}
& & \ln \left( \prod_{t=0}^{T-1} \prod_{i=1}^m \max \left\{ \frac{\margvar{i}{\Dv,t} }{\Dmarg{i}{t}}, k \right\} \right), \\
& \text{subject to} 
& & \MSE{\Dv, T} \leq \text{MSE}^*, \\
& & &  \Dmarg{i}{t} > 0, \eightspace \forall i,t ,
\end{aligned}
\end{equation}
where the constants $\frac{1}{2 \ln(2)}$ and $R_c$ are omitted, because they make no difference to the optimum point of the minimization. We will now introduce GGP and show that the optimization~\eqref{eq:cfun} reduces to such a problem.

\subsection{Basics of generalized geometric programming} The following information can be found in the book {\em Convex Optimization} by Boyd and Vandenberghe~\cite{BoydVandenberghe2004}. For this tutorial, we stay close to the authors' original notation. In the language of geometric programming (GP), a function of the form
\begin{align}
f(x) = c x_1^{a_1}x_2^{a_2}\cdots x_n^{a_n}, \fourspace c >0, x_i > 0 \fourspace \forall i, a_i \in \reals \fourspace \forall i,
\end{align}
is called a {\em monomial}~\cite{BoydVandenberghe2004,BoydKimVandenbergheHassibi2007}. Similarly, a function of the form
\begin{align}
f(x) = \sum_{i=1}^k g_i(x_1,\ldots,x_n),
\end{align}
where the $g_i(x_1,\ldots,x_n)$ are monomials, is termed a {\em posynomial} (not to be confused with the word {\em polynomial} from algebra)~\cite{BoydKimVandenbergheHassibi2007}. The word posynomial is intended to convey that the coefficients and values of the variables are constrained to be positive~\cite{BoydKimVandenbergheHassibi2007}. Boyd and Vandenberghe~\cite{BoydVandenberghe2004} state a certain number of closure properties:\\
\begin{quote}
  Posynomials are closed under addition, multiplication, and nonnegative scaling. Monomials are closed under multiplication and division. If a posynomial is multiplied by a monomial, the result is a posynomial; similarly, a posynomial can be divided by a monomial, with the result a posynomial~\autocite[\pno~161]{BoydVandenberghe2004}. \\
\end{quote}

By introducing dummy variables and additional constraints, more general cost and constraint functions, called generalized posynomials, can be converted into posynomials and used to formulate GPs~\cite{BoydKimVandenbergheHassibi2007}. A {\em generalized posynomial} is a function formed from posynomials using addition, multiplication, the maximum operation, and positive exponentiation~\cite{BoydKimVandenbergheHassibi2007}. Generalized posynomials are closed under addition, multiplication, division by monomials, and the maximum operation~\cite{BoydKimVandenbergheHassibi2007,BoydVandenberghe2004}. A standard inequality-constrained generalized GGP is of the form
\begin{equation}
\begin{aligned}
&\underset{x_1,\ldots,x_n}{\text{minimize}}
& & C(x_1,\ldots,x_n), \\
& \text{subject to} 
& & \fourspace f_i(x_1,\ldots,x_n) \leq 1, \eightspace \forall i \in \{ 1,\ldots,n_f \}, \\
& & & \fourspace g_i(x_1, \ldots, x_n) = 1, \eightspace \forall i \in \{ 1,\ldots,n_g \},
\end{aligned}
\end{equation}
where the cost $C(x_1, \ldots, x_n)$ and all the inequality constraints $f_i(x_1, \ldots, x_n)$ are generalized posynomials, and all the equality constraints $g_i(x_1, \ldots, x_n)$ are monomials~\cite{BoydVandenberghe2004}. By a logarithmic change of variables and objective function transformation, GGPs can be cast in convex form and efficiently solved using numerical techniques~\cite{BoydKimVandenbergheHassibi2007}. Note that there is an implicit constraint that $x_i > 0 \fourspace \forall i \in \{1, \ldots, n \}$~\cite{BoydKimVandenbergheHassibi2007}.

We now turn to the problem of posing the optimization~\eqref{eq:cfun} as a GGP. Boyd and Vandenberghe~\cite{BoydVandenberghe2004} state that applying a monotonically increasing function to the objective results in an equivalent problem.
Therefore, by applying the exponential function to the cost function~\eqref{eq:cfun}, we obtain the equivalent problem,
\begin{equation}
\begin{aligned}
\label{eq:cfun-final}
&\underset{\Dv}{\text{minimize}}
& & \prod_{t=0}^{T-1} \prod_{i=1}^m \max \left\{ \frac{\margvar{i}{\Dv,t} }{\Dmarg{i}{t}}, k \right\}, \\
& \text{subject to} 
& & \MSE{\Dv, T} \leq \text{MSE}^*, \\
& & &  \Dmarg{i}{t} > 0, \eightspace \forall i,t .
\end{aligned}
\end{equation}
\subsection{Proof that the optimization is a GGP}
Until now, the analysis has resembled the prior art. The state evolution equations~\cref{\seeqs} were derived for a different update equation than ours by Yildiz and Scaglione~\cite{YildizScaglione2008}. Huang and Hua~\cite{HuangHua2011} extended analysis of consensus to vectors, although they also used a different state update. However, neither of these works optimize with respect to exact MSE values~\eqref{eq:node-mse-def}, and Huang and Hua did not account for the possibilities of different coding rates at each node and iteration or entropy coding.

The main original contribution of this work is to pose the coding rate optimization problem with node- and iteration-dependent rates as a GGP, which can solve quickly and reliably for the global optimum coding rate vector $\Rhat_{\text{opt}}$. The following theorem proves that the optimization~\eqref{eq:cfun-final} can be cast as such a problem. \\

\hspace{-17pt}{\bf Theorem}\hspace{6pt}{\em The optimization problem~\eqref{eq:cfun-final} is a GGP.}

\hspace{-17pt}{\em Proof:}
To show that~\eqref{eq:cfun-final} is a GGP, our argument takes the following steps: ({\em i}) show that $\margvar{i}{\Dv, t}$ is a posynomial, ({\em ii}) show that $\frac{\margvar{i}{\Dv, t}}{\Dmarg{i}{t}}$ is a posynomial, ({\em iii}) show that $\max \{ \frac{\margvar{i}{\Dv, t}}{\Dmarg{i}{t}}, k \}$ is a generalized posynomial, ({\em iv}) show that $\prod_{t=0}^{T-1} \prod_{i=1}^m \max \{ \frac{\margvar{i}{\Dv, t}}{\Dmarg{i}{t}}, k \}$ is a generalized posynomial, and ({\em v}) show that $\frac{1}{\text{MSE}^*} \MSE{\Dv, t}$ is a posynomial.

({\em i}) To show that the variance $\margvar{i}{\Dv, t}$ is a posynomial, it is necessary to express $\margvar{i}{\Dv, t}$ in terms of the individual distortions $\Dmarg{i}{t}$. From the state-evolution equation~\eqref{eq:se-var},
\begin{align}
\margvar{i}{\Dv, t} = \left[ \W{t} \Sig{\mathbf{z}}{0} \W{t} + \sum_{s=0}^{t-1} \W{t-s-1} \WsubI \Sig{\qesu}{s} \WsubI \W{t-s-1} \right]_{ii} .
\end{align}
Recall that $\Sig{\qesu}{t}$ is a diagonal matrix by Assumption~\ref{assumption:quantization-noise-properties} and~\eqref{eq:qe-definition},
\begin{align}
\left[ \Sig{\qesu}{t} \right]_{ij} = \begin{cases}
\Dmarg{i}{t}, \fourspace i = j \\
0, \fourspace \text{otherwise}
\end{cases} .
\end{align}
Let $\What{t,s} = \W{t-s-1} \WsubI$, and denote the $i^{\text{th}}$ column of $\What{t,s}$ as $\what{i}{t,s} \in \reals^m$. Each term $\What{t,s} \Sig{\boldsymbol{\qesu}}{s} \Whatt{t,s}$ in the sum is then equal to
\begin{align}
\What{t,s} \Sig{\qesu}{s} \Whatt{t,s} = \sum_{k=1}^m \Dmarg{k}{s} \what{k}{t,s} \whatt{k}{t,s}.
\end{align}
Therefore,
\begin{align}
\margvar{i}{\Dv, t} = \left[ \W{t} \Sig{\mathbf{z}}{0} \W{t} \right]_{ii} + \sum_{s=0}^{t-1} \sum_{k=1}^m \Dmarg{k}{s} \left[ \what{k}{t,s} \whatt{k}{t,s} \right]_{ii} .
\end{align}
Because the diagonals of the outer product of a real vector with itself are always nonnegative, each coefficient in the above summation is nonnegative. 
The first term, $\left[ \W{t} \Sig{\mathbf{z}}{0} \W{t} \right]_{ii}$ is a monomial. It is clear that even if one or more of the $\left[ \what{k}{t,s} \whatt{k}{t,s} \right]_{ii}$ terms from the summation is zero, $\margvar{i}{\Dv, t}$  can still be considered a posynomial by omitting those terms from the summation. Therefore, $\margvar{i}{\Dv,t}$ is a posynomial.

({\em ii}) $\Dmarg{i}{t}$ is a monomial, and a posynomial divided by a monomial is a posynomial. Therefore, $\frac{\margvar{i}{\Dv, t}}{\Dmarg{i}{t}}$ is a posynomial.

({\em iii}) The maximum of a posynomial and a monomial is a generalized posynomial. Because $k$ is positive, it is trivially a monomial. Therefore, $\max \{ \frac{\margvar{i}{\Dv, t}}{\Dmarg{i}{t}}, k \}$ is a generalized posynomial.

({\em iv}) Generalized posynomials are closed under multiplication. Therefore, $\prod_{i,t} \max \{ \frac{\margvar{i}{\Dv, t}}{\Dmarg{i}{t}}, k \}$ is a generalized posynomial.

({\em v}) Define $\Wcar{t,s} = \M \W{t-s-1} \WsubI$ and let $\wcar{i}{t,s} \in \reals^m$ denote the $i^{\text{th}}$ column of $\Wcar{t,s}$. The MSE at each node~\eqref{eq:node-mse-def} is given by
\begin{align}
\mathrm{MSE}_i(\Dv, t) &= \left[ \vphantom{\sum_{s=1}^t} \M \W{t} \Sig{\mathbf{e}}{0} \W{t} \M \right.  \\
&+ \left. \sum_{s=0}^{t-1} \Wcar{t,s} \Sig{\qesu}{s} \Wcart{t,s} + \mean{\mathbf{e}}{t}\meant{\mathbf{e}}{t} \right]_{ii} .
\end{align}
As in the proof of ({\em i}), the above equation reduces to
\begin{align}
\mathrm{MSE}_i(\Dv,t) &= \left[ \vphantom{\sum_{s=1}^t} \M \W{t} \Sig{\mathbf{e}}{0} \W{t} \M \right. \\
&+ \left. \sum_{s=0}^{t-1} \sum_{k=1}^m \Dmarg{k}{s} \wcar{k}{t,s} \wcart{k}{t,s} +\mean{\mathbf{e}}{t}\meant{\mathbf{e}}{t} \right]_{ii} .
\end{align}
By the same argument as ({\em i}), $\mathrm{MSE}_i(\Dv,t)$ is a posynomial. Likewise, the network MSE, $\mathrm{MSE}(\Dv,t) = \frac{1}{m} \sum_{i=1}^m \mathrm{MSE}_i(\Dv,t)$ is a posynomial.

Because the objective and constraint functions are posynomials, and the variables $\Dmarg{i}{t}$ are constrained to be positive, the optimization~\eqref{eq:cfun-final} is a GGP. \QED

Two aspects of the above proof should be highlighted. Because the constraints are allowed to be generalized posynomials, one could also optimize with respect to a constraint on the maximum node MSE, for example 
\begin{align}
\max_i \,\, \{\text{MSE}_i(\Dv,T)\}_{i=1}^m \leq \text{MSE}^*,
\end{align} 
or constraints on each of the node MSE values, 
\begin{align}
\text{MSE}_i(\Dv,T) \leq \text{MSE}^*_i, \fourspace \forall i \in \{1,\ldots,m\},
\end{align}
where $\text{MSE}_i(\Dv,t)$ is defined in~\eqref{eq:se-mse}. Also, in its most general form, the optimization allows each node to use a different rate or distortion. In the interest of designing a distributed protocol (or for computational efficiency), one may wish to constrain the rates or distortions at each node to be the same. The constraint that all distortions be the same is a straightforward modification of~\eqref{eq:cfun-final} and is also provably a GGP.

\section{Efficient implementation of GGP optimization}
Solving the exact optimization problem~\eqref{eq:cfun-final} naively requires explicit representation of all $mT$ distortions $D_i(t)$ and all the coefficients of the log-sum-exp (LSE) model\footnote{GGPs are converted to convex LSE form for solution~\cite{cvx,gb08}.} required to compute $\text{MSE}_i(\Dv,t)$ and $\margvar{i}{\Dv,t}$ from $\Dv$. The result of this explicit representation is large time and memory complexity. In this section, we explore a simplification of the optimization problem to combat these issues, and we present a heuristic to apply the GGP optimization method to fixed-rate coding schemes.

\subsection{Equal-distortion simplification and implementation}
According to Boyd and coauthors~\cite{BoydVandenberghe2004,BoydKimVandenbergheHassibi2007}, GGPs can be converted into convex optimization problems by a change of variables and a logarithmic transformation of the objective function. For a detailed treatment in the case of GP, the reader is referred to the above references. GGPs can be converted in a similar fashion; however, the conversion is somewhat more complicated. In some software packages, such as CVX~\cite{cvx,gb08}, GGPs can be specified directly, and they are automatically converted into corresponding convex programs~\cite{BoydKimVandenbergheHassibi2007}. This approach guarantees that the global optimum can be found with little additional effort by the programmer.

In practice, for large networks, (i.e., $m > 20$ and $T \geq 6$) the memory and time requirements of the optimization~\eqref{eq:cfun-final} seem to grow very quickly. Furthermore, using CVX~\cite{cvx,gb08}, all compatible solvers tested displayed poor stability for large networks. Attempts to use alternative modeling frameworks, such as GPkit~\cite{gpkit}, CVXPY~\cite{cvxpy}, and YALMIP~\cite{Lofberg2004}, failed due to the high memory requirements of the model. Manually building a GGP model is memory- and time-intensive; if explicit model representation can be avoided, however, it may be possible to apply other convex optimization methods without these scaling issues. In particular, avoiding the explicit representation of posynomial coefficients and exponents would probably greatly improve performance for large problem sizes $m$ and $T$. To provide a program that is more easily solved in practice, we make two simplifications. First, we constrain the distortions to be equal at each node, which is well motivated for the end goal of designing a truly distributed protocol. Next, the program can be cast as
\begin{equation}
\begin{aligned}
\label{eq:cfun-mod}
&\underset{\Dv}{\text{minimize}}
& & \prod_{t=0}^{T-1} \prod_{i=1}^m  \max{  \left\{ \frac{ \margvar{i}{t} }{D(t)}, k \right\}  }, \\
& \text{subject to} 
& & \MSE{\Dv, T} \leq \text{MSE}^*, \\
& & &  \D{t} > 0, \eightspace \forall t .
\end{aligned}
\end{equation}
In the following chapter, the results of solving this simplified problem~\eqref{eq:cfun-mod} are compared to the solutions of the exact program~\eqref{eq:cfun-final} and the prior art~\cite{Thanou2013,YildizScaglione2008}. Surprisingly, the above approximation provides competitive results for random geometric networks~\cite{Penrose2003}, with significant reduction in memory and run-time requirements.

\subsection{Efficient search heuristic for fixed-rate quantizers}
\label{sec:heuristic}
Using fixed-length coding with uniform scalar quantization, the rates to be used are integer-valued~\cite{GrayNeuhoff1998}, and arbitrary distortions are not achievable for a given source variance. To account for this constraint, we suggest a search heuristic. Given a particular $T$ and $\text{MSE}^*$, there exist a finite number of rate vectors $\Rhat \in \mathbb{Z}_{>0}^{mT}$~\eqref{eq:r-vector}. Clearly, if each node is allowed to use different rates, then the size of the search space becomes too large to be practical, and we constrain our attention to the case that rates are constant over nodes for a particular iteration: $R_i(t)=R_j(t) = R(t)$, $\forall i,j \in \{1,\ldots,m\}$, $t \in \{0,\ldots,T-1\}$. Even in the node-constant rate case, the number of possible sequences such that $\min_{i,t} R_i(t) = 1$ and $\max_{i,t} R_i(t) = R_{\text{max}}$ grows combinatorially in $T$.

To find an efficient (though perhaps suboptimal) sequence, we propose to use the output of the GGP~\eqref{eq:cfun-final} as a starting point for the optimization. Because constraining the rates to be equal at each node is not straightforward, we instead constrain the distortions to be equal at each node as in~\eqref{eq:cfun-mod}. The optimal rates $R_i(t)$ from the solution of~\eqref{eq:cfun-mod} are then averaged into a single sequence $\left\{R_{\text{GGP}}(t)\right\}_{t=0}^{T-1}$,
\begin{align}
R_{\text{GGP}}(t) := \frac{1}{m} \sum_{i=1}^m R_i(t).
\end{align}
A neighborhood is formed as follows. To the original sequence 
\begin{align}
\RhatGGP := [\Rggp{0},\ldots,\Rggp{T-1}]\tp,
\end{align} 
we add $\bll \in \{-1,+1\}^T$. Each possible $\bll$ is added, and the result is stored, until all $2^T$ possible transformations of $\RhatGGP$ have been computed. Because the minimum fixed rate is one bit per symbol, all rates less than one bit per symbol from the previous step are re-assigned to one bit per symbol. Next, we apply the operators $A \in \{\lfloor \cdot \rfloor, \lceil \cdot \rceil \}^T$ to the resulting sequences, where, for example, $[\lfloor \cdot \rfloor, \lceil \cdot \rceil, \lfloor \cdot \rfloor] \,\, [x, y, z]\tp := [\lfloor x \rfloor, \lceil y \rceil, \lfloor z \rfloor]\tp$. Finally, we eliminate duplicate sequences. This results in a ``trellis'' containing the integers about each $R_{\text{GGP}}(t)$. The maximum possible total number of sequences in the search space is thus $4^T$. Brute-force search is conducted on this subset of sequences to find the one that satisfies the MSE constraint with the minimum aggregate rate $R_{\text{agg}} = m \sum_{t=0}^{T-1} R(t)$. In the case that multiple sequences achieve the MSE constraint with equal aggregate rate, the one that produces the lowest MSE is selected.

\chapter{NUMERICAL RESULTS}
\label{chap-four}
\begin{figure}
  \begin{center}
    \includegraphics[width=\textwidth,height=\figpctg\textheight,keepaspectratio]{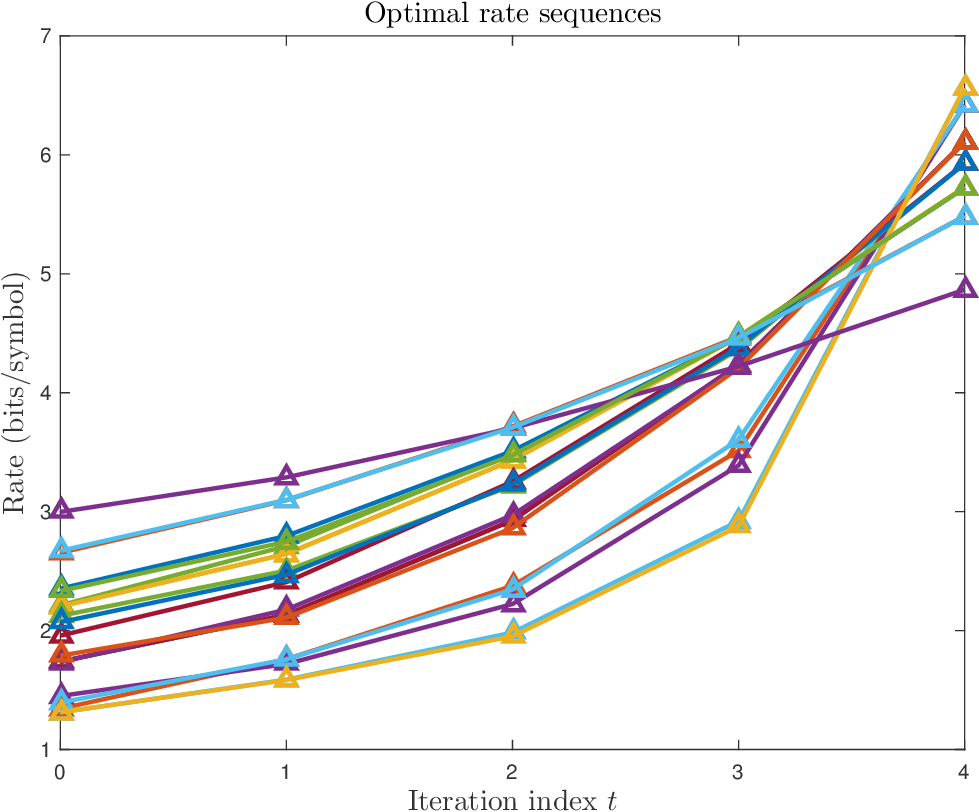} \\
    \vspace{0.4cm}
    \includegraphics[width=\textwidth,height=\figpctg\textheight,keepaspectratio]{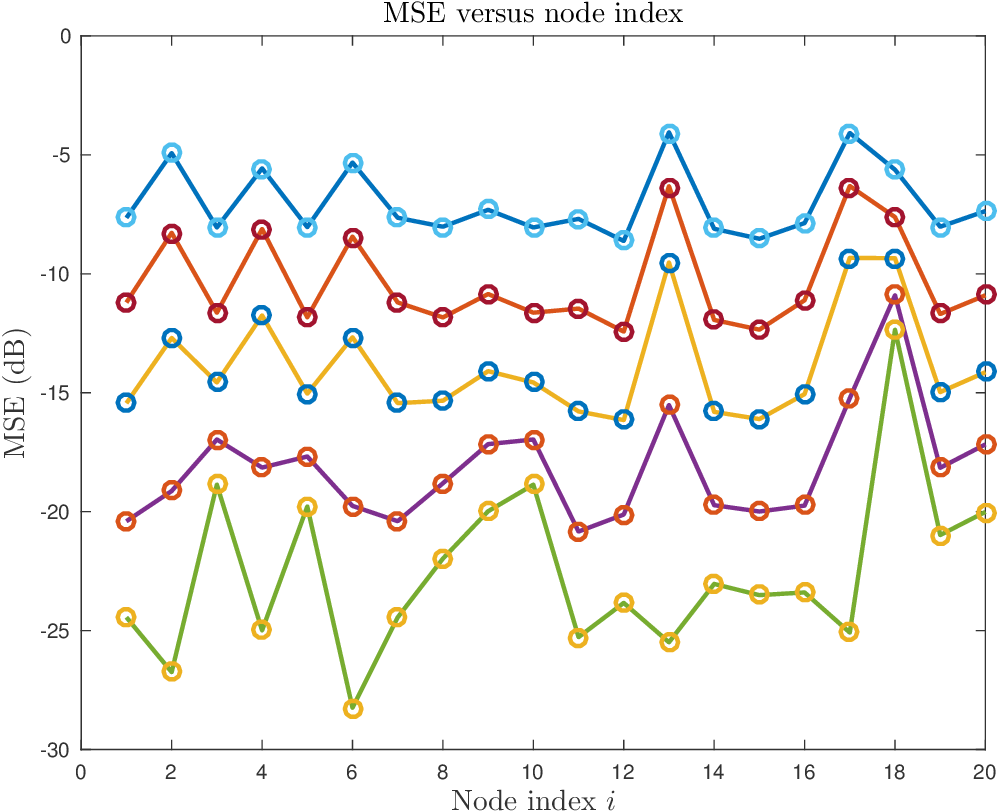}
    \caption{Optimal rates and MSE sequences from the solution of~\eqref{eq:cfun-final} ($T=5$, $\rho_c=0.35$, $\sigma_x^2=1$, $\sigma_n^2=0.5$, $m=20$). {\em Top:} Optimal rate sequences for the variable-distortion optimization problem. The rates are plotted against iteration indices, and each line represents the rates used by a different sensor. {\em Bottom:} MSE sequence corresponding to the above rate sequence. The MSE values are plotted against node indices, and each line represents a different iteration. In this case, the MSE values decrease monotonically for all nodes. The lines represent values predicted by the state evolution equations~\cref{\seeqs}, and the overlaid circles represent the simulated values.}
    \label{fig:varD-optimal-rates}
  \end{center}
\end{figure}

In this chapter, we present numerical results that provide further insight into the optimal rate sequences and their relationship to the target mean square error (MSE) and the network topology. We further compare the performance of the proposed generalized geometric program (GGP) and heuristic optimizations to the prior art. To test the effectiveness of the proposed approach, we used MATLAB~\cite{MATLAB} for simulation. We used the CVX toolbox~\cite{cvx,gb08} to solve the GGP presented in Chapter~\ref{chap-three}, which converts the GGP model to a convex problem automatically. 

In the results presented, the networks were generated by random geometric graph models~\cite{Penrose2003}, and each node state $\zv{i}{0}$ was initialized with the same independent and identically distributed (i.i.d.) variance-$\sigma_x^2$ zero-mean Gaussian vector $\x \in \reals^L$ corrupted by a different variance-$\sigma_n^2$, zero-mean Gaussian noise $\mathbf{n}_i \in \reals^L$, $i \in \{1,\ldots, m\}$. More precisely,
\begin{align}
\label{eq:rv-setting}
\zv{i}{0} = \x + \mathbf{n}_i, \fourspace \forall i \in \{1,\ldots,m\},
\end{align}
where $\x \sim \mathcal{N}(\mathbf{0}, \sigma_x^2 \I)$ and $\mathbf{n}_i \sim \mathcal{N}(\mathbf{0}, \sigma_n^2 \I)$, $\forall i \in \{1,\ldots, m\}$.
The random geometric graphs were generated on the unit torus (i.e., edge effects were neglected by ``wrapping'' the top and bottom and left and right edges of $[0,1]^2$)~\cite{Penrose2003,Torus,SquareTorus}.

In the case of very low rates using entropy-coded uniform scalar quantization (ECSQ), nothing would be communicated (all elements transmitted/received will be zeros). To prevent this behavior, the maximum normalized distortion $D_{\text{max}}$ allowed was set such that 1\% of the elements were nonzero.

\begin{figure}
  \begin{center}
    \includegraphics[width=\textwidth,height=\figpctg\textheight,keepaspectratio]{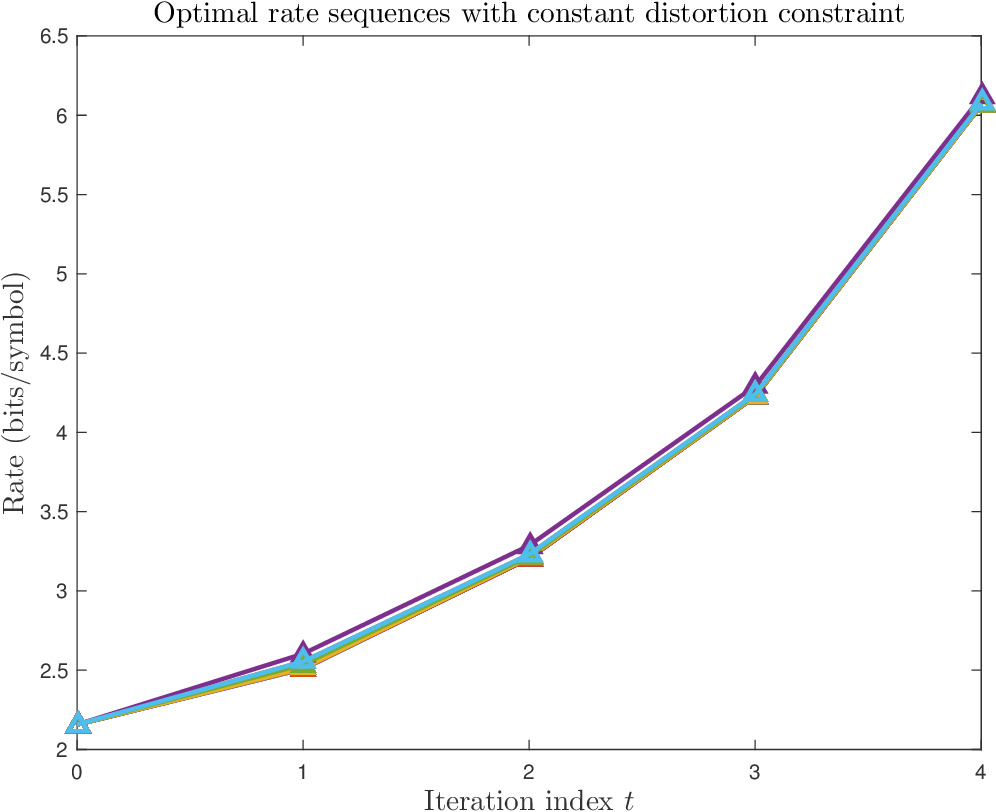} \\
    \vspace{0.4cm}
    \includegraphics[width=\textwidth,height=\figpctg\textheight,keepaspectratio]{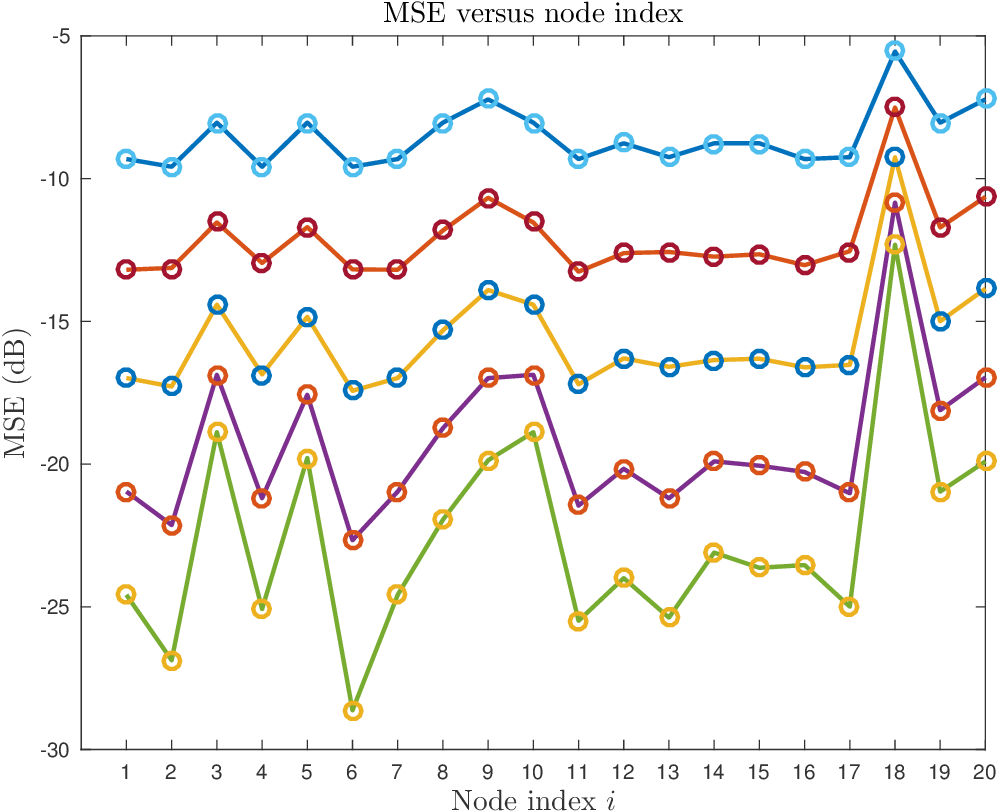}
    \caption{Optimal rates and MSE sequences from the solution of~\eqref{eq:cfun-mod} ($T=5$, $\rho_c=0.35$, $\sigma_x^2=1$, $\sigma_n^2=0.5$, $m=20$). {\em Top:} Optimal rate sequences for the constant-distortion optimization problem. The rates are plotted against iteration indices, and each line represents the rates used by a different sensor. {\em Bottom:} MSE sequence corresponding to the above rate sequence. The MSE values are plotted against node indices, and each line represents a different iteration. In this case, the MSE values decrease monotonically at all nodes. The lines represent values predicted by the state evolution equations~\cref{\seeqs}, and the overlaid circles represent the simulated values.}
    \label{fig:constD-optimal-rates}
  \end{center}
\end{figure}

\section{Structure of solutions}
To verify the state-evolution equations~\eqref{eq:se-data-mean-simple}--\eqref{eq:se-error-simple} and better understand the structure of the solutions to the optimization problems~\eqref{eq:cfun-final}~and~\eqref{eq:cfun-mod}, we present some anecdotal simulation results. Each of these results is taken from a different single instantiation of the optimization problem.

The optimal rate sequences $\{R_i(t)\}_{t=0}^{T-1}$, $i \in \{1,\ldots,m\}$ for both the variable-distortion \eqref{eq:cfun-final} and constant-distortion \eqref{eq:cfun-mod} problems typically exhibit monotonically nondecreasing structure, with an increasing rate of change toward the final iterations. In the constant-distortion case, the rates $R_i(t) \approx \frac{1}{2} \log_2\left(\frac{\margvar{i}{\Dv, t}}{D_i(t)} \right) + R_c$ are similar because the ratios $\frac{\margvar{i}{\Dv,t}}{D_i(t)}$ in the Gaussian operational rate-distortion (RD) relationship~\eqref{eq:pattern} are similar across the network. Examples of the optimal rate sequences and corresponding MSE state evolution are provided for both variants of the optimization problem~\cref{eq:cfun-final,eq:cfun-mod} in Figures~\ref{fig:varD-optimal-rates}~and~\ref{fig:constD-optimal-rates}.

The reader may wonder why the rate sequences of Figures~\ref{fig:varD-optimal-rates}~and~\ref{fig:constD-optimal-rates} look different. In Figure~\ref{fig:varD-optimal-rates}, the distortions $\{D_i(t)\}_{i=1,t=0}^{m,T-1}$ corresponding to the optimization of~\eqref{eq:cfun-final} are allowed to vary across both the node index $i \in \{1,\ldots,m\}$ and the iteration index $t \in \{0,\ldots,T-1\}$. This optimization is more flexible with respect to the choice of rates $\{R_i(t)\}_{i=1,t=0}^{m,T-1}$ corresponding to $\{D_i(t)\}_{i=1,t=0}^{m,T-1}$. In Figure~\ref{fig:constD-optimal-rates}, however, the distortion is constrained to be the same the network for a particular iteration (i.e., $\{D_i(t)\}_{i=1,t=0}^{m,T-1}$ varies only across $t \in \{1,\ldots,T-1\}$). This corresponds to the solution of~\eqref{eq:cfun-mod}.

The pattern of these rate sequences is intuitive and mirrors the results of Zhu and coauthors' study of multiprocessor approximate message passing~\cite{ZhuBaronMPAMP2016ArXiv,ZhuBeiramiBaron2016ISIT,ZhuPilgrimBaron2017}. As the estimate of the sample mean at each node increases in precision, higher-resolution messages must be exchanged among processors to achieve increasing estimate quality. In the case of coding without side information, this implies the use of larger coding rates in the later iterations of the algorithm.

\begin{figure}
  \begin{center}
    \includegraphics[width=\textwidth,height=\figpctgtwo\textheight,keepaspectratio]{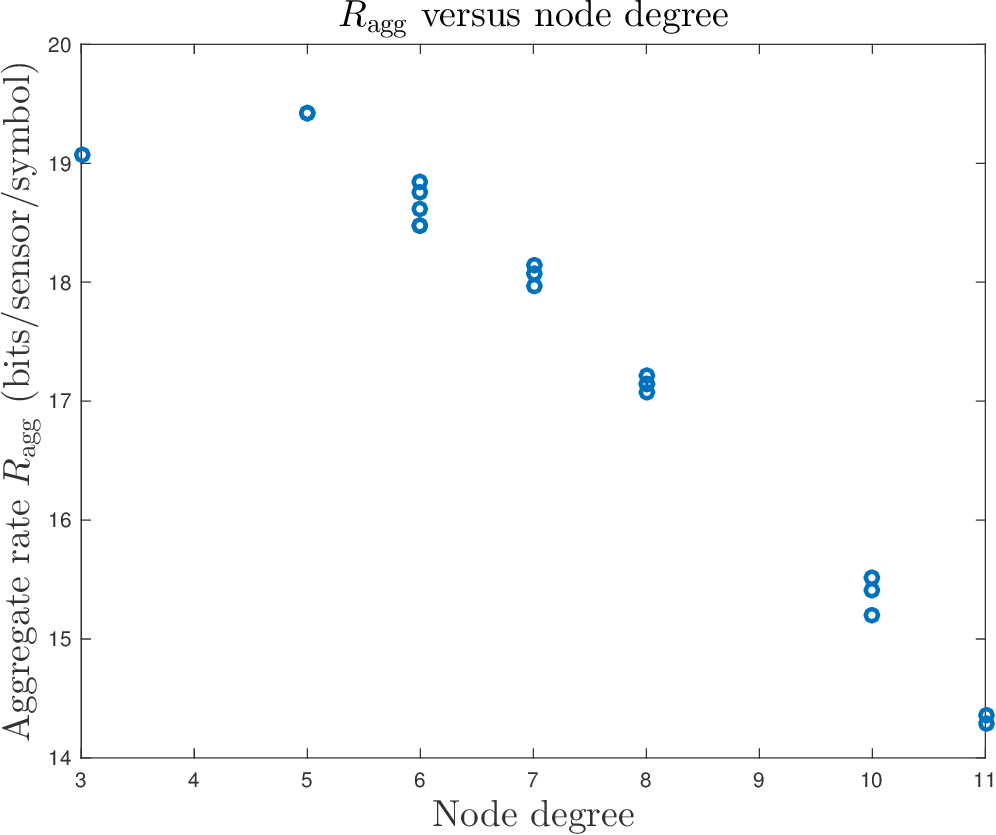} \\
    \vspace{0.4cm}
    \includegraphics[width=\textwidth,height=\figpctgtwo\textheight,keepaspectratio]{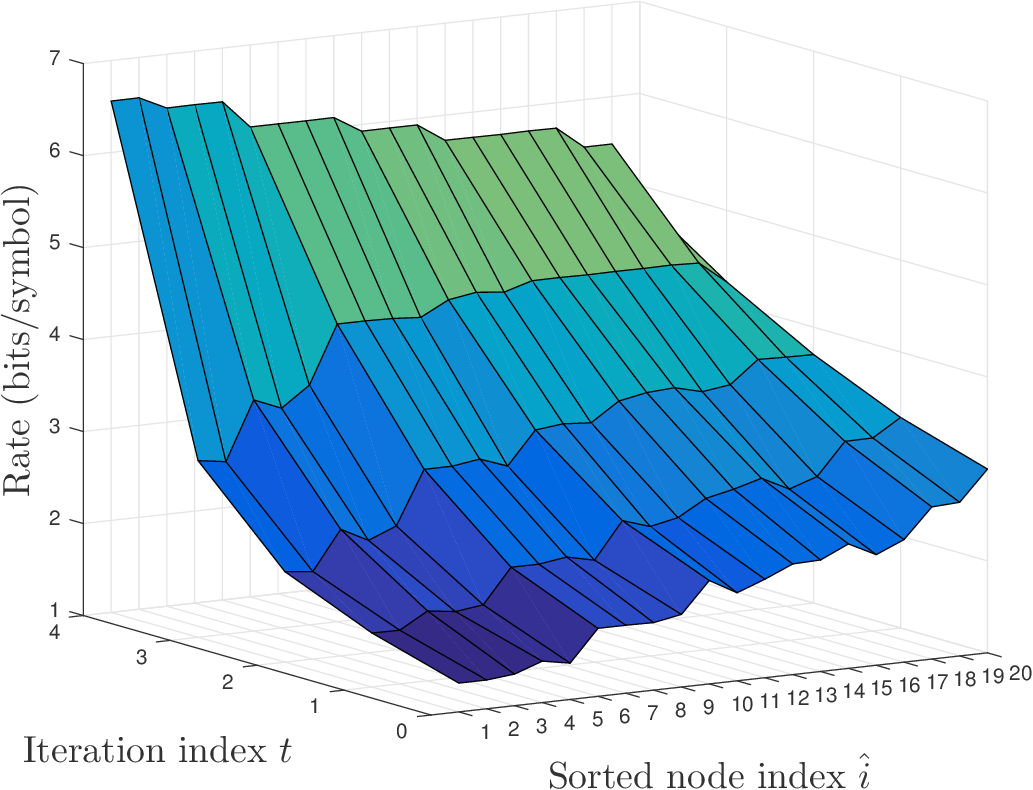}
    \caption{Influence of connectivity on the optimal rates from the solution of~\eqref{eq:cfun-mod} ($T=5$, $\rho_c=0.35$, $\sigma_x^2=1$, $\sigma_n^2=0.5$, $m=20$). {\em Top:} Scatter plot of aggregate rate used at each node versus node degree. {\em Bottom:} Surface of optimal rate sequences at each node, where the node indices are sorted by node degree in decreasing order.}
    \label{fig:sortedRateSurface}
  \end{center}
\end{figure}

The results of the variable-distortion problem~\eqref{eq:cfun-final} also exhibit structure based on the connectivity of each node. Figure~\ref{fig:sortedRateSurface} contains a scatter plot of the aggregate rate versus the node degree, which is a proxy for the ``connectedness'' of each node of the network. There is a negative correlation between the node degree and the aggregate rate used by each node. Figure~\ref{fig:sortedRateSurface} also contains a surface demonstrating the changing shape of the rate sequence as the node degree decreases. In general, the experimental results suggest that nodes with higher degree tend to have flatter optimal rate sequences, whereas the nodes with lower degree ramp up their coding rates more aggressively from beginning to end. 

\begin{figure}
  \begin{center}
    \includegraphics[width=\textwidth,height=\figpctgtwo\textheight,keepaspectratio]{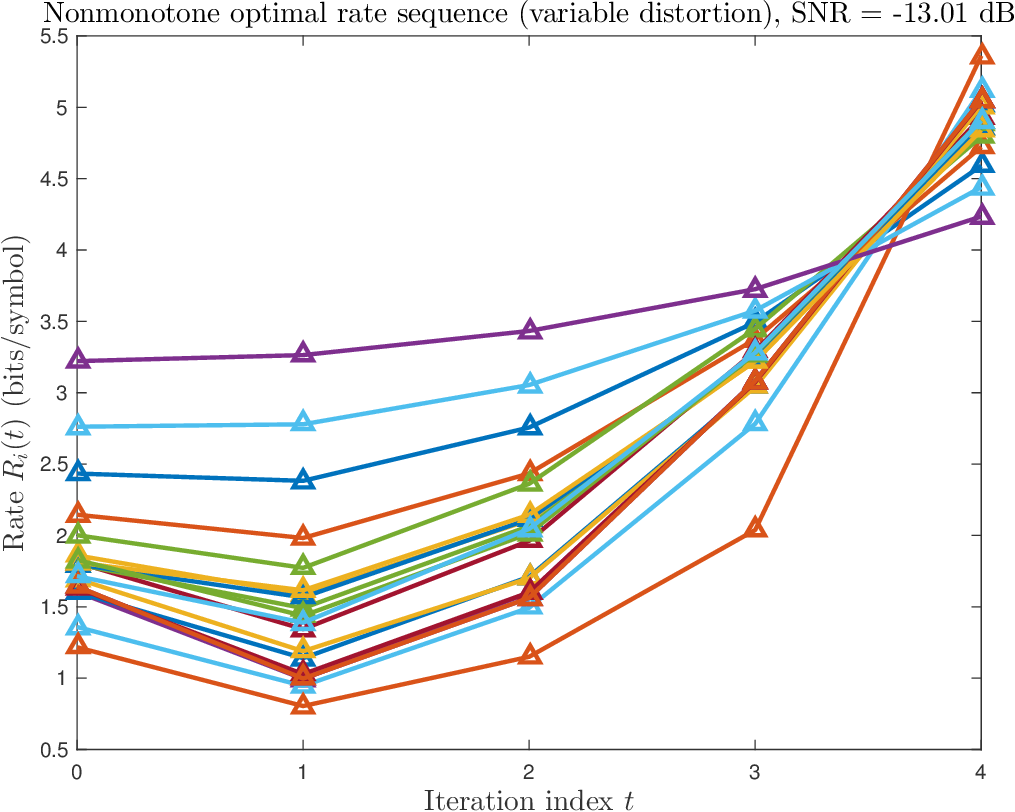} \\
    \vspace{0.4cm}
    \includegraphics[width=\textwidth,height=\figpctgtwo\textheight,keepaspectratio]{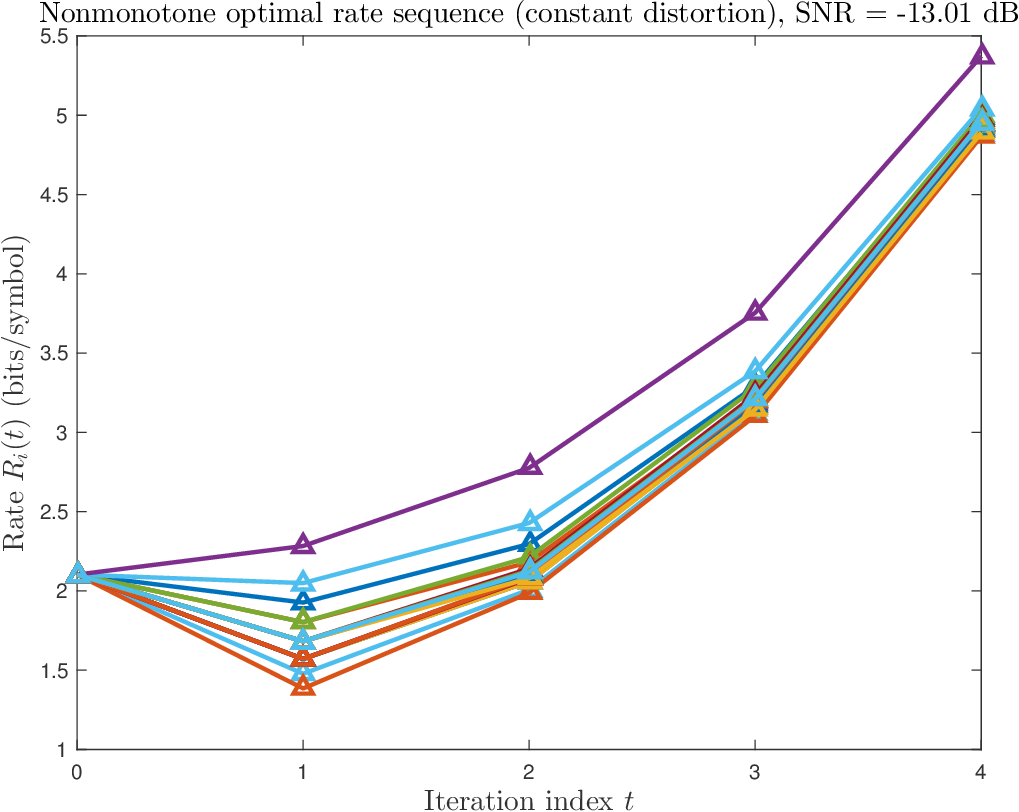}
    \caption{Nonmonotone optimal rate sequences for the low-SNR, low-MSE setting ($T=5$, $\rho_c = 0.35$, $\sigma_x^2 = 1$, $\sigma_n^2 = 20$, $m=20$). Each line corresponds to the rate sequence used by a different node. {\em Top:} Variable-distortion optimization~\eqref{eq:cfun-final} result. {\em Bottom:} Constant-distortion optimization~\eqref{eq:cfun-mod} result.} 
    \label{fig:dip}
  \end{center}
\end{figure}

Intuitively, this pattern can be explained by considering the overall goal of consensus and the impact of quantization on the information content~\cite{YangGroverKar2016} of the network's sample mean estimates. The more poorly connected nodes must use greater rates at each iteration to diffuse their information throughout the network, due to topological bottleneck effects~\cite{AyasoShahDahleh2010}. The more highly connected nodes, however, can save coding rate by exploiting their topological advantage, as the information content of their messages will be diffused more quickly.

In all cases simulated, the differences in aggregate rate between the exact optimization problem~\eqref{eq:cfun-final} and the constant-distortion problem~\eqref{eq:cfun-mod} were small, on the order of 5\%. For the sequences plotted in Figures~\ref{fig:varD-optimal-rates}~and~\ref{fig:constD-optimal-rates}, the difference was 4.45\%. This similarity may be an effect of the highly regular random geometric graph structure. For random geometric networks on squares rather than tori, the difference may be more significant due to the irregularity arising from edge effects.


\subsection{Nonmonotone behavior}
Intuitively, it seems reasonable to expect the MSE to decrease monotonically, and the rates to increase monotonically, since greater precision is required to encode the increasingly precise estimates of the sample average available to each node. However, in certain edge cases, neither of these patterns holds. In this subsection, we discuss these behaviors and present some hypothesized reasons for these nonintuitive results.

Figure~\ref{fig:dip} demonstrates a case in which the optimal rate sequences are nonmonotone. In this case, the signal-to-noise ratio (SNR),
\begin{align}
\text{SNR} := 10 \log_{10} \frac{\sigma_x^2}{\sigma_n^2},
\end{align} 
is low (-13.01 dB), which introduces a trade-off for the rates $R_i(t)$ corresponding to the first few iterations of the consensus algorithm. Recall that the simulation initializes each node to have access to the same zero-mean Gaussian signal plus its own realization of additive white Gaussian noise (AWGN). In this case, the covariance matrix~\eqref{eq:se-data-simple} has the form 
\begin{align}
\label{eq:awgn-form}
  \Sig{\mathbf{z}}{t=0} = \sigma_x^2 \ones \ones \tp + \sigma_n^2 \I.
\end{align}
If the signal variance is set to zero (corresponding to the case of averaging independent AWGN processes), then the diagonals of the covariance matrix~\eqref{eq:se-data-simple} will converge to $\frac{\sigma_n^2}{m}$ as the iteration index $t \rightarrow \infty$, which is significantly less than the original variance $\sigma_n^2$ for large networks.

\begin{figure}
  \begin{center}
    \includegraphics[width=\textwidth,height=\figpctgtwo\textheight,keepaspectratio]{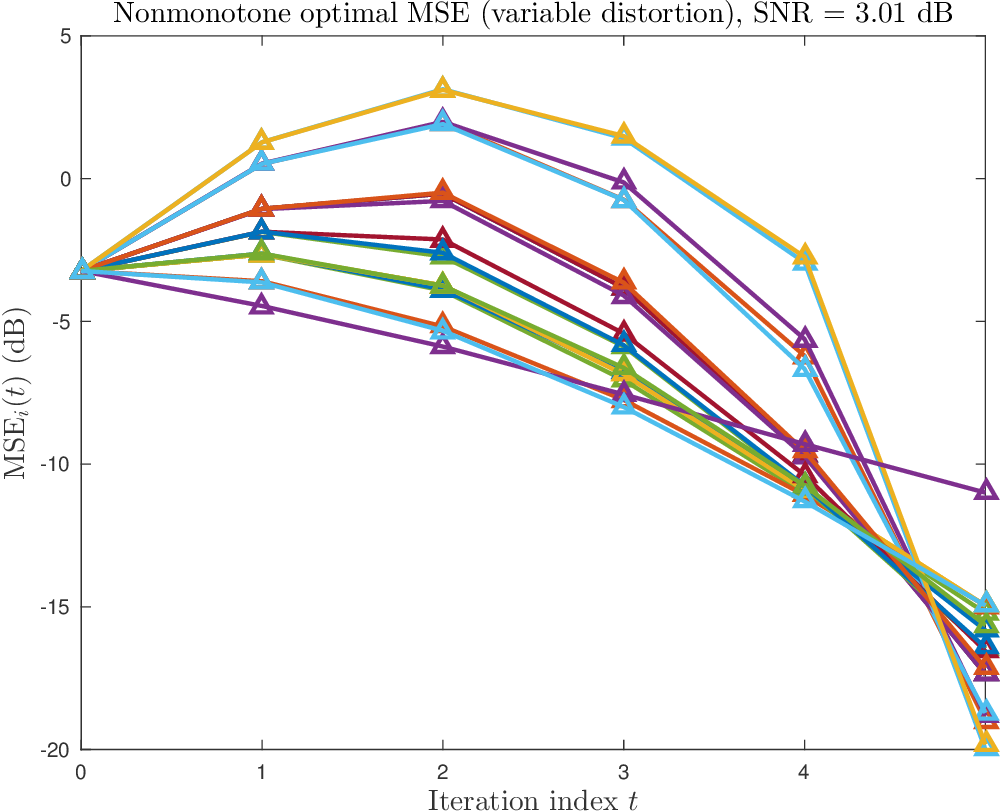} \\
    \vspace{0.4cm}
    \includegraphics[width=\textwidth,height=\figpctgtwo\textheight,keepaspectratio]{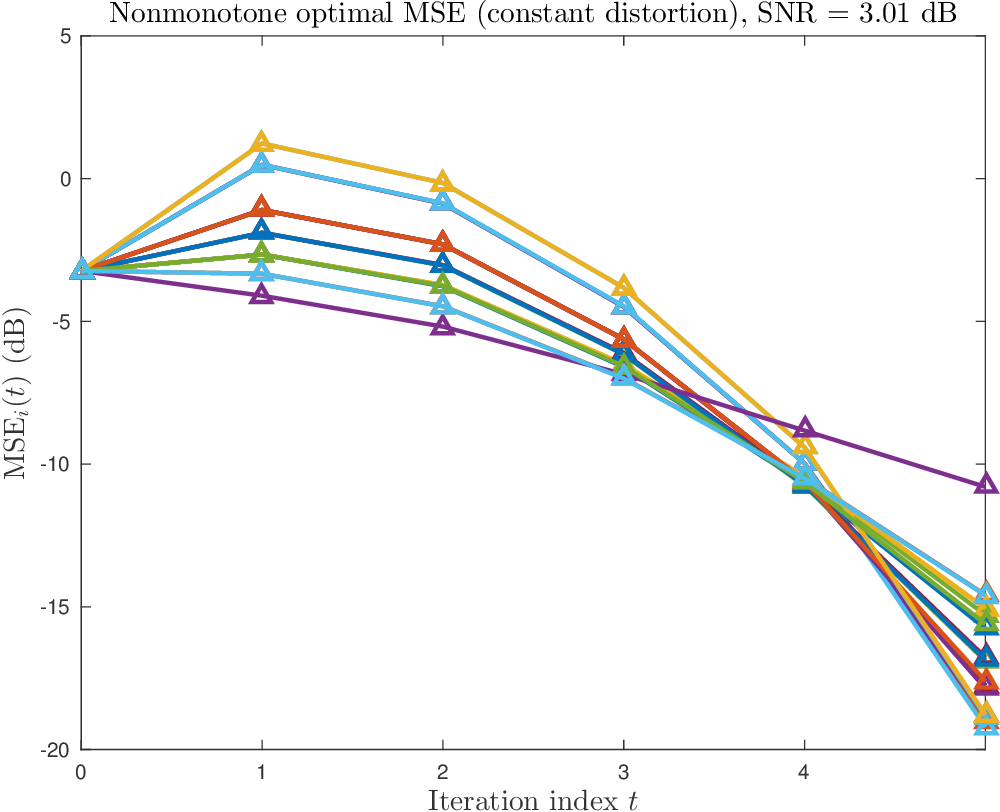}
    \caption{Nonmonotone MSE sequences. Each line corresponds to the MSE sequence of a different node ($T=5$, $\rho_c=0.35$, $\sigma_x^2=1$, $\sigma_n^2=0.5$). {\em Top:} MSE evolution for the variable-distortion problem \eqref{eq:cfun-final}. {\em Bottom:} MSE evolution for the constant-distortion problem \eqref{eq:cfun-mod}.}
    \label{fig:bump}
  \end{center}
\end{figure}

Therefore, for low SNR, the impact of quantization error on the variance becomes significant. The variance state-evolution equation is~\eqref{eq:se-data-simple}
\begin{align}
\margvar{i}{t} &= \left[  \W{t} \Sig{\mathbf{z}}{0} \W{t} + \sum_{s=0}^{t-1} \W{t-s-1} \WsubI \Sig{\qesu}{s} \WsubI \W{t-s-1} \right]_{ii}.
\end{align}
It can be seen from the above equation that both the original covariance matrix and the distortions introduced in past iterations play a role in the variance at any iteration $t \in \mathbb{Z}_{>0}$. Furthermore, recall the family of operational RD relationships we study~\eqref{eq:pattern},
\begin{align}
R_i(\Dv, t)  \approx \frac{1}{2}  \log_2 \left( \max \left\{ \frac{\margvar{i}{\Dv,t}}{\Dmarg{i}{t}}, k \right\} \right) + R_c ,
\end{align}
which indicates that, holding the distortion $D_i(t)$ constant, larger coding rates are required if the source has higher variance $\margvar{i}{t}$. From~\eqref{eq:se-var} and~\eqref{eq:awgn-form}, we can see that introducing larger distortions will increase the source variance for all future iterations, and thus increase the communication cost. Therefore, using higher coding rates (introducing lower distortions) in the initial iterations may be preferable in settings where the source variance makes a significant contribution to the overall cost, as in the low-SNR setting explored here.

It seems intuitive to expect the MSE to decrease monotonically in iterations as $t \rightarrow \infty$, but this has been observed not to be true in general. If the target $\text{MSE}^*$ is much higher than the $\MSE{\Dv, T}$~\eqref{eq:se-mse-net} achievable in the lossless case (i.e., $R_i(t) \rightarrow \infty, \forall i \in \{1,\ldots,m\},$ $t \in \{0,\ldots,T-1\}$), then the MSE at each node might increase initially, followed by a rapid decrease in the later iterations. This behavior is illustrated in Figure~\ref{fig:bump}. We hypothesize that this nonmonotonicity is an effect of the different treatment of states and quantization errors in the state update. Recall the iteration we use~\eqref{eq:lossy-iteration}, borrowed from Frasca \etal~\cite{Frasca2008}:
\begin{align}
\ze{t+1} &= \ze{t} + \left(\Om{} - \I\right)\q{\ze{t}}.
\end{align} 
This equation can be rewritten as~\eqref{eq:lossy-alt},
\begin{align}
\ze{t+1} &= \Om \ze{t} + \left(\Om{} - \I\right)\qe{t},
\end{align}
where the reader is reminded that $\ze{t} \in \reals^{mL}$ is the state supervector~\eqref{eq:network-state-def} that contains the vector-valued states $\zv{i}{t}$ at each node $i \in \{1,\ldots,m\}$, and $\qe{t} \in \reals^{mL}$ is the quantization error~\eqref{eq:quant-err-def} corresponding to each element of $\ze{t}$. Because the state $\ze{t}$ and quantization error $\qe{t}$ are incorporated into the state update differently, they have different impacts on the MSE. In our case, this difference appears to be favorable in the sense that it allows the impact of distortions to decay quickly.

Mathematically, the reason for the nonmonotone optimal MSE behavior may be discerned from the state evolution equations. The reader is reminded of the MSE state evolution relationship~\eqref{eq:se-mse},
\begin{align}
\mathrm{MSE}_i(t) &= \left[ \M \W{t} \Sig{\mathbf{e}}{0} \W{t} \M \right. \nonumber \\
&+ \left. \sum_{s=0}^{t-1} \M \W{t-s-1} \WsubI \Sig{\qesu}{s} \WsubI \W{t-s-1} \M \right]_{ii} ,
\label{eq:t}
\end{align}
where the $\mean{\mathbf{e}}{t} \meant{\mathbf{e}}{t}$ term is omitted because the simulated signal and noise are zero-mean.

To simplify the explanation of the nonmonotone MSE behavior, we will consider the constant-distortion optimization, for which the above~\eqref{eq:t} reduces to
\begin{align}
\mathrm{MSE}_i(t) &= \left[ \M \W{t} \Sig{\mathbf{e}}{0} \W{t} \M \right. \\
&+ \left. \sum_{s=0}^{t-1} D(s) \M \W{t-s-1} \WsubI^2 \W{t-s-1} \M \right]_{ii} .
\end{align}

\begin{figure}
  \begin{center}
    \includegraphics[width=\textwidth,height=\textheight,keepaspectratio]{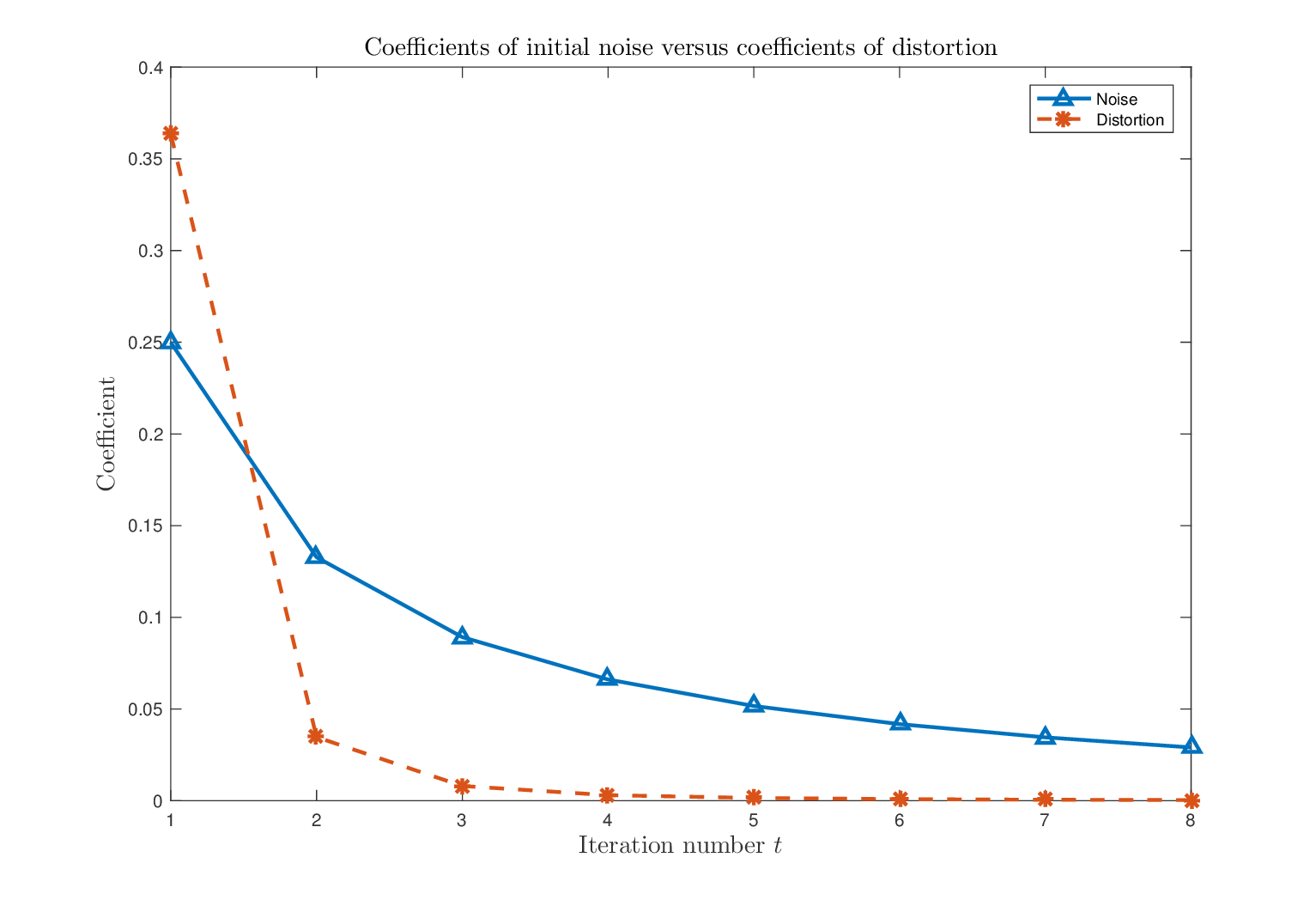}
    \caption{Coefficients of the noise variance $\sigma_n^2$ (solid blue curve) and distortion $D(0)$ (dashed red curve) in the equation for the MSE~\eqref{eq:se-mse} when the states are initialized as a Gaussian random vector corrupted by AWGN ($\rho_c=0.35$, $m=20$).}
    \label{fig:coeffs}
  \end{center}
\end{figure}

In the Gaussian-signal-plus-noise setting~\eqref{eq:rv-setting} mentioned at the beginning of this chapter, the initial error covariance matrix~\eqref{eq:se-error-simple} is $\Sig{\mathbf{e}}{0} = \sigma_n^2 (\I - \avg)$. Because $\I - \avg$ is idempotent,\footnote{An idempotent matrix $\A$ is one for which $\A^2 = \A$~\cite{HoffmanKunze71}.} the first term in the above equation reduces to $\sigma_n^2 \M \W{2t} \M$, and we can assess the different contributions of noise variance $\sigma_n^2$ and distortion $\D{t}$ to the network MSE, $\MSE{\Dv, t}$, by examining the traces of the matrices
\begin{align}
\M \W{2t} \M
\end{align} and 
\begin{align}
\M \W{t} \WsubI^2 \W{t} \M,
\end{align} 
for $t \in \{0,\ldots,T-1\}$. The reader is reminded that the trace of a matrix $\A \in \reals^{n\times n}$, written $\tr{\A}$, is the sum of its diagonal elements (i.e., $\tr{\A} := \sum_{i=1}^n a_{ii}$).

These diagonal components are the coefficients of the noise variance $\sigma_n^2$ and the distortions $D_i(t)$ in the equation for the MSE. Figure~\ref{fig:coeffs} plots the diagonal values for the above matrices, $\forall t \in \{0,\ldots,T-1\}$. The coefficients of $\D{t}$ decrease sharply after a single iteration, meaning that the distortion contribution to the MSE vanishes quickly compared to the contribution of the initial state variance.

\section{Comparison to prior art}
To compare our work to the prior art~\cite{YildizScaglione2008,Thanou2013}, we generated 32 random geometric networks~\cite{Penrose2003} with connectivity radius\footnote{A random geometric network is one for which each node $V_i \in \mathcal{V}$ is associated with a geographic coordinate $\mathbf{v}_i$. Nodes $V_i, V_j$ are connected if $\norm{\mathbf{v}_i-\mathbf{v}_j}_2 \leq \rho_c$~\cite{BoydMixingTimes}.} $\rho_c \in \{0.35, 0.45\}$ on a two-dimensional unit torus. For each of these networks, consensus was run on 1000 realizations of the initial states, which were length-10000 i.i.d. Gaussian vectors $\mathbf{z}_i(0) = \x + \mathbf{n}_i,$ $\forall i \in \{1,\ldots,m\}$, $\x \sim \mathcal{N}(\mathbf{0}, \I), \mathbf{n}_i \sim \mathcal{N}(\mathbf{0}, 0.5 \I)$ (SNR = 2, which is 3.01 dB).

We simulated ProgQ~\cite{Thanou2013} and order-one predictive coding~\cite{YildizScaglione2008}, using initial rates $R_i(0) \in \{4,\ldots,7\} \fourspace \forall i$ and $R_i(0) \in \{3,\ldots,6\} \fourspace \forall i$, respectively. The measured final MSE~\eqref{eq:se-mse} for these schemes were set as the target values for the proposed GGP and heuristic optimizations.

For all schemes, $\MSE{\Dv, T}$ and $\Ragg$ were computed. These values were averaged over all 32 realizations of each ($\rho_c$, $R_i(0)$, $T$) setting, and the resulting averages were plotted against each other. Assuming an additive noise model for the quantization error means that the MSE for quantized communication will always be higher than the MSE of the unquantized algorithm. We therefore introduce two terms to define the MSE performance relative to the ideal, unquantized algorithm. To compensate for the effect of network topology on the MSE, we define the {\em lossless MSE},
\begin{align}
\text{MSE}_{\text{lossless}}(t) := \MSE{\Dv,t} \Big\vert_{\Dv=\mathbf{0}}.
\end{align}
We then define the {\em excess MSE} (EMSE) as
\begin{align}
\text{EMSE} := 10 \log_{10} \frac{\text{MSE}(\Dv,T)}{\text{MSE}_{\text{lossless}}(T)},
\end{align}
which represents the increase in MSE over lossless consensus resulting from distortion.
The EMSE was used for the generation of RD trade-off curves.

\subsection{Implementation details}
To generate the weight matrix $\W{}$, the max-degree heuristic~\cite{XiaoBoyd2004,Xiao2005} was used, so that
\begin{align}
\label{eq:max-degree}
\W{} &= \I - \alpha \mathbf{L} = \I - \frac{0.9}{\max_i \{ \deg i \}_{i=1}^m} \mathbf{L},
\end{align}
where $\deg i$ is the degree of node $i$ and $\mathbf{L}$ is the graph Laplacian~\autocite[Sec. 10.9]{GraphsOptimizationAndAlgos}. This weighting scheme was selected for simplicity, following the implementation of Yildiz and Scaglione~\cite{YildizScaglione2008}. The factor of 0.9 is used to ensure numerical stability, since the coefficient $\alpha$ of $\mathbf{L}$ in~\eqref{eq:max-degree} must satisfy $\alpha \in (0,1/\max_i \{ \deg i \}_{i=1}^m)$ to guarantee convergence~\cite{XiaoBoyd2004}. Based on \eqref{eq:max-degree}, $\W{}$ is symmetric and satisfies all the requirements of the matrix properties discussed in Chapter~\ref{chap-two}. 

The implementation of Yildiz and Scaglione's order-one predictive coding~\cite{YildizScaglione2008}, which was graciously provided by the authors, was modified to use fixed-rate uniform quantization but allow for the rate to vary with iteration and node indices. This capability was implemented by running two rate update recursions---one to keep track of the ideal (real-valued) rates given by the quantization noise variance recursion~\cite{YildizScaglione2008}, and another to perform the predictive coding using rates that were rounded to the nearest integral value. Interested readers can access the code, which will be posted online, for further details.

The optimization modeler for the GGP and its associated solvers suffered from poor stability at high iteration count, and poor scaling in the network size and number of iterations for the exact program~\eqref{eq:cfun-final}. Therefore, only the fixed-distortion problem~\eqref{eq:cfun-mod} was solved for comparison to the prior art.

The bin size of all fixed-rate uniform quantizers was set to 12 times the standard deviation of the Gaussian data to prevent clipping.

\subsection{Discussion of comparison results}

\begin{figure}
  \begin{center}
    \includegraphics[width=\textwidth,height=\figpctgtwo\textheight,keepaspectratio]{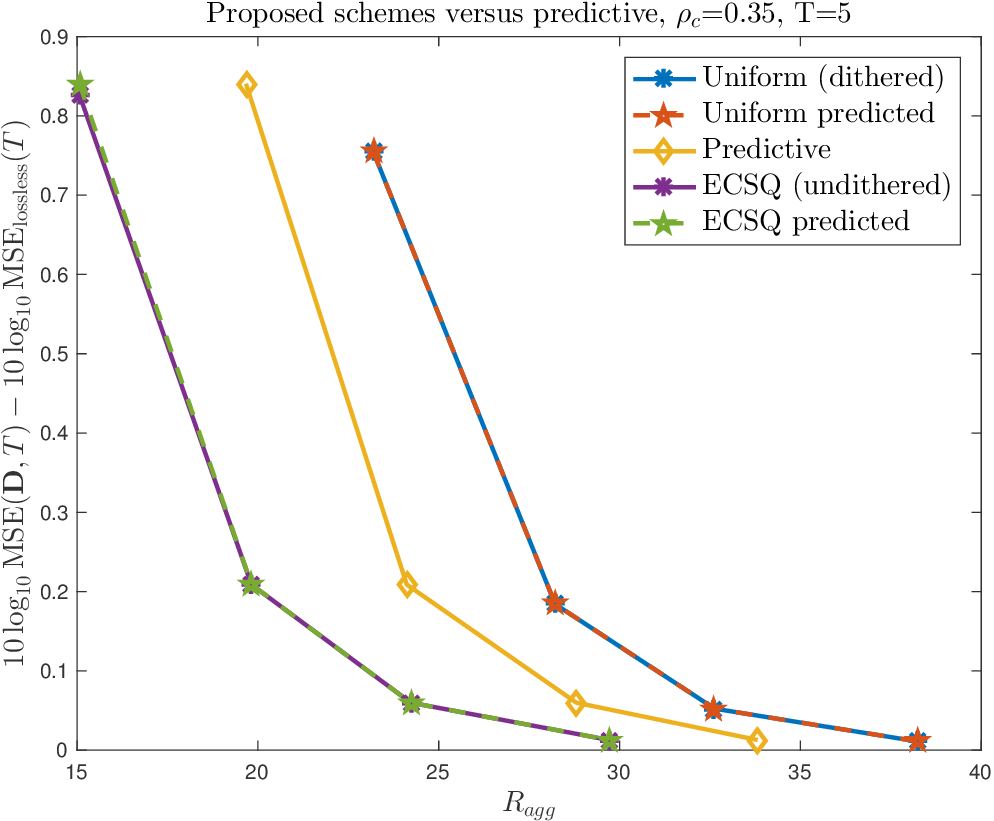} \\
    \vspace{0.4cm}
    \includegraphics[width=\textwidth,height=\figpctgtwo\textheight,keepaspectratio]{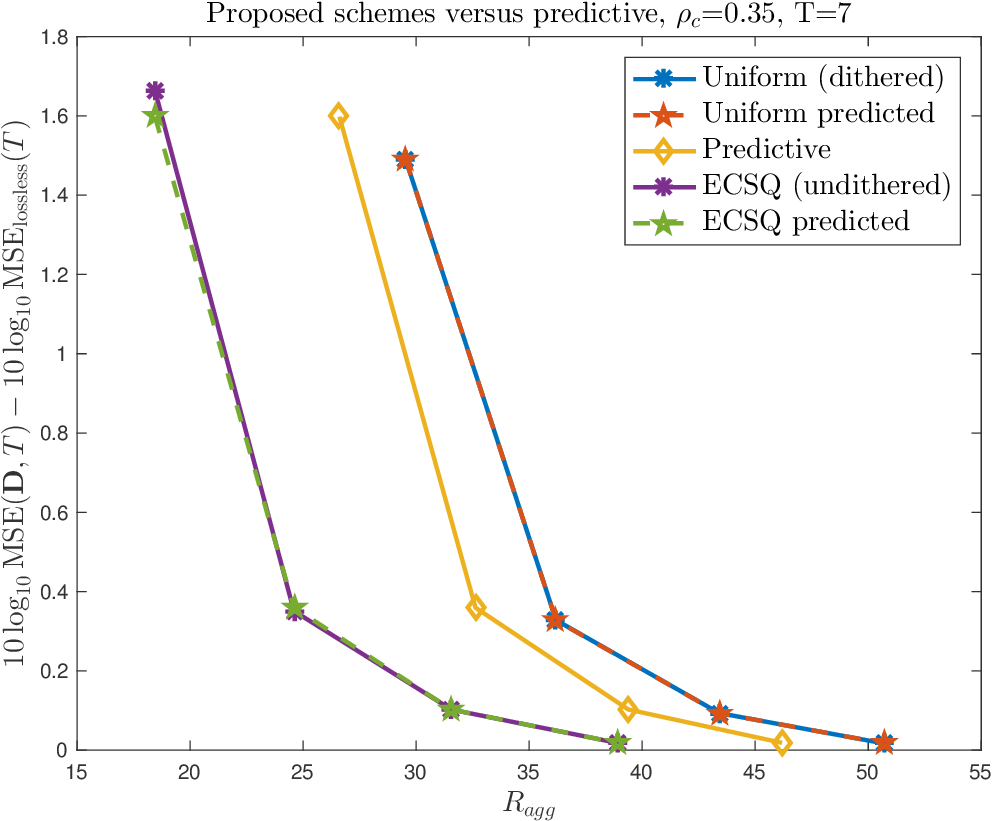}
    \caption{RD trade-off curves for the proposed schemes versus order-one predictive coding~\cite{YildizScaglione2008}. All schemes were simulated for random geometric graphs on a unit torus with connectivity radius $\rho_c=0.35$. {\em Top}: number of iterations $T=5$. {\em Bottom}: number of iterations $T=7$.}
    \label{fig:tradeoff-curve-ysp1}
  \end{center}
\end{figure}

\begin{figure}
\begin{center}
\includegraphics[width=\textwidth,height=\figpctgtwo\textheight,keepaspectratio]{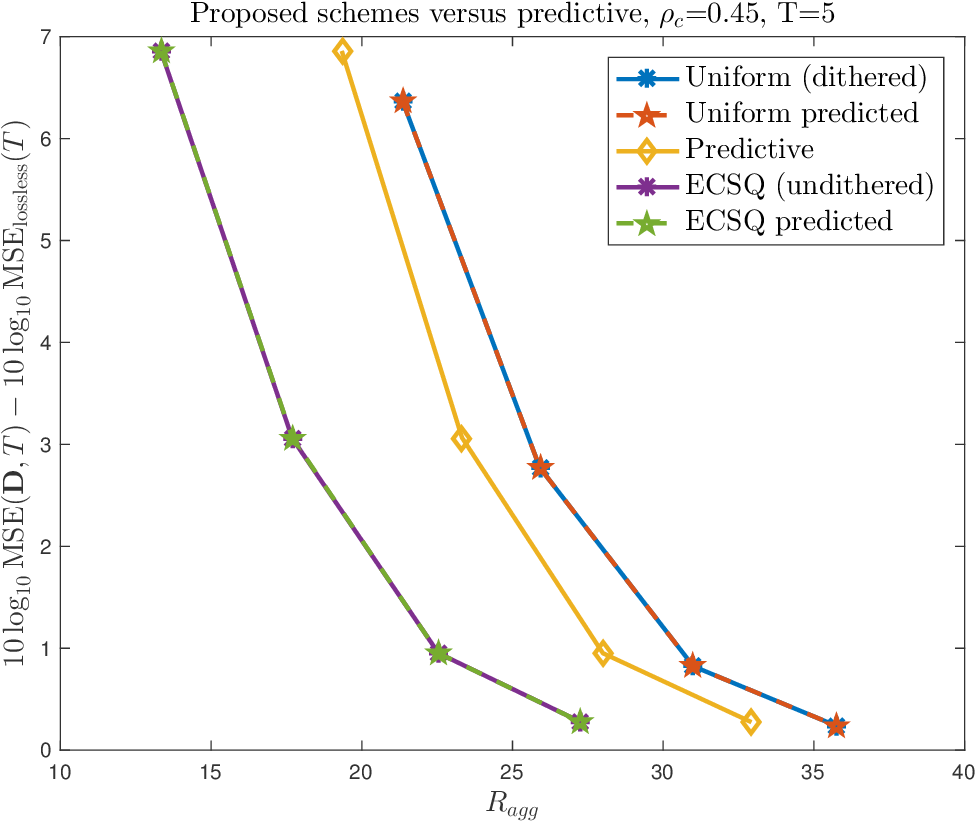} \\
\vspace{0.4cm}
\includegraphics[width=\textwidth,height=\figpctgtwo\textheight,keepaspectratio]{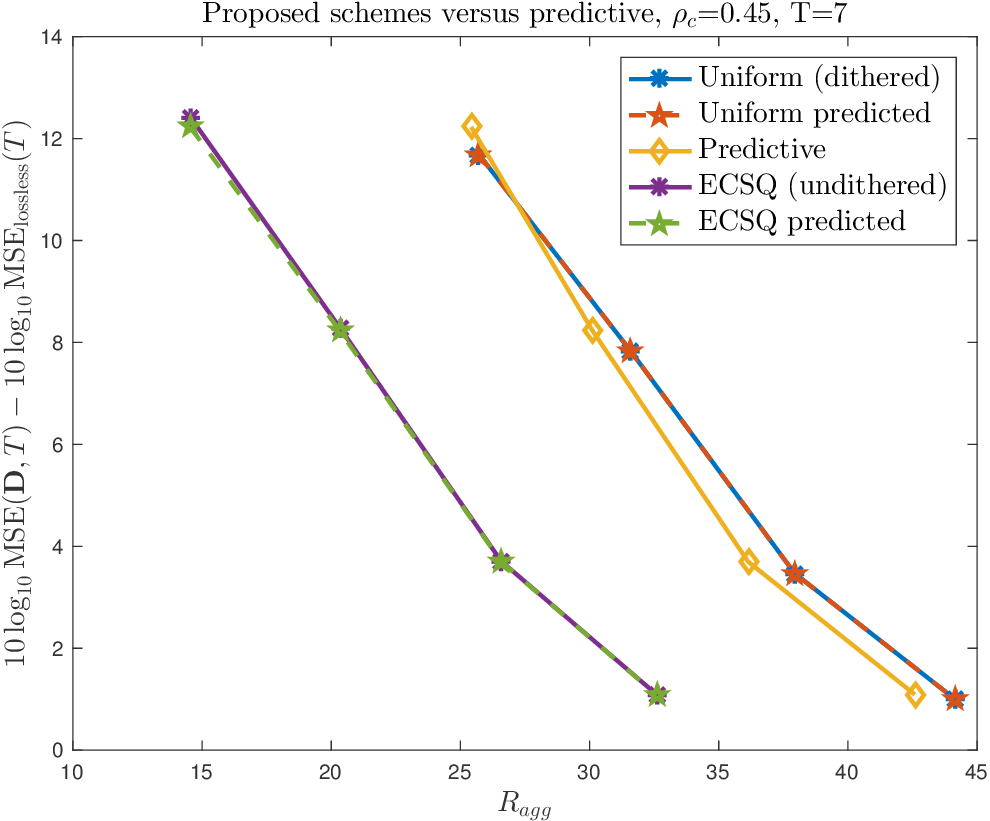}
\caption{RD trade-off curves for the proposed schemes versus order-one predictive coding~\cite{YildizScaglione2008}. All schemes were simulated for random geometric graphs on a unit torus with connectivity radius $\rho_c=0.45$. {\em Top}: number of iterations $T=5$. {\em Bottom}: number of iterations $T=7$.}
\label{fig:tradeoff-curve-ysp2}
\end{center}
\end{figure}

To adequately discuss the RD results, it is first necessary to comment on some properties of each of the schemes presented. The ProgQ algorithm of Thanou \etal~\cite{Thanou2013} uses a time- and node-invariant fixed-rate uniform quantizer (i.e., $R_i(t) = R$, $\forall i \in \{1,\ldots,m\}, t \in \{0, \ldots, T-1\}$), whereas Yildiz and Scaglione~\cite{YildizScaglione2008} allow the use of different rates at each node and iteration. 

Thanou \etal~\cite{Thanou2013} use the same state update as ours~\eqref{eq:lossy-iteration}, but Yildiz and Scaglione~\cite{YildizScaglione2008} use a different update that is incapable of truly converging in the presence of quantization error. The final asymptotic MSE for the predictive scheme depends on the sum of distortions $D_i(t), t \in \mathbb{Z}_{\geq 0}$. If these distortions are chosen to form a convergent series, then the MSE will converge to a nonzero, but bounded, value. Because of this limitation, the predictive scheme~\cite{YildizScaglione2008} is heavily dependent on the starting rates $R_i(0)$.

\begin{figure}
  \begin{center}
    \includegraphics[width=\textwidth,height=\figpctgtwo\textheight,keepaspectratio]{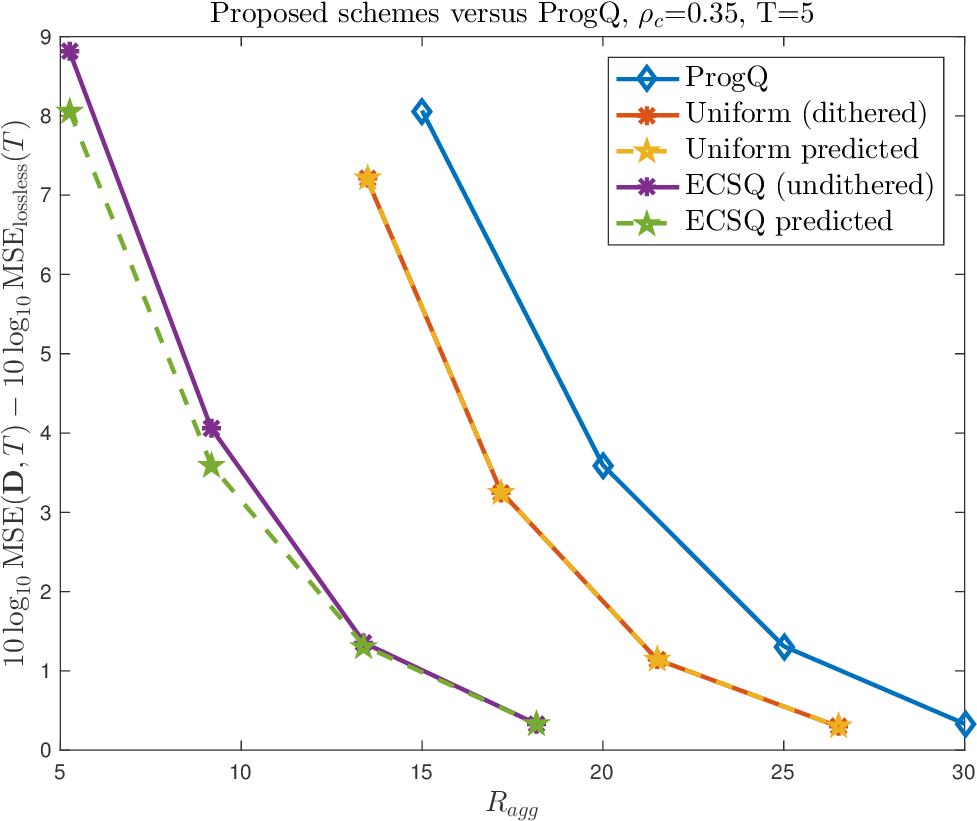} \\
    \vspace{0.4cm}
    \includegraphics[width=\textwidth,height=\figpctgtwo\textheight,keepaspectratio]{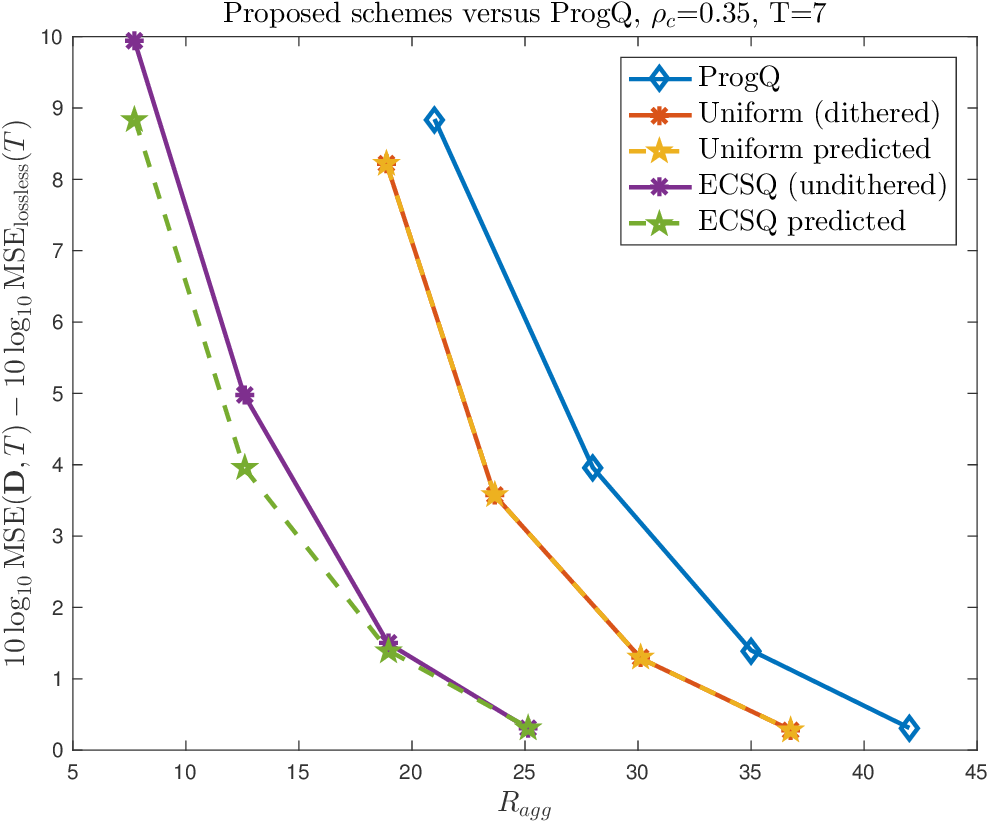}
    \caption{RD trade-off curves for the proposed schemes versus ProgQ~\cite{Thanou2013}. All schemes were simulated for random geometric graphs on a unit torus with connectivity radius $\rho_c=0.35$. {\em Top}: number of iterations $T=5$. {\em Bottom}: number of iterations $T=7$.}
    \label{fig:tradeoff-curve-progq1}
  \end{center}
\end{figure}

\begin{figure}
  \begin{center}
    \includegraphics[width=\textwidth,height=\figpctgtwo\textheight,keepaspectratio]{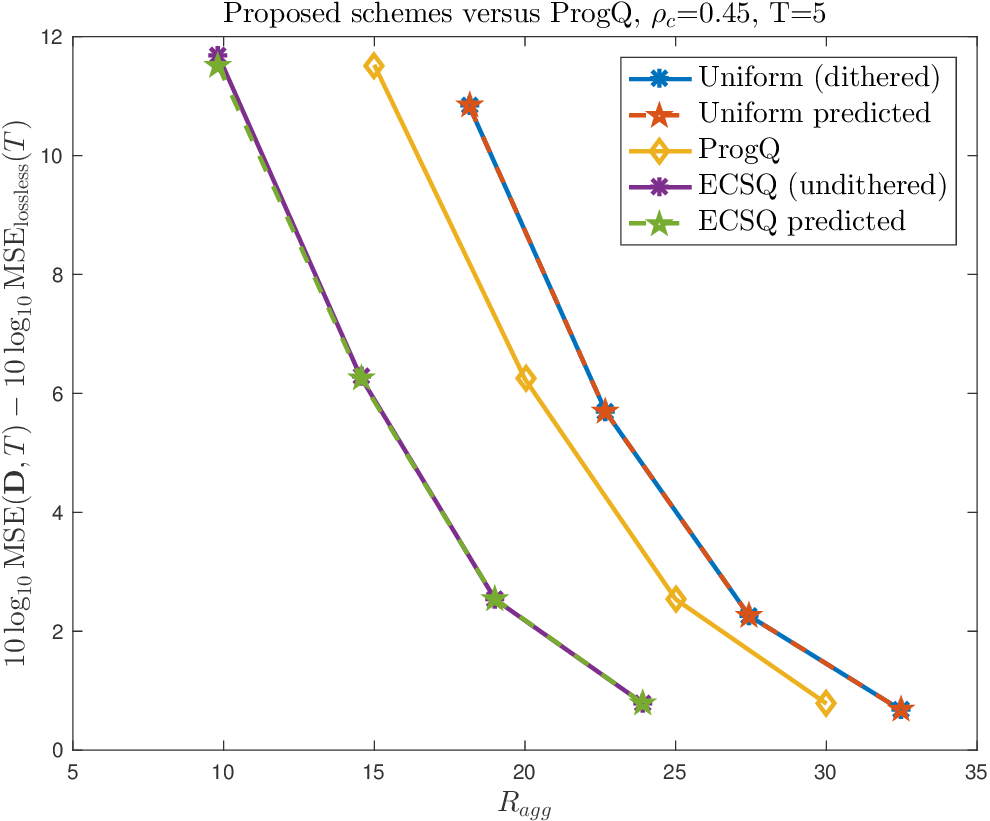} \\
    \vspace{0.4cm}
    \includegraphics[width=\textwidth,height=\figpctgtwo\textheight,keepaspectratio]{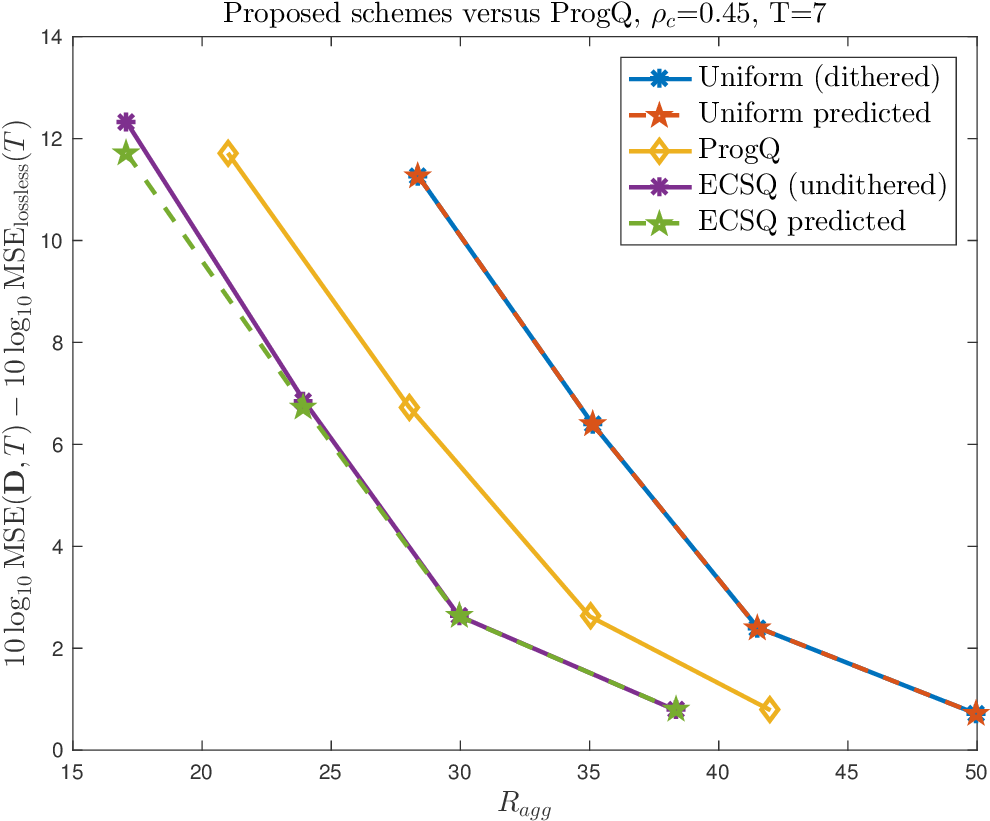}
    \caption{RD trade-off curves for the proposed schemes versus ProgQ~\cite{Thanou2013}. All schemes were simulated for random geometric graphs on a unit torus with connectivity radius $\rho_c=0.45$. {\em Top}: number of iterations $T=5$. {\em Bottom}: number of iterations $T=7$.}
    \label{fig:tradeoff-curve-progq2}
  \end{center}
\end{figure}

For each fixed-rate uniform quantizer sequence in the prior art comparisons, the result of the heuristic from Sec.~3.4 was compared to the optimum obtained by an exhaustive search over all possible rate sequences with a maximum aggregate rate value per node of $\sum_{t=0}^{T-1} R_{\text{GGP}}(t)$. In every case, the heuristic was able to find the optimal result. These results seem to justify our heuristic search. Despite the reduced search complexity (exponential instead of combinatorial) of this heuristic, the simulated performance was no worse. Therefore, at the least, this heuristic seems to be a promising approach whenever fixed-rate coding is used. The RD performances of the proposed optimization schemes are compared to the predictive coding scheme of Yildiz and Scaglione~\cite{YildizScaglione2008} in Figures~\ref{fig:tradeoff-curve-ysp1} and \ref{fig:tradeoff-curve-ysp2} and to the ProgQ algorithm of Thanou \etal~\cite{Thanou2013} in Figures~\ref{fig:tradeoff-curve-progq1} and \ref{fig:tradeoff-curve-progq2}.

These RD trade-off curves (Figures~\ref{fig:tradeoff-curve-ysp1}--\ref{fig:tradeoff-curve-progq2}) contain the performance of both ECSQ and fixed-rate uniform quantization. For both of our schemes, the predicted RD performance (as computed by the state evolution equations~\cref{\seeqs}) is compared to the actual performance. The distortion (measured by the EMSE) is given by the $y$-axis, and the aggregate rate $R_{\text{agg}}$ by the $x$-axis. The predicted performance is denoted by a dashed line, while the simulated performance is represented as a solid line. The legend disambiguates the many curves that are plotted. Alongside our approaches, we plot the RD performance of one of the comparators. A curve closer to the bottom left corner of these figures indicates better performance, meaning lower aggregate rate $R_{\text{agg}}$ to achieve the same EMSE, or alternatively lower EMSE for a particular $R_{\text{agg}}$ investment.

In some cases, the predicted performance and measured performance of ECSQ do not match. Because the ECSQ used in the simulations is not dithered, the additive quantization model only holds approximately. As the aggregate rate increases, the performance improves, and better adherence to predicted performance can be accomplished using dithering. 

In addition to the ECSQ and fixed-rate uniform quantizers, Lloyd-Max quantization~\cite{Max60,Lloyd82} was studied. However, its empirical MSE performance did not match predictions, because its quantization error $\qe{t}$ is correlated with the source $\ze{t}$. Due to this correlation, the use of Lloyd-Max quantization in our framework seems inappropriate.

\subsection{Comparison to predictive coding scheme} 
If we neglect the effect of entropy coding on the RD curves in Figures~\ref{fig:tradeoff-curve-ysp1}~and~\ref{fig:tradeoff-curve-ysp2}, then we can more easily assess the impact of predictive coding on the aggregate rate. In each of these cases, the predictive coding scheme requires lower aggregate rate than the proposed fixed-rate uniform quantization strategy. However, when ECSQ is used in place of fixed-rate uniform quantization, the results improve significantly. 

The performance advantage of the predictive scheme stems from its use of linear minimum MSE (LMMSE) prediction~\cite{GershoGray1993} to reduce the source variance prior to quantization. As the connectivity parameter $\rho_c$ and total iteration count $T$ increase, there is no significant change in performance relative to the proposed schemes. This lack of improvement may be due to the previously discussed nonzero asymptotic MSE. Clearly, the advantages of the predictive quantization strategy combined with entropy coding and the modified state update of Frasca \etal~\cite{Frasca2008} would produce better results, and it is worthwhile to explore whether LMMSE prediction can be incorporated into our framework.

\subsection{Comparison to the ProgQ scheme} 
Unlike the predictive scheme~\cite{YildizScaglione2008}, ProgQ~\cite{Thanou2013} converges to the true sample average in the limit $t \rightarrow \infty$. In the low-connectivity ($\rho_c = 0.35$) setting, the ProgQ scheme is outdone by our fixed-rate uniform quantizer. Increasing the connectivity to $\rho_c = 0.45$ also increases the speed of convergence. Because the ProgQ scheme uses differential coding, faster convergence leads to lower source variance, and hence lower distortion for a given coding rate~\eqref{eq:rd}. Therefore, the performance of the ProgQ scheme improves relative to our approaches with higher $\rho_c$. Furthermore, from Figure~\ref{fig:tradeoff-curve-progq2}, it is clear that this advantage becomes more significant as the iteration count $T$ increases. Even with constant coding rates, the ProgQ scheme can converge in the limit; however, it is clear from Figure~\ref{fig:constD-optimal-rates} that the optimal sequences for our schemes grow with $t$. Therefore, asymptotically, we expect that ProgQ will outdo our proposed schemes, despite our constrained optimization approach.

\chapter{CONCLUSION}
\label{chap-five}

In this thesis, we presented a new approach to reducing the communication cost in distributed average consensus using lossy compression and generalized geometric programming (GGP) modeling in the case of initial Gaussian states with entropy-coded scalar uniform quantization (ECSQ), or the case of an arbitrary distribution using fixed-rate uniform quantization. The main results in Chapter~\ref{chap-three} of this thesis are ({\em i}) the proof that the problem of minimizing the coding rate for a particular number of consensus iterations $T$, subject to a constraint on the mean square error (MSE), can be posed as a GGP, and ({\em ii}) to a lesser extent, the state evolution equations for consensus using Frasca's modified iteration~\cite{Frasca2008}. This analysis allows the minimization of communication cost to be tackled with a structured and principled approach that may be promising for future research.

 Numerical results in Chapter~\ref{chap-four} verify the state-evolution equations and hint at properties of the optimal rate sequences in the absence of side information at the encoder and decoder. Additionally, in the case that fixed-rate coding is used, a heuristic is presented in Section~\ref{sec:heuristic} that reduces the optimization complexity substantially. The main advantage of the proposed approach is the ability to specify fine trade-offs among accuracy, run time, and communication cost as measured by the aggregate coding rate.

From the comparisons to prior art in Chapter 4, it can be seen that even simple incorporation of side information into the coding strategy can result in significant performance improvements in terms of aggregate rate. Using differential coding, as in the approach of Thanou \etal~\cite{Thanou2013}, is simple. Exploring its effect on performance would make an interesting future research topic.

Unfortunately, the optimization~\eqref{eq:cfun-final} does not scale well, and it may be worthwhile to study the properties of the optimal rate sequences to design some theoretically motivated heuristics, or to investigate alternative optimization implementations that do not rely on explicit GGP modeling. In the case of uniform quantization with fixed-rate coding, numerical results showcased the effectiveness of our proposed heuristic approach. For each simulation conducted, the reduced complexity search heuristic found the same result as an exhaustive search, with a significant reduction in run time.

On the applied front, it would be interesting to explore potential uses of this optimized form of consensus in existing distributed algorithms. One potential path forward is to use consensus to perform the pseudodata fusion step in row-partitioned multiprocessor approximate message passing~\cite{ZhuBaronMPAMP2016ArXiv,ZhuBeiramiBaron2016ISIT,ZhuDissertation2017,HanZhuNiuBaron2016ICASSP,ZhuPilgrimBaron2017}. Because of the nonhomogeneous MSE evolution at each node, this becomes a nontrivial problem. The distributed nature of signal acquisition in sensor networks may make approximate message passing in the multimeasurement vector framework~\cite{ZhuBaronKrzakala2017IEEE} appropriate. Other potential areas for application include dictionary learning~\cite{Raja2016} and computer vision~\cite{Tron2011}.

Another possible application of this work is in information theory. With relaxations of some of the state-update assumptions, it may be possible to numerically compute the operational rate distortion (RD) relationship for synchronous-time consensus for Gaussian data using our optimization~\eqref{eq:cfun-final}. This would represent an upper bound on the operational RD relationship in consensus, and it might offer insight into the continuing information-theoretic investigation of source coding for distributed average consensus.


%
%
\begingroup
\makeatletter
\@ifundefined{ver@biblatex.sty}
  {\@latex@error
     {Missing 'biblatex' package}
     {The bibliography requires the 'biblatex' package.}
      \aftergroup }
  {}
\endgroup

\datalist[entry]{none/global//global/global}
  \entry{pottie2000}{article}{}
    \name{author}{2}{}{%
      {{hash=PGJ}{%
         family={Pottie},
         familyi={P\bibinitperiod},
         given={G.\bibnamedelima J.},
         giveni={G\bibinitperiod\bibinitdelim J\bibinitperiod},
      }}%
      {{hash=KWJ}{%
         family={Kaiser},
         familyi={K\bibinitperiod},
         given={W.\bibnamedelima J.},
         giveni={W\bibinitperiod\bibinitdelim J\bibinitperiod},
      }}%
    }
    \strng{namehash}{PGJKWJ1}
    \strng{fullhash}{PGJKWJ1}
    \field{labelnamesource}{author}
    \field{labeltitlesource}{title}
    \field{number}{5}
    \field{pages}{51\bibrangedash 58}
    \field{title}{{Wireless integrated network sensors}}
    \field{volume}{43}
    \field{journaltitle}{Commun. ACM}
    \field{month}{05}
    \field{year}{2000}
  \endentry

  \entry{estrin2002}{article}{}
    \name{author}{4}{}{%
      {{hash=ED}{%
         family={Estrin},
         familyi={E\bibinitperiod},
         given={D.},
         giveni={D\bibinitperiod},
      }}%
      {{hash=CD}{%
         family={Culler},
         familyi={C\bibinitperiod},
         given={D.},
         giveni={D\bibinitperiod},
      }}%
      {{hash=PK}{%
         family={Pister},
         familyi={P\bibinitperiod},
         given={K.},
         giveni={K\bibinitperiod},
      }}%
      {{hash=SG}{%
         family={Sukhatme},
         familyi={S\bibinitperiod},
         given={G.},
         giveni={G\bibinitperiod},
      }}%
    }
    \strng{namehash}{EDCDPKSG1}
    \strng{fullhash}{EDCDPKSG1}
    \field{labelnamesource}{author}
    \field{labeltitlesource}{title}
    \field{number}{1}
    \field{pages}{59\bibrangedash 69}
    \field{title}{Connecting the physical world with pervasive networks}
    \field{volume}{1}
    \field{journaltitle}{IEEE Pervasive Comput.}
    \field{month}{01}
    \field{year}{2002}
  \endentry

  \entry{EC2}{misc}{}
    \field{labeltitlesource}{title}
    \field{note}{https://aws.amazon.com/ec2/}
    \field{title}{Amazon {E}{C}2}
  \endentry

  \entry{NoorshamsWainwright2011}{article}{}
    \name{author}{2}{}{%
      {{hash=NN}{%
         family={Noorshams},
         familyi={N\bibinitperiod},
         given={N.},
         giveni={N\bibinitperiod},
      }}%
      {{hash=WMJ}{%
         family={Wainwright},
         familyi={W\bibinitperiod},
         given={M.\bibnamedelima J.},
         giveni={M\bibinitperiod\bibinitdelim J\bibinitperiod},
      }}%
    }
    \strng{namehash}{NNWMJ1}
    \strng{fullhash}{NNWMJ1}
    \field{labelnamesource}{author}
    \field{labeltitlesource}{title}
    \field{number}{4}
    \field{pages}{833\bibrangedash 844}
    \field{title}{Non-Asymptotic Analysis of an Optimal Algorithm for
  Network-Constrained Averaging With Noisy Links}
    \field{volume}{5}
    \field{journaltitle}{IEEE J. Sel. Topics Signal Process.}
    \field{month}{08}
    \field{year}{2011}
  \endentry

  \entry{Dimakis2010}{article}{}
    \name{author}{5}{}{%
      {{hash=DAG}{%
         family={Dimakis},
         familyi={D\bibinitperiod},
         given={A.\bibnamedelima G.},
         giveni={A\bibinitperiod\bibinitdelim G\bibinitperiod},
      }}%
      {{hash=KS}{%
         family={Kar},
         familyi={K\bibinitperiod},
         given={S.},
         giveni={S\bibinitperiod},
      }}%
      {{hash=MJMF}{%
         family={Moura},
         familyi={M\bibinitperiod},
         given={J.\bibnamedelima M.\bibnamedelima F.},
         giveni={J\bibinitperiod\bibinitdelim M\bibinitperiod\bibinitdelim
  F\bibinitperiod},
      }}%
      {{hash=RMG}{%
         family={Rabbat},
         familyi={R\bibinitperiod},
         given={M.\bibnamedelima G.},
         giveni={M\bibinitperiod\bibinitdelim G\bibinitperiod},
      }}%
      {{hash=SA}{%
         family={Scaglione},
         familyi={S\bibinitperiod},
         given={A.},
         giveni={A\bibinitperiod},
      }}%
    }
    \strng{namehash}{DAGKSMJMFRMGSA1}
    \strng{fullhash}{DAGKSMJMFRMGSA1}
    \field{labelnamesource}{author}
    \field{labeltitlesource}{title}
    \field{number}{11}
    \field{pages}{1847\bibrangedash 1864}
    \field{title}{Gossip Algorithms for Distributed Signal Processing}
    \field{volume}{98}
    \field{journaltitle}{Proc. IEEE}
    \field{month}{11}
    \field{year}{2010}
  \endentry

  \entry{Olfati-Saber2007}{article}{}
    \name{author}{3}{}{%
      {{hash=OSR}{%
         family={Olfati-Saber},
         familyi={O\bibinithyphendelim S\bibinitperiod},
         given={R.},
         giveni={R\bibinitperiod},
      }}%
      {{hash=FJA}{%
         family={Fax},
         familyi={F\bibinitperiod},
         given={J.\bibnamedelima A.},
         giveni={J\bibinitperiod\bibinitdelim A\bibinitperiod},
      }}%
      {{hash=MRM}{%
         family={Murray},
         familyi={M\bibinitperiod},
         given={R.\bibnamedelima M.},
         giveni={R\bibinitperiod\bibinitdelim M\bibinitperiod},
      }}%
    }
    \strng{namehash}{OSRFJAMRM1}
    \strng{fullhash}{OSRFJAMRM1}
    \field{labelnamesource}{author}
    \field{labeltitlesource}{title}
    \field{number}{1}
    \field{pages}{215\bibrangedash 233}
    \field{title}{{Consensus and cooperation in networked multi-agent systems}}
    \field{volume}{95}
    \field{journaltitle}{Proc. IEEE}
    \field{month}{01}
    \field{year}{2007}
  \endentry

  \entry{Nokleby2013}{article}{}
    \name{author}{4}{}{%
      {{hash=NM}{%
         family={Nokleby},
         familyi={N\bibinitperiod},
         given={M.},
         giveni={M\bibinitperiod},
      }}%
      {{hash=BWU}{%
         family={Bajwa},
         familyi={B\bibinitperiod},
         given={W.\bibnamedelima U.},
         giveni={W\bibinitperiod\bibinitdelim U\bibinitperiod},
      }}%
      {{hash=CR}{%
         family={Calderbank},
         familyi={C\bibinitperiod},
         given={R.},
         giveni={R\bibinitperiod},
      }}%
      {{hash=AB}{%
         family={Aazhang},
         familyi={A\bibinitperiod},
         given={B.},
         giveni={B\bibinitperiod},
      }}%
    }
    \strng{namehash}{NMBWUCRAB1}
    \strng{fullhash}{NMBWUCRAB1}
    \field{labelnamesource}{author}
    \field{labeltitlesource}{title}
    \field{number}{2}
    \field{pages}{284\bibrangedash 295}
    \field{title}{Toward Resource-Optimal Consensus Over the Wireless Medium}
    \field{volume}{7}
    \field{journaltitle}{IEEE J. Sel. Topics Signal Process.}
    \field{month}{04}
    \field{year}{2013}
  \endentry

  \entry{Tron2011}{inproceedings}{}
    \name{author}{2}{}{%
      {{hash=TT}{%
         family={Tron},
         familyi={T\bibinitperiod},
         given={T.},
         giveni={T\bibinitperiod},
      }}%
      {{hash=VR}{%
         family={Vidal},
         familyi={V\bibinitperiod},
         given={R.},
         giveni={R\bibinitperiod},
      }}%
    }
    \strng{namehash}{TTVR1}
    \strng{fullhash}{TTVR1}
    \field{labelnamesource}{author}
    \field{labeltitlesource}{title}
    \field{booktitle}{Proc. IEEE Conf. Comput. Vision and Pattern Recognition
  (CVPR)}
    \field{pages}{57\bibrangedash 63}
    \field{title}{Distributed Computer Vision Algorithms Through Distributed
  Averaging}
    \list{location}{1}{%
      {Providence, RI}%
    }
    \field{month}{06}
    \field{year}{2011}
  \endentry

  \entry{Raja2016}{article}{}
    \name{author}{2}{}{%
      {{hash=RH}{%
         family={Raja},
         familyi={R\bibinitperiod},
         given={H.},
         giveni={H\bibinitperiod},
      }}%
      {{hash=BWU}{%
         family={Bajwa},
         familyi={B\bibinitperiod},
         given={W.\bibnamedelima U.},
         giveni={W\bibinitperiod\bibinitdelim U\bibinitperiod},
      }}%
    }
    \strng{namehash}{RHBWU1}
    \strng{fullhash}{RHBWU1}
    \field{labelnamesource}{author}
    \field{labeltitlesource}{title}
    \field{number}{1}
    \field{pages}{173\bibrangedash 188}
    \field{title}{Cloud {{K-SVD}}: A Collaborative Dictionary Learning
  Algorithm for Big, Distributed Data}
    \field{volume}{64}
    \field{journaltitle}{IEEE Trans. Signal Process.}
    \field{month}{01}
    \field{year}{2016}
  \endentry

  \entry{DeGroot74}{article}{}
    \name{author}{1}{}{%
      {{hash=DMH}{%
         family={DeGroot},
         familyi={D\bibinitperiod},
         given={M.\bibnamedelima H.},
         giveni={M\bibinitperiod\bibinitdelim H\bibinitperiod},
      }}%
    }
    \strng{namehash}{DMH1}
    \strng{fullhash}{DMH1}
    \field{labelnamesource}{author}
    \field{labeltitlesource}{title}
    \field{number}{345}
    \field{pages}{118\bibrangedash 121}
    \field{title}{Reaching a consensus}
    \field{volume}{69}
    \field{journaltitle}{J. Amer. Statist. Assoc.}
    \field{year}{1974}
  \endentry

  \entry{BorkarVaraiya82}{article}{}
    \name{author}{2}{}{%
      {{hash=BV}{%
         family={Borkar},
         familyi={B\bibinitperiod},
         given={V.},
         giveni={V\bibinitperiod},
      }}%
      {{hash=VP}{%
         family={Varaiya},
         familyi={V\bibinitperiod},
         given={P.},
         giveni={P\bibinitperiod},
      }}%
    }
    \strng{namehash}{BVVP1}
    \strng{fullhash}{BVVP1}
    \field{labelnamesource}{author}
    \field{labeltitlesource}{title}
    \field{number}{3}
    \field{pages}{650\bibrangedash 655}
    \field{title}{Asymptotic agreement in distributed estimation}
    \field{volume}{AC-27}
    \field{journaltitle}{IEEE Trans. Autom. Control}
    \field{month}{06}
    \field{year}{1982}
  \endentry

  \entry{Tsitsiklis84}{thesis}{}
    \name{author}{1}{}{%
      {{hash=TJN}{%
         family={Tsitsiklis},
         familyi={T\bibinitperiod},
         given={J.\bibnamedelima N.},
         giveni={J\bibinitperiod\bibinitdelim N\bibinitperiod},
      }}%
    }
    \strng{namehash}{TJN1}
    \strng{fullhash}{TJN1}
    \field{labelnamesource}{author}
    \field{labeltitlesource}{title}
    \field{title}{Problems in decentralized decision making and computation}
    \list{location}{1}{%
      {Cambridge, MA}%
    }
    \list{institution}{1}{%
      {Massachussetts Inst. Technol.}%
    }
    \field{type}{phdthesis}
    \field{month}{11}
    \field{year}{1984}
  \endentry

  \entry{TsitsiklisBertsekasAthans86}{article}{}
    \name{author}{3}{}{%
      {{hash=TJ}{%
         family={Tsitsiklis},
         familyi={T\bibinitperiod},
         given={J.},
         giveni={J\bibinitperiod},
      }}%
      {{hash=BD}{%
         family={Bertsekas},
         familyi={B\bibinitperiod},
         given={D.},
         giveni={D\bibinitperiod},
      }}%
      {{hash=AM}{%
         family={Athans},
         familyi={A\bibinitperiod},
         given={M.},
         giveni={M\bibinitperiod},
      }}%
    }
    \strng{namehash}{TJBDAM1}
    \strng{fullhash}{TJBDAM1}
    \field{labelnamesource}{author}
    \field{labeltitlesource}{title}
    \field{number}{9}
    \field{pages}{803\bibrangedash 812}
    \field{title}{Distributed asynchronous deterministic and stochastic
  gradient optimization algorithms}
    \field{volume}{31}
    \field{journaltitle}{IEEE Trans. Autom. Control}
    \field{month}{09}
    \field{year}{1986}
  \endentry

  \entry{Frasca2008}{article}{}
    \name{author}{4}{}{%
      {{hash=FP}{%
         family={Frasca},
         familyi={F\bibinitperiod},
         given={P.},
         giveni={P\bibinitperiod},
      }}%
      {{hash=CR}{%
         family={Carli},
         familyi={C\bibinitperiod},
         given={R.},
         giveni={R\bibinitperiod},
      }}%
      {{hash=FF}{%
         family={Fagnani},
         familyi={F\bibinitperiod},
         given={F.},
         giveni={F\bibinitperiod},
      }}%
      {{hash=ZS}{%
         family={Zampieri},
         familyi={Z\bibinitperiod},
         given={S.},
         giveni={S\bibinitperiod},
      }}%
    }
    \strng{namehash}{FPCRFFZS1}
    \strng{fullhash}{FPCRFFZS1}
    \field{labelnamesource}{author}
    \field{labeltitlesource}{title}
    \field{number}{16}
    \field{pages}{1787\bibrangedash 1816}
    \field{title}{Average consensus on networks with quantized communication}
    \field{volume}{19}
    \field{journaltitle}{Int. J. Robust Nonlinear Control}
    \field{month}{11}
    \field{year}{2008}
  \endentry

  \entry{CarliBulloZampieri2009}{article}{}
    \name{author}{3}{}{%
      {{hash=CR}{%
         family={Carli},
         familyi={C\bibinitperiod},
         given={R.},
         giveni={R\bibinitperiod},
      }}%
      {{hash=BF}{%
         family={Bullo},
         familyi={B\bibinitperiod},
         given={F.},
         giveni={F\bibinitperiod},
      }}%
      {{hash=ZS}{%
         family={Zampieri},
         familyi={Z\bibinitperiod},
         given={S.},
         giveni={S\bibinitperiod},
      }}%
    }
    \strng{namehash}{CRBFZS1}
    \strng{fullhash}{CRBFZS1}
    \field{labelnamesource}{author}
    \field{labeltitlesource}{title}
    \field{number}{2}
    \field{pages}{156\bibrangedash 175}
    \field{title}{Quantized average consensus via dynamic coding/decoding
  schemes}
    \field{volume}{20}
    \field{journaltitle}{Int. J. Robust Nonlinear Control}
    \field{month}{05}
    \field{year}{2009}
  \endentry

  \entry{Chamie2014}{inproceedings}{}
    \name{author}{3}{}{%
      {{hash=CME}{%
         family={Chamie},
         familyi={C\bibinitperiod},
         given={M.\bibnamedelima El},
         giveni={M\bibinitperiod\bibinitdelim E\bibinitperiod},
      }}%
      {{hash=LJ}{%
         family={Liu},
         familyi={L\bibinitperiod},
         given={J.},
         giveni={J\bibinitperiod},
      }}%
      {{hash=BT}{%
         family={Başar},
         familyi={B\bibinitperiod},
         given={T.},
         giveni={T\bibinitperiod},
      }}%
    }
    \strng{namehash}{CMELJBT1}
    \strng{fullhash}{CMELJBT1}
    \field{labelnamesource}{author}
    \field{labeltitlesource}{title}
    \field{booktitle}{Proc. 53rd IEEE Conf. Decision Control}
    \field{pages}{3860\bibrangedash 3865}
    \field{title}{Design and analysis of distributed averaging with quantized
  communication}
    \field{month}{12}
    \field{year}{2014}
  \endentry

  \entry{CarliFagnaniFrascaZampieri2009}{article}{}
    \name{author}{4}{}{%
      {{hash=CR}{%
         family={Carli},
         familyi={C\bibinitperiod},
         given={R.},
         giveni={R\bibinitperiod},
      }}%
      {{hash=FF}{%
         family={Fagnani},
         familyi={F\bibinitperiod},
         given={F.},
         giveni={F\bibinitperiod},
      }}%
      {{hash=FP}{%
         family={Frasca},
         familyi={F\bibinitperiod},
         given={P.},
         giveni={P\bibinitperiod},
      }}%
      {{hash=ZS}{%
         family={Zampieri},
         familyi={Z\bibinitperiod},
         given={S.},
         giveni={S\bibinitperiod},
      }}%
    }
    \strng{namehash}{CRFFFPZS1}
    \strng{fullhash}{CRFFFPZS1}
    \field{labelnamesource}{author}
    \field{labeltitlesource}{title}
    \field{title}{Efficient quantization for average consensus}
    \field{journaltitle}{Arxiv preprint arXiv:0903.1337}
    \field{month}{03}
    \field{year}{2009}
  \endentry

  \entry{FangLi2009}{inproceedings}{}
    \name{author}{2}{}{%
      {{hash=FJ}{%
         family={Fang},
         familyi={F\bibinitperiod},
         given={J.},
         giveni={J\bibinitperiod},
      }}%
      {{hash=LH}{%
         family={Li},
         familyi={L\bibinitperiod},
         given={H.},
         giveni={H\bibinitperiod},
      }}%
    }
    \strng{namehash}{FJLH1}
    \strng{fullhash}{FJLH1}
    \field{labelnamesource}{author}
    \field{labeltitlesource}{title}
    \field{booktitle}{Proc. IEEE Int. Conf. Acoust., Speech, Signal Process.
  (ICASSP)}
    \field{pages}{2777\bibrangedash 2780}
    \field{title}{An adaptive quantization scheme for distributed consensus}
    \field{month}{04}
    \field{year}{2009}
  \endentry

  \entry{AysalCoatesRabbat2008}{article}{}
    \name{author}{3}{}{%
      {{hash=ATC}{%
         family={Aysal},
         familyi={A\bibinitperiod},
         given={T.\bibnamedelima C.},
         giveni={T\bibinitperiod\bibinitdelim C\bibinitperiod},
      }}%
      {{hash=CMJ}{%
         family={Coates},
         familyi={C\bibinitperiod},
         given={M.\bibnamedelima J.},
         giveni={M\bibinitperiod\bibinitdelim J\bibinitperiod},
      }}%
      {{hash=RMG}{%
         family={Rabbat},
         familyi={R\bibinitperiod},
         given={M.\bibnamedelima G.},
         giveni={M\bibinitperiod\bibinitdelim G\bibinitperiod},
      }}%
    }
    \strng{namehash}{ATCCMJRMG1}
    \strng{fullhash}{ATCCMJRMG1}
    \field{labelnamesource}{author}
    \field{labeltitlesource}{title}
    \field{number}{10}
    \field{pages}{4905\bibrangedash 4918}
    \field{title}{Distributed average consensus with dithered quantization}
    \field{volume}{56}
    \field{journaltitle}{IEEE Trans. Signal Process.}
    \field{month}{10}
    \field{year}{2008}
  \endentry

  \entry{Thanou2013}{article}{}
    \name{author}{4}{}{%
      {{hash=TD}{%
         family={Thanou},
         familyi={T\bibinitperiod},
         given={D.},
         giveni={D\bibinitperiod},
      }}%
      {{hash=KE}{%
         family={Kokiopoulou},
         familyi={K\bibinitperiod},
         given={E.},
         giveni={E\bibinitperiod},
      }}%
      {{hash=PY}{%
         family={Pu},
         familyi={P\bibinitperiod},
         given={Y.},
         giveni={Y\bibinitperiod},
      }}%
      {{hash=FP}{%
         family={Frossard},
         familyi={F\bibinitperiod},
         given={P.},
         giveni={P\bibinitperiod},
      }}%
    }
    \strng{namehash}{TDKEPYFP1}
    \strng{fullhash}{TDKEPYFP1}
    \field{labelnamesource}{author}
    \field{labeltitlesource}{title}
    \field{number}{1}
    \field{pages}{194\bibrangedash 205}
    \field{title}{Distributed average consensus with quantization refinement}
    \field{volume}{61}
    \field{journaltitle}{IEEE Trans. Signal Process.}
    \field{month}{01}
    \field{year}{2013}
  \endentry

  \entry{YildizScaglione2008}{article}{}
    \name{author}{2}{}{%
      {{hash=YME}{%
         family={Yildiz},
         familyi={Y\bibinitperiod},
         given={M.\bibnamedelima E.},
         giveni={M\bibinitperiod\bibinitdelim E\bibinitperiod},
      }}%
      {{hash=SA}{%
         family={Scaglione},
         familyi={S\bibinitperiod},
         given={A.},
         giveni={A\bibinitperiod},
      }}%
    }
    \strng{namehash}{YMESA1}
    \strng{fullhash}{YMESA1}
    \field{labelnamesource}{author}
    \field{labeltitlesource}{title}
    \field{number}{8}
    \field{pages}{3753\bibrangedash 3764}
    \field{title}{Coding with Side Information for Rate-Constrained Consensus}
    \field{volume}{56}
    \field{journaltitle}{IEEE Trans. Signal Process.}
    \field{month}{08}
    \field{year}{2008}
  \endentry

  \entry{YildizScaglione2008b}{inproceedings}{}
    \name{author}{2}{}{%
      {{hash=YME}{%
         family={Yildiz},
         familyi={Y\bibinitperiod},
         given={M.\bibnamedelima E.},
         giveni={M\bibinitperiod\bibinitdelim E\bibinitperiod},
      }}%
      {{hash=SA}{%
         family={Scaglione},
         familyi={S\bibinitperiod},
         given={A.},
         giveni={A\bibinitperiod},
      }}%
    }
    \strng{namehash}{YMESA1}
    \strng{fullhash}{YMESA1}
    \field{labelnamesource}{author}
    \field{labeltitlesource}{title}
    \field{booktitle}{Proc. IEEE Int. Conf. Acoust., Speech, Signal Process.
  (ICASSP)}
    \field{pages}{2717\bibrangedash 2720}
    \field{title}{Limiting Rate Behavior and Rate Allocation Strategies for
  Average Consensus Problems with Bounded Convergence}
    \field{month}{04}
    \field{year}{2008}
  \endentry

  \entry{MosqueraLopezValcarceJayaweera2010}{article}{}
    \name{author}{3}{}{%
      {{hash=MC}{%
         family={Mosquera},
         familyi={M\bibinitperiod},
         given={C.},
         giveni={C\bibinitperiod},
      }}%
      {{hash=LVR}{%
         family={López-Valcarce},
         familyi={L\bibinithyphendelim V\bibinitperiod},
         given={R.},
         giveni={R\bibinitperiod},
      }}%
      {{hash=JSK}{%
         family={Jayaweera},
         familyi={J\bibinitperiod},
         given={S.\bibnamedelima K.},
         giveni={S\bibinitperiod\bibinitdelim K\bibinitperiod},
      }}%
    }
    \strng{namehash}{MCLVRJSK1}
    \strng{fullhash}{MCLVRJSK1}
    \field{labelnamesource}{author}
    \field{labeltitlesource}{title}
    \field{number}{2}
    \field{pages}{169\bibrangedash 172}
    \field{title}{Stepsize Sequence Design for Distributed Average Consensus}
    \field{volume}{17}
    \field{journaltitle}{IEEE Signal Process. Lett.}
    \field{month}{02}
    \field{year}{2010}
  \endentry

  \entry{RajagopalWainwright2011}{article}{}
    \name{author}{2}{}{%
      {{hash=RR}{%
         family={Rajagopal},
         familyi={R\bibinitperiod},
         given={R.},
         giveni={R\bibinitperiod},
      }}%
      {{hash=WMJ}{%
         family={Wainwright},
         familyi={W\bibinitperiod},
         given={M.\bibnamedelima J.},
         giveni={M\bibinitperiod\bibinitdelim J\bibinitperiod},
      }}%
    }
    \strng{namehash}{RRWMJ1}
    \strng{fullhash}{RRWMJ1}
    \field{labelnamesource}{author}
    \field{labeltitlesource}{title}
    \field{number}{1}
    \field{pages}{373\bibrangedash 385}
    \field{title}{Network-Based Consensus Averaging With General Noisy
  Channels}
    \field{volume}{59}
    \field{journaltitle}{IEEE Trans. Signal Process.}
    \field{month}{01}
    \field{year}{2011}
  \endentry

  \entry{Nedic2009}{article}{}
    \name{author}{4}{}{%
      {{hash=NA}{%
         family={Nedić},
         familyi={N\bibinitperiod},
         given={A.},
         giveni={A\bibinitperiod},
      }}%
      {{hash=OA}{%
         family={Olshevsky},
         familyi={O\bibinitperiod},
         given={A.},
         giveni={A\bibinitperiod},
      }}%
      {{hash=OA}{%
         family={Ozdaglar},
         familyi={O\bibinitperiod},
         given={A.},
         giveni={A\bibinitperiod},
      }}%
      {{hash=TJN}{%
         family={Tsitsiklis},
         familyi={T\bibinitperiod},
         given={J.\bibnamedelima N.},
         giveni={J\bibinitperiod\bibinitdelim N\bibinitperiod},
      }}%
    }
    \strng{namehash}{NAOAOATJN1}
    \strng{fullhash}{NAOAOATJN1}
    \field{labelnamesource}{author}
    \field{labeltitlesource}{title}
    \field{number}{11}
    \field{pages}{2506\bibrangedash 2517}
    \field{title}{On Distributed Averaging Algorithms and Quantization Effects}
    \field{volume}{54}
    \field{journaltitle}{IEEE Trans. Autom. Control}
    \field{month}{11}
    \field{year}{2009}
  \endentry

  \entry{Zhu2015}{inproceedings}{}
    \name{author}{5}{}{%
      {{hash=ZS}{%
         family={Zhu},
         familyi={Z\bibinitperiod},
         given={S.},
         giveni={S\bibinitperiod},
      }}%
      {{hash=LS}{%
         family={Liu},
         familyi={L\bibinitperiod},
         given={S.},
         giveni={S\bibinitperiod},
      }}%
      {{hash=XJ}{%
         family={Xu},
         familyi={X\bibinitperiod},
         given={J.},
         giveni={J\bibinitperiod},
      }}%
      {{hash=SYC}{%
         family={Soh},
         familyi={S\bibinitperiod},
         given={Y.\bibnamedelima C.},
         giveni={Y\bibinitperiod\bibinitdelim C\bibinitperiod},
      }}%
      {{hash=XL}{%
         family={Xie},
         familyi={X\bibinitperiod},
         given={L.},
         giveni={L\bibinitperiod},
      }}%
    }
    \strng{namehash}{ZSLSXJSYCXL1}
    \strng{fullhash}{ZSLSXJSYCXL1}
    \field{labelnamesource}{author}
    \field{labeltitlesource}{title}
    \field{booktitle}{Proc. IEEE Ann. Conf. Decision Control (CDC)}
    \field{pages}{6245\bibrangedash 6250}
    \field{title}{Averaging Based Distributed Estimation Algorithm for
  Rate-Constrained Sensor Networks with Additive Quantization Model}
    \list{location}{1}{%
      {Osaka, Japan}%
    }
    \field{month}{12}
    \field{year}{2015}
  \endentry

  \entry{KarMoura2010}{article}{}
    \name{author}{2}{}{%
      {{hash=KS}{%
         family={Kar},
         familyi={K\bibinitperiod},
         given={S.},
         giveni={S\bibinitperiod},
      }}%
      {{hash=MJMF}{%
         family={Moura},
         familyi={M\bibinitperiod},
         given={J.\bibnamedelima M.\bibnamedelima F.},
         giveni={J\bibinitperiod\bibinitdelim M\bibinitperiod\bibinitdelim
  F\bibinitperiod},
      }}%
    }
    \strng{namehash}{KSMJMF1}
    \strng{fullhash}{KSMJMF1}
    \field{labelnamesource}{author}
    \field{labeltitlesource}{title}
    \field{number}{3}
    \field{pages}{1383\bibrangedash 1400}
    \field{title}{Distributed Consensus Algorithms in Sensor Networks:
  Quantized Data and Random Link Failures}
    \field{volume}{58}
    \field{journaltitle}{IEEE Trans. Signal Process.}
    \field{month}{03}
    \field{year}{2010}
  \endentry

  \entry{HuangHua2011}{article}{}
    \name{author}{2}{}{%
      {{hash=HY}{%
         family={Huang},
         familyi={H\bibinitperiod},
         given={Y.},
         giveni={Y\bibinitperiod},
      }}%
      {{hash=HY}{%
         family={Hua},
         familyi={H\bibinitperiod},
         given={Y.},
         giveni={Y\bibinitperiod},
      }}%
    }
    \strng{namehash}{HYHY1}
    \strng{fullhash}{HYHY1}
    \field{labelnamesource}{author}
    \field{labeltitlesource}{title}
    \field{number}{8}
    \field{pages}{3863\bibrangedash 3875}
    \field{title}{On Energy for Progressive and Consensus Estimation in
  Multihop Sensor Networks}
    \field{volume}{59}
    \field{journaltitle}{IEEE Trans. Signal Process.}
    \field{month}{08}
    \field{year}{2011}
  \endentry

  \entry{Berger71}{book}{}
    \name{author}{1}{}{%
      {{hash=BT}{%
         family={Berger},
         familyi={B\bibinitperiod},
         given={T.},
         giveni={T\bibinitperiod},
      }}%
    }
    \list{publisher}{1}{%
      {Prentice-Hall}%
    }
    \strng{namehash}{BT1}
    \strng{fullhash}{BT1}
    \field{labelnamesource}{author}
    \field{labeltitlesource}{title}
    \field{title}{Rate Distortion Theory: Mathematical Basis for Data
  Compression}
    \list{location}{1}{%
      {Englewood Cliffs, NJ}%
    }
    \field{type}{Book}
    \field{year}{1971}
  \endentry

  \entry{Cover06}{book}{}
    \name{author}{2}{}{%
      {{hash=CTM}{%
         family={Cover},
         familyi={C\bibinitperiod},
         given={T.\bibnamedelima M.},
         giveni={T\bibinitperiod\bibinitdelim M\bibinitperiod},
      }}%
      {{hash=TJA}{%
         family={Thomas},
         familyi={T\bibinitperiod},
         given={J.\bibnamedelima A.},
         giveni={J\bibinitperiod\bibinitdelim A\bibinitperiod},
      }}%
    }
    \list{publisher}{1}{%
      {New York, NY, USA: Wiley-Interscience}%
    }
    \strng{namehash}{CTMTJA1}
    \strng{fullhash}{CTMTJA1}
    \field{labelnamesource}{author}
    \field{labeltitlesource}{title}
    \field{title}{Elements of Information Theory}
    \field{year}{1991}
  \endentry

  \entry{BoydVandenberghe2004}{book}{}
    \name{author}{2}{}{%
      {{hash=BS}{%
         family={Boyd},
         familyi={B\bibinitperiod},
         given={S.},
         giveni={S\bibinitperiod},
      }}%
      {{hash=VL}{%
         family={Vandenberghe},
         familyi={V\bibinitperiod},
         given={L.},
         giveni={L\bibinitperiod},
      }}%
    }
    \list{publisher}{1}{%
      {Cambridge, UK: Cambridge University Press}%
    }
    \strng{namehash}{BSVL1}
    \strng{fullhash}{BSVL1}
    \field{labelnamesource}{author}
    \field{labeltitlesource}{title}
    \field{title}{Convex Optimization}
    \field{year}{2004}
  \endentry

  \entry{XiaoBoydKim2007}{article}{}
    \name{author}{3}{}{%
      {{hash=XL}{%
         family={Xiao},
         familyi={X\bibinitperiod},
         given={L.},
         giveni={L\bibinitperiod},
      }}%
      {{hash=BS}{%
         family={Boyd},
         familyi={B\bibinitperiod},
         given={S.},
         giveni={S\bibinitperiod},
      }}%
      {{hash=KSJ}{%
         family={Kim},
         familyi={K\bibinitperiod},
         given={S.-J.},
         giveni={S\bibinithyphendelim J\bibinitperiod},
      }}%
    }
    \strng{namehash}{XLBSKSJ1}
    \strng{fullhash}{XLBSKSJ1}
    \field{labelnamesource}{author}
    \field{labeltitlesource}{title}
    \field{number}{1}
    \field{pages}{33\bibrangedash 46}
    \field{title}{Distributed average consensus with least-mean-square
  deviation}
    \field{volume}{67}
    \field{journaltitle}{J. Parallel Distributed Comput.}
    \field{month}{01}
    \field{year}{2007}
  \endentry

  \entry{Rego2015}{inproceedings}{}
    \name{author}{5}{}{%
      {{hash=RFFC}{%
         family={Rego},
         familyi={R\bibinitperiod},
         given={F.\bibnamedelima F.\bibnamedelima C.},
         giveni={F\bibinitperiod\bibinitdelim F\bibinitperiod\bibinitdelim
  C\bibinitperiod},
      }}%
      {{hash=PY}{%
         family={Pu},
         familyi={P\bibinitperiod},
         given={Y.},
         giveni={Y\bibinitperiod},
      }}%
      {{hash=AA}{%
         family={Alessandretti},
         familyi={A\bibinitperiod},
         given={A.},
         giveni={A\bibinitperiod},
      }}%
      {{hash=AAP}{%
         family={Aguiar},
         familyi={A\bibinitperiod},
         given={A.\bibnamedelima P.},
         giveni={A\bibinitperiod\bibinitdelim P\bibinitperiod},
      }}%
      {{hash=JCN}{%
         family={Jones},
         familyi={J\bibinitperiod},
         given={C.\bibnamedelima N.},
         giveni={C\bibinitperiod\bibinitdelim N\bibinitperiod},
      }}%
    }
    \strng{namehash}{RFFCPYAAAAPJCN1}
    \strng{fullhash}{RFFCPYAAAAPJCN1}
    \field{labelnamesource}{author}
    \field{labeltitlesource}{title}
    \field{booktitle}{Proc. 53rd Annu. Allerton Conf. on Commun., Control, and
  Comput.}
    \field{pages}{488\bibrangedash 495}
    \field{title}{A consensus algorithm for networks with process noise and
  quantization error}
    \list{location}{1}{%
      {Allerton House, UIUC, IL}%
    }
    \field{month}{10}
    \field{year}{2015}
  \endentry

  \entry{Li2011}{article}{}
    \name{author}{4}{}{%
      {{hash=LT}{%
         family={Li},
         familyi={L\bibinitperiod},
         given={T.},
         giveni={T\bibinitperiod},
      }}%
      {{hash=FM}{%
         family={Fu},
         familyi={F\bibinitperiod},
         given={M.},
         giveni={M\bibinitperiod},
      }}%
      {{hash=XL}{%
         family={Xie},
         familyi={X\bibinitperiod},
         given={L.},
         giveni={L\bibinitperiod},
      }}%
      {{hash=ZJF}{%
         family={Zhang},
         familyi={Z\bibinitperiod},
         given={J.-F.},
         giveni={J\bibinithyphendelim F\bibinitperiod},
      }}%
    }
    \strng{namehash}{LTFMXLZJF1}
    \strng{fullhash}{LTFMXLZJF1}
    \field{labelnamesource}{author}
    \field{labeltitlesource}{title}
    \field{number}{2}
    \field{pages}{279\bibrangedash 292}
    \field{title}{Distributed Consensus With Limited Communication Data Rate}
    \field{volume}{56}
    \field{journaltitle}{IEEE Trans. Autom. Control}
    \field{month}{02}
    \field{year}{2011}
  \endentry

  \entry{GershoGray1993}{book}{}
    \name{author}{2}{}{%
      {{hash=GA}{%
         family={Gersho},
         familyi={G\bibinitperiod},
         given={A.},
         giveni={A\bibinitperiod},
      }}%
      {{hash=GRM}{%
         family={Gray},
         familyi={G\bibinitperiod},
         given={R.\bibnamedelima M.},
         giveni={R\bibinitperiod\bibinitdelim M\bibinitperiod},
      }}%
    }
    \list{publisher}{1}{%
      {Kluwer}%
    }
    \strng{namehash}{GAGRM1}
    \strng{fullhash}{GAGRM1}
    \field{labelnamesource}{author}
    \field{labeltitlesource}{title}
    \field{title}{Vector Quantization and Signal Compression}
    \list{location}{1}{%
      {Norwell, MA}%
    }
    \field{year}{1993}
  \endentry

  \entry{YildizScaglione2007}{inproceedings}{}
    \name{author}{2}{}{%
      {{hash=YME}{%
         family={Yildiz},
         familyi={Y\bibinitperiod},
         given={M.\bibnamedelima E.},
         giveni={M\bibinitperiod\bibinitdelim E\bibinitperiod},
      }}%
      {{hash=SA}{%
         family={Scaglione},
         familyi={S\bibinitperiod},
         given={A.},
         giveni={A\bibinitperiod},
      }}%
    }
    \strng{namehash}{YMESA1}
    \strng{fullhash}{YMESA1}
    \field{labelnamesource}{author}
    \field{labeltitlesource}{title}
    \field{booktitle}{Proc. 6th Int. Symp. Inform. Process. Sensor Networks
  (IPSN)}
    \field{pages}{89\bibrangedash 98}
    \field{title}{Differential Nested Lattice Encoding for Consensus Problems}
    \list{location}{1}{%
      {Cambridge, MA}%
    }
    \field{month}{04}
    \field{year}{2007}
  \endentry

  \entry{XiaoBoyd2004}{article}{}
    \name{author}{2}{}{%
      {{hash=XL}{%
         family={Xiao},
         familyi={X\bibinitperiod},
         given={L.},
         giveni={L\bibinitperiod},
      }}%
      {{hash=BS}{%
         family={Boyd},
         familyi={B\bibinitperiod},
         given={S.},
         giveni={S\bibinitperiod},
      }}%
    }
    \strng{namehash}{XLBS1}
    \strng{fullhash}{XLBS1}
    \field{labelnamesource}{author}
    \field{labeltitlesource}{title}
    \field{number}{1}
    \field{pages}{65\bibrangedash 78}
    \field{title}{Fast linear iterations for distributed averaging}
    \field{volume}{53}
    \field{journaltitle}{Syst. Control Lett.}
    \field{month}{09}
    \field{year}{2004}
  \endentry

  \entry{ZhuSohXie2015}{article}{}
    \name{author}{3}{}{%
      {{hash=ZS}{%
         family={Zhu},
         familyi={Z\bibinitperiod},
         given={S.},
         giveni={S\bibinitperiod},
      }}%
      {{hash=SYC}{%
         family={Soh},
         familyi={S\bibinitperiod},
         given={Y.\bibnamedelima C.},
         giveni={Y\bibinitperiod\bibinitdelim C\bibinitperiod},
      }}%
      {{hash=XL}{%
         family={Xie},
         familyi={X\bibinitperiod},
         given={L.},
         giveni={L\bibinitperiod},
      }}%
    }
    \strng{namehash}{ZSSYCXL1}
    \strng{fullhash}{ZSSYCXL1}
    \field{labelnamesource}{author}
    \field{labeltitlesource}{title}
    \field{number}{17}
    \field{pages}{4634\bibrangedash 4646}
    \field{title}{Distributed parameter estimation with quantized communication
  via running average}
    \field{volume}{63}
    \field{journaltitle}{IEEE Trans. Signal Process.}
    \field{month}{09}
    \field{year}{2015}
  \endentry

  \entry{Thanou2010}{inproceedings}{}
    \name{author}{4}{}{%
      {{hash=TD}{%
         family={Thanou},
         familyi={T\bibinitperiod},
         given={D.},
         giveni={D\bibinitperiod},
      }}%
      {{hash=PH}{%
         family={Park},
         familyi={P\bibinitperiod},
         given={H.},
         giveni={H\bibinitperiod},
      }}%
      {{hash=KE}{%
         family={Kokiopoulou},
         familyi={K\bibinitperiod},
         given={E.},
         giveni={E\bibinitperiod},
      }}%
      {{hash=FP}{%
         family={Frossard},
         familyi={F\bibinitperiod},
         given={P.},
         giveni={P\bibinitperiod},
      }}%
    }
    \strng{namehash}{TDPHKEFP1}
    \strng{fullhash}{TDPHKEFP1}
    \field{labelnamesource}{author}
    \field{labeltitlesource}{title}
    \field{booktitle}{Proc. 18th European Signal Process. Conf. (EUSIPCO)}
    \field{pages}{184\bibrangedash 188}
    \field{title}{Polynomial Filter Design for Quantized Consensus}
    \list{location}{1}{%
      {Aalborg, Denmark}%
    }
    \field{month}{08}
    \field{year}{2010}
  \endentry

  \entry{FangLi2010}{article}{}
    \name{author}{2}{}{%
      {{hash=FJ}{%
         family={Fang},
         familyi={F\bibinitperiod},
         given={J.},
         giveni={J\bibinitperiod},
      }}%
      {{hash=LH}{%
         family={Li},
         familyi={L\bibinitperiod},
         given={H.},
         giveni={H\bibinitperiod},
      }}%
    }
    \strng{namehash}{FJLH1}
    \strng{fullhash}{FJLH1}
    \field{labelnamesource}{author}
    \field{labeltitlesource}{title}
    \field{number}{2}
    \field{pages}{944\bibrangedash 948}
    \field{title}{Distributed Consensus With Quantized Data via Sequence
  Averaging}
    \field{volume}{58}
    \field{journaltitle}{IEEE Trans. Signal Process.}
    \field{month}{02}
    \field{year}{2010}
  \endentry

  \entry{AyasoShahDahleh2010}{article}{}
    \name{author}{3}{}{%
      {{hash=AO}{%
         family={Ayaso},
         familyi={A\bibinitperiod},
         given={O.},
         giveni={O\bibinitperiod},
      }}%
      {{hash=SD}{%
         family={Shah},
         familyi={S\bibinitperiod},
         given={D.},
         giveni={D\bibinitperiod},
      }}%
      {{hash=DMA}{%
         family={Dahleh},
         familyi={D\bibinitperiod},
         given={M.\bibnamedelima A.},
         giveni={M\bibinitperiod\bibinitdelim A\bibinitperiod},
      }}%
    }
    \strng{namehash}{AOSDDMA1}
    \strng{fullhash}{AOSDDMA1}
    \field{labelnamesource}{author}
    \field{labeltitlesource}{title}
    \field{number}{12}
    \field{pages}{6020\bibrangedash 6039}
    \field{title}{Information Theoretic Bounds for Distributed Computation Over
  Networks of Point-to-Point Channels}
    \field{volume}{56}
    \field{journaltitle}{IEEE Trans. Inf. Theory}
    \field{month}{12}
    \field{year}{2010}
  \endentry

  \entry{XuRaginsky2014}{inproceedings}{}
    \name{author}{2}{}{%
      {{hash=XA}{%
         family={Xu},
         familyi={X\bibinitperiod},
         given={A.},
         giveni={A\bibinitperiod},
      }}%
      {{hash=RM}{%
         family={Raginsky},
         familyi={R\bibinitperiod},
         given={M.},
         giveni={M\bibinitperiod},
      }}%
    }
    \strng{namehash}{XARM1}
    \strng{fullhash}{XARM1}
    \field{labelnamesource}{author}
    \field{labeltitlesource}{title}
    \field{booktitle}{Proc. IEEE Int. Symp. Inform. Theory (ISIT)}
    \field{pages}{2227\bibrangedash 2231}
    \field{title}{A New Information-Theoretic Lower Bound for Distributed
  Function Computation}
    \list{location}{1}{%
      {Honolulu, HI}%
    }
    \field{month}{06}
    \field{year}{2014}
  \endentry

  \entry{SuGamal2010}{article}{}
    \name{author}{2}{}{%
      {{hash=SHI}{%
         family={Su},
         familyi={S\bibinitperiod},
         given={H.-I.},
         giveni={H\bibinithyphendelim I\bibinitperiod},
      }}%
      {{hash=EA}{%
         family={{{El Gamal}}},
         familyi={E\bibinitperiod},
         given={A.},
         giveni={A\bibinitperiod},
      }}%
    }
    \strng{namehash}{SHIEA1}
    \strng{fullhash}{SHIEA1}
    \field{labelnamesource}{author}
    \field{labeltitlesource}{title}
    \field{number}{7}
    \field{pages}{3422\bibrangedash 3437}
    \field{title}{Distributed Lossy Averaging}
    \field{volume}{56}
    \field{journaltitle}{IEEE Trans. Inf. Theory}
    \field{month}{07}
    \field{year}{2010}
  \endentry

  \entry{YangGroverKar2016}{article}{}
    \name{author}{3}{}{%
      {{hash=YY}{%
         family={Yang},
         familyi={Y\bibinitperiod},
         given={Y.},
         giveni={Y\bibinitperiod},
      }}%
      {{hash=GP}{%
         family={Grover},
         familyi={G\bibinitperiod},
         given={P.},
         giveni={P\bibinitperiod},
      }}%
      {{hash=KS}{%
         family={Kar},
         familyi={K\bibinitperiod},
         given={S.},
         giveni={S\bibinitperiod},
      }}%
    }
    \strng{namehash}{YYGPKS1}
    \strng{fullhash}{YYGPKS1}
    \field{labelnamesource}{author}
    \field{labeltitlesource}{title}
    \field{number}{99}
    \field{pages}{1\bibrangedash 29}
    \field{title}{Rate Distortion for Lossy In-network Function Computation:
  Information Dissipation and Sequential Reverse Water-Filling}
    \field{volume}{{PP}}
    \field{journaltitle}{IEEE Trans. Inf. Theory}
    \field{month}{05}
    \field{year}{2017}
  \endentry

  \entry{Penrose2003}{book}{}
    \name{author}{1}{}{%
      {{hash=PM}{%
         family={Penrose},
         familyi={P\bibinitperiod},
         given={M.},
         giveni={M\bibinitperiod},
      }}%
    }
    \list{publisher}{1}{%
      {Oxford University Press}%
    }
    \strng{namehash}{PM1}
    \strng{fullhash}{PM1}
    \field{labelnamesource}{author}
    \field{labeltitlesource}{title}
    \field{title}{Random Geometric Graphs}
    \list{location}{1}{%
      {New York, NY}%
    }
    \field{year}{2003}
  \endentry

  \entry{BoydMixingTimes}{inproceedings}{}
    \name{author}{4}{}{%
      {{hash=BS}{%
         family={Boyd},
         familyi={B\bibinitperiod},
         given={S.},
         giveni={S\bibinitperiod},
      }}%
      {{hash=GA}{%
         family={Ghosh},
         familyi={G\bibinitperiod},
         given={A.},
         giveni={A\bibinitperiod},
      }}%
      {{hash=PB}{%
         family={Prabhakar},
         familyi={P\bibinitperiod},
         given={B.},
         giveni={B\bibinitperiod},
      }}%
      {{hash=SD}{%
         family={Shah},
         familyi={S\bibinitperiod},
         given={D.},
         giveni={D\bibinitperiod},
      }}%
    }
    \strng{namehash}{BSGAPBSD1}
    \strng{fullhash}{BSGAPBSD1}
    \field{labelnamesource}{author}
    \field{labeltitlesource}{title}
    \field{booktitle}{ALENEX/ANALCO}
    \field{pages}{240\bibrangedash 249}
    \field{title}{Mixing Times for Random Walks on Geometric Random Graphs}
    \list{location}{1}{%
      {Vancouver, British Columbia, Canada}%
    }
    \field{month}{01}
    \field{year}{2005}
  \endentry

  \entry{ZhuBaronMPAMP2016ArXiv}{article}{}
    \name{author}{3}{}{%
      {{hash=ZJ}{%
         family={Zhu},
         familyi={Z\bibinitperiod},
         given={J.},
         giveni={J\bibinitperiod},
      }}%
      {{hash=BD}{%
         family={Baron},
         familyi={B\bibinitperiod},
         given={D.},
         giveni={D\bibinitperiod},
      }}%
      {{hash=BA}{%
         family={Beirami},
         familyi={B\bibinitperiod},
         given={A.},
         giveni={A\bibinitperiod},
      }}%
    }
    \strng{namehash}{ZJBDBA1}
    \strng{fullhash}{ZJBDBA1}
    \field{labelnamesource}{author}
    \field{labeltitlesource}{title}
    \field{title}{Optimal trade-offs in multi-processor approximate message
  passing}
    \field{journaltitle}{Arxiv preprint arXiv:1601.03790}
    \field{month}{11}
    \field{year}{2016}
  \endentry

  \entry{ZhuBeiramiBaron2016ISIT}{inproceedings}{}
    \name{author}{3}{}{%
      {{hash=ZJ}{%
         family={Zhu},
         familyi={Z\bibinitperiod},
         given={J.},
         giveni={J\bibinitperiod},
      }}%
      {{hash=BA}{%
         family={Beirami},
         familyi={B\bibinitperiod},
         given={A.},
         giveni={A\bibinitperiod},
      }}%
      {{hash=BD}{%
         family={Baron},
         familyi={B\bibinitperiod},
         given={D.},
         giveni={D\bibinitperiod},
      }}%
    }
    \strng{namehash}{ZJBABD1}
    \strng{fullhash}{ZJBABD1}
    \field{labelnamesource}{author}
    \field{labeltitlesource}{title}
    \field{booktitle}{Proc. IEEE Int. Symp. Inform. Theory (ISIT)}
    \field{pages}{680\bibrangedash 684}
    \field{title}{Performance Trade-Offs in Multi-Processor Approximate Message
  Passing}
    \list{location}{1}{%
      {Barcelona, Spain}%
    }
    \field{month}{07}
    \field{year}{2016}
  \endentry

  \entry{ZhuDissertation2017}{thesis}{}
    \name{author}{1}{}{%
      {{hash=ZJ}{%
         family={Zhu},
         familyi={Z\bibinitperiod},
         given={J.},
         giveni={J\bibinitperiod},
      }}%
    }
    \strng{namehash}{ZJ1}
    \strng{fullhash}{ZJ1}
    \field{labelnamesource}{author}
    \field{labeltitlesource}{title}
    \field{title}{Statistical Physics and Information Theory Perspectives on
  Linear Inverse Problems}
    \list{location}{1}{%
      {Raleigh, NC}%
    }
    \list{institution}{1}{%
      {North Carolina State University}%
    }
    \field{type}{phdthesis}
    \field{month}{01}
    \field{year}{2017}
  \endentry

  \entry{HanZhuNiuBaron2016ICASSP}{inproceedings}{}
    \name{author}{4}{}{%
      {{hash=HP}{%
         family={Han},
         familyi={H\bibinitperiod},
         given={P.},
         giveni={P\bibinitperiod},
      }}%
      {{hash=ZJ}{%
         family={Zhu},
         familyi={Z\bibinitperiod},
         given={J.},
         giveni={J\bibinitperiod},
      }}%
      {{hash=NR}{%
         family={Niu},
         familyi={N\bibinitperiod},
         given={R.},
         giveni={R\bibinitperiod},
      }}%
      {{hash=BD}{%
         family={Baron},
         familyi={B\bibinitperiod},
         given={D.},
         giveni={D\bibinitperiod},
      }}%
    }
    \strng{namehash}{HPZJNRBD1}
    \strng{fullhash}{HPZJNRBD1}
    \field{labelnamesource}{author}
    \field{labeltitlesource}{title}
    \field{booktitle}{Proc. IEEE Int. Conf. Acoust., Speech, Signal Process.
  (ICASSP)}
    \field{pages}{6240\bibrangedash 6244}
    \field{title}{Multi-processor approximate message passing using lossy
  compression}
    \list{location}{1}{%
      {Shanghai, China}%
    }
    \field{month}{03}
    \field{year}{2016}
  \endentry

  \entry{ZhuPilgrimBaron2017}{inproceedings}{}
    \name{author}{3}{}{%
      {{hash=ZJ}{%
         family={Zhu},
         familyi={Z\bibinitperiod},
         given={J.},
         giveni={J\bibinitperiod},
      }}%
      {{hash=PR}{%
         family={Pilgrim},
         familyi={P\bibinitperiod},
         given={R.},
         giveni={R\bibinitperiod},
      }}%
      {{hash=BD}{%
         family={Baron},
         familyi={B\bibinitperiod},
         given={D.},
         giveni={D\bibinitperiod},
      }}%
    }
    \strng{namehash}{ZJPRBD1}
    \strng{fullhash}{ZJPRBD1}
    \field{labelnamesource}{author}
    \field{labeltitlesource}{title}
    \field{booktitle}{Proc. 51st IEEE Conf. Inform. Sci. Syst.}
    \field{title}{An Overview of Multi-Processor Approximate Message Passing}
    \list{location}{1}{%
      {Baltimore, MD}%
    }
    \field{month}{03}
    \field{year}{2017}
  \endentry

  \entry{WidrowKollar2008}{book}{}
    \name{author}{2}{}{%
      {{hash=WB}{%
         family={Widrow},
         familyi={W\bibinitperiod},
         given={B.},
         giveni={B\bibinitperiod},
      }}%
      {{hash=KI}{%
         family={Kollár},
         familyi={K\bibinitperiod},
         given={I.},
         giveni={I\bibinitperiod},
      }}%
    }
    \list{publisher}{1}{%
      {Cambridge University Press}%
    }
    \strng{namehash}{WBKI1}
    \strng{fullhash}{WBKI1}
    \field{labelnamesource}{author}
    \field{labeltitlesource}{title}
    \field{title}{Quantization Noise: Roundoff Error in Digital Computation,
  Signal Processing, Control, and Communications}
    \list{location}{1}{%
      {New York, NY}%
    }
    \field{year}{2008}
  \endentry

  \entry{Lipshitz1992}{article}{}
    \name{author}{3}{}{%
      {{hash=LSP}{%
         family={Lipshitz},
         familyi={L\bibinitperiod},
         given={S.\bibnamedelima P.},
         giveni={S\bibinitperiod\bibinitdelim P\bibinitperiod},
      }}%
      {{hash=WRA}{%
         family={Wannamaker},
         familyi={W\bibinitperiod},
         given={R.\bibnamedelima A.},
         giveni={R\bibinitperiod\bibinitdelim A\bibinitperiod},
      }}%
      {{hash=VJ}{%
         family={Vanderkooy},
         familyi={V\bibinitperiod},
         given={J.},
         giveni={J\bibinitperiod},
      }}%
    }
    \strng{namehash}{LSPWRAVJ1}
    \strng{fullhash}{LSPWRAVJ1}
    \field{labelnamesource}{author}
    \field{labeltitlesource}{title}
    \field{number}{5}
    \field{pages}{355\bibrangedash 375}
    \field{title}{Quantization and Dither: A Theoretical Survey}
    \field{volume}{40}
    \field{journaltitle}{J. Audio Eng. Soc.}
    \field{year}{1992}
  \endentry

  \entry{GraphsOptimizationAndAlgos}{book}{}
    \name{editor}{4}{}{%
      {{hash=BA}{%
         family={Brandstädt},
         familyi={B\bibinitperiod},
         given={A.},
         giveni={A\bibinitperiod},
      }}%
      {{hash=NT}{%
         family={Nishizeki},
         familyi={N\bibinitperiod},
         given={T.},
         giveni={T\bibinitperiod},
      }}%
      {{hash=TK}{%
         family={Thulasiraman},
         familyi={T\bibinitperiod},
         given={K.},
         giveni={K\bibinitperiod},
      }}%
      {{hash=AS}{%
         family={Arumugam},
         familyi={A\bibinitperiod},
         given={S.},
         giveni={S\bibinitperiod},
      }}%
    }
    \list{publisher}{2}{%
      {Chapman}%
      {Hall/CRC}%
    }
    \strng{namehash}{BANTTKAS1}
    \strng{fullhash}{BANTTKAS1}
    \field{labelnamesource}{editor}
    \field{labeltitlesource}{title}
    \field{series}{Chapman and Hall/CRC Computer and Information Science
  Series}
    \field{title}{Handbook of Graph Theory, Combinatorial Optimization, and
  Algorithms}
    \field{volume}{34}
    \list{location}{1}{%
      {Boca Raton, FL}%
    }
    \field{year}{2016}
  \endentry

  \entry{Shannon48}{article}{}
    \name{author}{1}{}{%
      {{hash=SCE}{%
         family={Shannon},
         familyi={S\bibinitperiod},
         given={C.\bibnamedelima E.},
         giveni={C\bibinitperiod\bibinitdelim E\bibinitperiod},
      }}%
    }
    \strng{namehash}{SCE1}
    \strng{fullhash}{SCE1}
    \field{labelnamesource}{author}
    \field{labeltitlesource}{title}
    \field{pages}{379\bibrangedash 423,623\bibrangedash 656}
    \field{title}{A mathematical theory of communication}
    \field{journaltitle}{Bell Syst. Tech. J.}
    \field{year}{1948}
  \endentry

  \entry{TU-Berlin-lecture}{online}{}
    \name{author}{2}{}{%
      {{hash=WT}{%
         family={Wiegand},
         familyi={W\bibinitperiod},
         given={T.},
         giveni={T\bibinitperiod},
      }}%
      {{hash=SH}{%
         family={Schwarz},
         familyi={S\bibinitperiod},
         given={H.},
         giveni={H\bibinitperiod},
      }}%
    }
    \strng{namehash}{WTSH1}
    \strng{fullhash}{WTSH1}
    \field{labelnamesource}{author}
    \field{labeltitlesource}{title}
    \field{howpublished}{Lecture Notes}
    \field{title}{Quantization}
    \verb{url}
    \verb https://www.ic.tu-berlin.de/fileadmin/fg121/Source-Coding_WS12/05_Qua
    \verb ntization-WS12.pdf
    \endverb
    \list{institution}{1}{%
      {TU Berlin}%
    }
    \field{month}{12}
    \field{year}{2012}
  \endentry

  \entry{Arimoto72}{article}{}
    \name{author}{1}{}{%
      {{hash=AS}{%
         family={Arimoto},
         familyi={A\bibinitperiod},
         given={S.},
         giveni={S\bibinitperiod},
      }}%
    }
    \strng{namehash}{AS1}
    \strng{fullhash}{AS1}
    \field{labelnamesource}{author}
    \field{labeltitlesource}{title}
    \field{number}{1}
    \field{pages}{14\bibrangedash 20}
    \field{title}{An algorithm for computing the capacity of an arbitrary
  discrete memoryless channel}
    \field{volume}{18}
    \field{journaltitle}{IEEE Trans. Inf. Theory}
    \field{month}{01}
    \field{year}{1972}
  \endentry

  \entry{Blahut72}{article}{}
    \name{author}{1}{}{%
      {{hash=BRE}{%
         family={Blahut},
         familyi={B\bibinitperiod},
         given={R.\bibnamedelima E.},
         giveni={R\bibinitperiod\bibinitdelim E\bibinitperiod},
      }}%
    }
    \strng{namehash}{BRE1}
    \strng{fullhash}{BRE1}
    \field{labelnamesource}{author}
    \field{labeltitlesource}{title}
    \field{number}{4}
    \field{pages}{460\bibrangedash 473}
    \field{title}{Computation of channel capacity and rate-distortion
  functions}
    \field{volume}{18}
    \field{journaltitle}{IEEE Trans. Inf. Theory}
    \field{month}{07}
    \field{year}{1972}
  \endentry

  \entry{Rose94}{article}{}
    \name{author}{1}{}{%
      {{hash=RK}{%
         family={Rose},
         familyi={R\bibinitperiod},
         given={K.},
         giveni={K\bibinitperiod},
      }}%
    }
    \strng{namehash}{RK1}
    \strng{fullhash}{RK1}
    \field{labelnamesource}{author}
    \field{labeltitlesource}{title}
    \field{number}{6}
    \field{pages}{1939\bibrangedash 1952}
    \field{title}{A mapping approach to rate-distortion computation and
  analysis}
    \field{volume}{40}
    \field{journaltitle}{IEEE Trans. Inf. Theory}
    \field{month}{11}
    \field{year}{1994}
  \endentry

  \entry{Lloyd82}{article}{}
    \name{author}{1}{}{%
      {{hash=LSP}{%
         family={Lloyd},
         familyi={L\bibinitperiod},
         given={S.\bibnamedelima P.},
         giveni={S\bibinitperiod\bibinitdelim P\bibinitperiod},
      }}%
    }
    \strng{namehash}{LSP1}
    \strng{fullhash}{LSP1}
    \field{labelnamesource}{author}
    \field{labeltitlesource}{title}
    \field{number}{2}
    \field{pages}{129\bibrangedash 137}
    \field{title}{Least squares quantization in {{PCM}}}
    \field{volume}{28}
    \field{journaltitle}{IEEE Trans. Inf. Theory}
    \field{month}{03}
    \field{year}{1982}
  \endentry

  \entry{Max60}{article}{}
    \name{author}{1}{}{%
      {{hash=MJ}{%
         family={Max},
         familyi={M\bibinitperiod},
         given={J.},
         giveni={J\bibinitperiod},
      }}%
    }
    \strng{namehash}{MJ1}
    \strng{fullhash}{MJ1}
    \field{labelnamesource}{author}
    \field{labeltitlesource}{title}
    \field{number}{1}
    \field{pages}{7\bibrangedash 12}
    \field{title}{Quantization for minimum distortion}
    \field{volume}{6}
    \field{journaltitle}{IRE Trans. Inf. Theory}
    \field{month}{03}
    \field{year}{1960}
  \endentry

  \entry{Huffman52}{article}{}
    \name{author}{1}{}{%
      {{hash=HDA}{%
         family={Huffman},
         familyi={H\bibinitperiod},
         given={D.\bibnamedelima A.},
         giveni={D\bibinitperiod\bibinitdelim A\bibinitperiod},
      }}%
    }
    \strng{namehash}{HDA1}
    \strng{fullhash}{HDA1}
    \field{labelnamesource}{author}
    \field{labeltitlesource}{title}
    \field{number}{40}
    \field{pages}{1098\bibrangedash 1101}
    \field{title}{A Method for the Construction of Minimum-Redundancy Codes}
    \field{volume}{9}
    \field{journaltitle}{Proc. Inst. Radio Eng.}
    \field{month}{09}
    \field{year}{1952}
  \endentry

  \entry{GishPierce68}{article}{}
    \name{author}{2}{}{%
      {{hash=GH}{%
         family={Gish},
         familyi={G\bibinitperiod},
         given={H.},
         giveni={H\bibinitperiod},
      }}%
      {{hash=PJ}{%
         family={Pierce},
         familyi={P\bibinitperiod},
         given={J.},
         giveni={J\bibinitperiod},
      }}%
    }
    \strng{namehash}{GHPJ1}
    \strng{fullhash}{GHPJ1}
    \field{labelnamesource}{author}
    \field{labeltitlesource}{title}
    \field{number}{5}
    \field{pages}{676\bibrangedash 683}
    \field{title}{Asymptotically efficient quantizing}
    \field{volume}{14}
    \field{journaltitle}{IEEE Trans. Inf. Theory}
    \field{month}{09}
    \field{year}{1968}
  \endentry

  \entry{GrayNeuhoff1998}{article}{}
    \name{author}{2}{}{%
      {{hash=GRM}{%
         family={Gray},
         familyi={G\bibinitperiod},
         given={R.\bibnamedelima M.},
         giveni={R\bibinitperiod\bibinitdelim M\bibinitperiod},
      }}%
      {{hash=NDL}{%
         family={Neuhoff},
         familyi={N\bibinitperiod},
         given={D.\bibnamedelima L.},
         giveni={D\bibinitperiod\bibinitdelim L\bibinitperiod},
      }}%
    }
    \strng{namehash}{GRMNDL1}
    \strng{fullhash}{GRMNDL1}
    \field{labelnamesource}{author}
    \field{labeltitlesource}{title}
    \field{pages}{2325\bibrangedash 2383}
    \field{title}{Quantization}
    \field{volume}{IT-44}
    \field{journaltitle}{IEEE Trans. Inf. Theory}
    \field{month}{10}
    \field{year}{1998}
  \endentry

  \entry{FarvardinModestino84}{article}{}
    \name{author}{2}{}{%
      {{hash=FN}{%
         family={Farvardin},
         familyi={F\bibinitperiod},
         given={N.},
         giveni={N\bibinitperiod},
      }}%
      {{hash=MJW}{%
         family={Modestino},
         familyi={M\bibinitperiod},
         given={J.\bibnamedelima W.},
         giveni={J\bibinitperiod\bibinitdelim W\bibinitperiod},
      }}%
    }
    \strng{namehash}{FNMJW1}
    \strng{fullhash}{FNMJW1}
    \field{labelnamesource}{author}
    \field{labeltitlesource}{title}
    \field{number}{3}
    \field{pages}{485\bibrangedash 497}
    \field{title}{Optimum Quantizer Performance for a Class of Non-Gaussian
  Memoryless Sources}
    \field{volume}{IT-30}
    \field{journaltitle}{IEEE Trans. Inf. Theory}
    \field{month}{05}
    \field{year}{1984}
  \endentry

  \entry{ZamirFeder96}{article}{}
    \name{author}{2}{}{%
      {{hash=ZR}{%
         family={Zamir},
         familyi={Z\bibinitperiod},
         given={R.},
         giveni={R\bibinitperiod},
      }}%
      {{hash=FM}{%
         family={Feder},
         familyi={F\bibinitperiod},
         given={M.},
         giveni={M\bibinitperiod},
      }}%
    }
    \strng{namehash}{ZRFM1}
    \strng{fullhash}{ZRFM1}
    \field{labelnamesource}{author}
    \field{labeltitlesource}{title}
    \field{number}{4}
    \field{pages}{1152\bibrangedash 1159}
    \field{title}{On Lattice Quantization Noise}
    \field{volume}{42}
    \field{journaltitle}{IEEE Trans. Inf. Theory}
    \field{month}{07}
    \field{year}{1996}
  \endentry

  \entry{SripadSnyder77}{article}{}
    \name{author}{2}{}{%
      {{hash=SAB}{%
         family={Sripad},
         familyi={S\bibinitperiod},
         given={A.\bibnamedelima B.},
         giveni={A\bibinitperiod\bibinitdelim B\bibinitperiod},
      }}%
      {{hash=SDL}{%
         family={Snyder},
         familyi={S\bibinitperiod},
         given={D.\bibnamedelima L.},
         giveni={D\bibinitperiod\bibinitdelim L\bibinitperiod},
      }}%
    }
    \strng{namehash}{SABSDL1}
    \strng{fullhash}{SABSDL1}
    \field{labelnamesource}{author}
    \field{labeltitlesource}{title}
    \field{number}{5}
    \field{pages}{442\bibrangedash 448}
    \field{title}{A Necessary and Sufficient Condition for Quantization Errors
  to be Uniform and White}
    \field{volume}{ASSP-25}
    \field{journaltitle}{IEEE Trans. Acoust., Speech, Signal Process.}
    \field{month}{10}
    \field{year}{1977}
  \endentry

  \entry{DirectSum}{online}{}
    \name{author}{1}{}{%
      {{hash=WEW}{%
         family={Weisstein},
         familyi={W\bibinitperiod},
         given={E.\bibnamedelima W.},
         giveni={E\bibinitperiod\bibinitdelim W\bibinitperiod},
      }}%
    }
    \strng{namehash}{WEW1}
    \strng{fullhash}{WEW1}
    \field{labelnamesource}{author}
    \field{labeltitlesource}{title}
    \field{note}{From MathWorld---a Wolfram Web Resource}
    \field{title}{Matrix Direct Sum}
    \verb{url}
    \verb http://mathworld.wolfram.com/MatrixDirectSum.html
    \endverb
    \field{year}{2017}
  \endentry

  \entry{Kabal2011}{inproceedings}{}
    \name{author}{1}{}{%
      {{hash=KP}{%
         family={Kabal},
         familyi={K\bibinitperiod},
         given={P.},
         giveni={P\bibinitperiod},
      }}%
    }
    \strng{namehash}{KP1}
    \strng{fullhash}{KP1}
    \field{labelnamesource}{author}
    \field{labeltitlesource}{title}
    \field{booktitle}{Proc. IEEE Int. Conf. Acoust., Speech, Signal Process.
  (ICASSP)}
    \field{pages}{5244\bibrangedash 5247}
    \field{title}{Correlation properties of quantization noise}
    \list{location}{1}{%
      {Prague, Czech Republic}%
    }
    \field{month}{05}
    \field{year}{2011}
  \endentry

  \entry{AykolRose2013}{article}{}
    \name{author}{2}{}{%
      {{hash=AE}{%
         family={Aykol},
         familyi={A\bibinitperiod},
         given={E.},
         giveni={E\bibinitperiod},
      }}%
      {{hash=RK}{%
         family={Rose},
         familyi={R\bibinitperiod},
         given={K.},
         giveni={K\bibinitperiod},
      }}%
    }
    \strng{namehash}{AERK1}
    \strng{fullhash}{AERK1}
    \field{labelnamesource}{author}
    \field{labeltitlesource}{title}
    \field{number}{13}
    \field{pages}{3291\bibrangedash 3302}
    \field{title}{On Constrained Randomized Quantization}
    \field{volume}{61}
    \field{journaltitle}{IEEE Trans. Signal Process.}
    \field{month}{07}
    \field{year}{2013}
  \endentry

  \entry{BoydKimVandenbergheHassibi2007}{article}{}
    \name{author}{4}{}{%
      {{hash=BS}{%
         family={Boyd},
         familyi={B\bibinitperiod},
         given={S.},
         giveni={S\bibinitperiod},
      }}%
      {{hash=KSJ}{%
         family={Kim},
         familyi={K\bibinitperiod},
         given={S.-J.},
         giveni={S\bibinithyphendelim J\bibinitperiod},
      }}%
      {{hash=VL}{%
         family={Vandenberghe},
         familyi={V\bibinitperiod},
         given={L.},
         giveni={L\bibinitperiod},
      }}%
      {{hash=HA}{%
         family={Hassibi},
         familyi={H\bibinitperiod},
         given={A.},
         giveni={A\bibinitperiod},
      }}%
    }
    \strng{namehash}{BSKSJVLHA1}
    \strng{fullhash}{BSKSJVLHA1}
    \field{labelnamesource}{author}
    \field{labeltitlesource}{title}
    \field{number}{67}
    \field{pages}{67\bibrangedash 127}
    \field{title}{A tutorial on geometric programming}
    \field{volume}{8}
    \field{journaltitle}{Optimization Eng.}
    \field{month}{03}
    \field{year}{2007}
  \endentry

  \entry{PanterDite51}{article}{}
    \name{author}{2}{}{%
      {{hash=PPF}{%
         family={Panter},
         familyi={P\bibinitperiod},
         given={P.\bibnamedelima F.},
         giveni={P\bibinitperiod\bibinitdelim F\bibinitperiod},
      }}%
      {{hash=DW}{%
         family={Dite},
         familyi={D\bibinitperiod},
         given={W.},
         giveni={W\bibinitperiod},
      }}%
    }
    \strng{namehash}{PPFDW1}
    \strng{fullhash}{PPFDW1}
    \field{labelnamesource}{author}
    \field{labeltitlesource}{title}
    \field{number}{1}
    \field{pages}{44\bibrangedash 48}
    \field{title}{Quantization Distortion in Pulse-Count Modulation with
  Nonuniform Spacing of Levels}
    \field{volume}{39}
    \field{journaltitle}{Proc. IRE}
    \field{month}{01}
    \field{year}{1951}
  \endentry

  \entry{cvx}{misc}{}
    \name{author}{2}{}{%
      {{hash=GM}{%
         family={Grant},
         familyi={G\bibinitperiod},
         given={M.},
         giveni={M\bibinitperiod},
      }}%
      {{hash=BS}{%
         family={Boyd},
         familyi={B\bibinitperiod},
         given={S.},
         giveni={S\bibinitperiod},
      }}%
    }
    \strng{namehash}{GMBS1}
    \strng{fullhash}{GMBS1}
    \field{labelnamesource}{author}
    \field{labeltitlesource}{title}
    \field{howpublished}{\url{http://cvxr.com/cvx}}
    \field{title}{{{CVX}}: {{MATLAB}} Software for Disciplined Convex
  Programming, version 2.1}
    \field{month}{03}
    \field{year}{2014}
  \endentry

  \entry{gb08}{incollection}{}
    \name{author}{2}{}{%
      {{hash=GM}{%
         family={Grant},
         familyi={G\bibinitperiod},
         given={M.},
         giveni={M\bibinitperiod},
      }}%
      {{hash=BS}{%
         family={Boyd},
         familyi={B\bibinitperiod},
         given={S.},
         giveni={S\bibinitperiod},
      }}%
    }
    \name{editor}{3}{}{%
      {{hash=BV}{%
         family={Blondel},
         familyi={B\bibinitperiod},
         given={V.},
         giveni={V\bibinitperiod},
      }}%
      {{hash=BS}{%
         family={Boyd},
         familyi={B\bibinitperiod},
         given={S.},
         giveni={S\bibinitperiod},
      }}%
      {{hash=KH}{%
         family={Kimura},
         familyi={K\bibinitperiod},
         given={H.},
         giveni={H\bibinitperiod},
      }}%
    }
    \list{publisher}{1}{%
      {Springer-Verlag Limited}%
    }
    \strng{namehash}{GMBS1}
    \strng{fullhash}{GMBS1}
    \field{labelnamesource}{author}
    \field{labeltitlesource}{title}
    \field{booktitle}{Recent Advances in Learning and Control}
    \field{note}{\url{http://stanford.edu/~boyd/graph_dcp.html}}
    \field{pages}{95\bibrangedash 110}
    \field{series}{Lecture Notes in Control and Information Sciences}
    \field{title}{Graph implementations for nonsmooth convex programs}
    \field{year}{2008}
  \endentry

  \entry{gpkit}{misc}{}
    \name{author}{2}{}{%
      {{hash=BE}{%
         family={Burnell},
         familyi={B\bibinitperiod},
         given={Edward},
         giveni={E\bibinitperiod},
      }}%
      {{hash=HW}{%
         family={Hoburg},
         familyi={H\bibinitperiod},
         given={Warren},
         giveni={W\bibinitperiod},
      }}%
    }
    \strng{namehash}{BEHW1}
    \strng{fullhash}{BEHW1}
    \field{labelnamesource}{author}
    \field{labeltitlesource}{title}
    \field{howpublished}{\url{https://github.com/hoburg/gpkit}}
    \field{note}{Version 0.5.2}
    \field{title}{{{GP}}kit software for geometric programming}
    \field{year}{2017}
  \endentry

  \entry{cvxpy}{article}{}
    \name{author}{2}{}{%
      {{hash=DS}{%
         family={Diamond},
         familyi={D\bibinitperiod},
         given={Steven},
         giveni={S\bibinitperiod},
      }}%
      {{hash=BS}{%
         family={Boyd},
         familyi={B\bibinitperiod},
         given={Stephen},
         giveni={S\bibinitperiod},
      }}%
    }
    \strng{namehash}{DSBS1}
    \strng{fullhash}{DSBS1}
    \field{labelnamesource}{author}
    \field{labeltitlesource}{title}
    \field{number}{83}
    \field{pages}{1\bibrangedash 5}
    \field{title}{{{CVXPY}}: A {P}ython-Embedded Modeling Language for Convex
  Optimization}
    \field{volume}{17}
    \field{journaltitle}{J. Mach. Learning Research}
    \field{year}{2016}
  \endentry

  \entry{Lofberg2004}{inproceedings}{}
    \name{author}{1}{}{%
      {{hash=LJ}{%
         family={Löfberg},
         familyi={L\bibinitperiod},
         given={J.},
         giveni={J\bibinitperiod},
      }}%
    }
    \strng{namehash}{LJ1}
    \strng{fullhash}{LJ1}
    \field{labelnamesource}{author}
    \field{labeltitlesource}{title}
    \field{booktitle}{Proc. CACSD Conf.}
    \field{title}{{{YALMIP}}: A Toolbox for Modeling and Optimization in
  {{MATLAB}}}
    \list{location}{1}{%
      {Taipei, Taiwan}%
    }
    \field{year}{2004}
  \endentry

  \entry{MATLAB}{book}{}
    \name{author}{1}{}{%
      {{hash=M}{%
         family={MATLAB},
         familyi={M\bibinitperiod},
      }}%
    }
    \list{publisher}{1}{%
      {The MathWorks, Inc.}%
    }
    \strng{namehash}{M1}
    \strng{fullhash}{M1}
    \field{labelnamesource}{author}
    \field{labeltitlesource}{title}
    \field{title}{{{r}}elease 2017a}
    \list{location}{1}{%
      {Natick, MA}%
    }
    \field{year}{2017}
  \endentry

  \entry{Torus}{online}{}
    \name{author}{1}{}{%
      {{hash=WEW}{%
         family={Weisstein},
         familyi={W\bibinitperiod},
         given={E.\bibnamedelima W.},
         giveni={E\bibinitperiod\bibinitdelim W\bibinitperiod},
      }}%
    }
    \strng{namehash}{WEW1}
    \strng{fullhash}{WEW1}
    \field{labelnamesource}{author}
    \field{labeltitlesource}{title}
    \field{note}{From MathWorld---a Wolfram Web Resource}
    \field{title}{Torus}
    \verb{url}
    \verb http://mathworld.wolfram.com/Torus.html
    \endverb
    \field{year}{2017}
  \endentry

  \entry{SquareTorus}{online}{}
    \name{author}{1}{}{%
      {{hash=WEW}{%
         family={Weisstein},
         familyi={W\bibinitperiod},
         given={E.\bibnamedelima W.},
         giveni={E\bibinitperiod\bibinitdelim W\bibinitperiod},
      }}%
    }
    \strng{namehash}{WEW1}
    \strng{fullhash}{WEW1}
    \field{labelnamesource}{author}
    \field{labeltitlesource}{title}
    \field{note}{From MathWorld---a Wolfram Web Resource}
    \field{title}{Square Torus}
    \verb{url}
    \verb http://mathworld.wolfram.com/SquareTorus.html
    \endverb
    \field{year}{2017}
  \endentry

  \entry{HoffmanKunze71}{book}{}
    \name{author}{2}{}{%
      {{hash=HK}{%
         family={Hoffman},
         familyi={H\bibinitperiod},
         given={K.},
         giveni={K\bibinitperiod},
      }}%
      {{hash=KR}{%
         family={Kunze},
         familyi={K\bibinitperiod},
         given={R.},
         giveni={R\bibinitperiod},
      }}%
    }
    \list{publisher}{1}{%
      {Prentice Hall, Inc.}%
    }
    \strng{namehash}{HKKR1}
    \strng{fullhash}{HKKR1}
    \field{labelnamesource}{author}
    \field{labeltitlesource}{title}
    \field{title}{Linear Algebra}
    \list{location}{1}{%
      {Englewood Cliffs, NJ}%
    }
    \field{year}{1971}
  \endentry

  \entry{Xiao2005}{inproceedings}{}
    \name{author}{3}{}{%
      {{hash=XL}{%
         family={Xiao},
         familyi={X\bibinitperiod},
         given={L.},
         giveni={L\bibinitperiod},
      }}%
      {{hash=BS}{%
         family={Boyd},
         familyi={B\bibinitperiod},
         given={S.},
         giveni={S\bibinitperiod},
      }}%
      {{hash=LS}{%
         family={Lall},
         familyi={L\bibinitperiod},
         given={S.},
         giveni={S\bibinitperiod},
      }}%
    }
    \strng{namehash}{XLBSLS1}
    \strng{fullhash}{XLBSLS1}
    \field{labelnamesource}{author}
    \field{labeltitlesource}{title}
    \field{booktitle}{Proc. 4th Int. Symp. Inform. Process. Sensor Networks
  (IPSN)}
    \field{pages}{63\bibrangedash 70}
    \field{title}{A scheme for robust distributed sensor fusion based on
  average consensus}
    \field{month}{04}
    \field{year}{2005}
  \endentry

  \entry{ZhuBaronKrzakala2017IEEE}{article}{}
    \name{author}{3}{}{%
      {{hash=ZJ}{%
         family={Zhu},
         familyi={Z\bibinitperiod},
         given={J.},
         giveni={J\bibinitperiod},
      }}%
      {{hash=BD}{%
         family={Baron},
         familyi={B\bibinitperiod},
         given={D.},
         giveni={D\bibinitperiod},
      }}%
      {{hash=KF}{%
         family={Krzakala},
         familyi={K\bibinitperiod},
         given={F.},
         giveni={F\bibinitperiod},
      }}%
    }
    \strng{namehash}{ZJBDKF1}
    \strng{fullhash}{ZJBDKF1}
    \field{labelnamesource}{author}
    \field{labeltitlesource}{title}
    \field{number}{9}
    \field{pages}{2444\bibrangedash 2454}
    \field{title}{Performance Limits for Noisy Multimeasurement Vector
  Problems}
    \field{volume}{65}
    \field{journaltitle}{IEEE Trans. Signal Process.}
    \field{month}{05}
    \field{year}{2017}
  \endentry
\enddatalist


\begin{spacing}{1}
\setlength\bibitemsep{11pt} 
\addcontentsline{toc}{chapter}{{\uppercase{\bibname}}} 
\titleformat{\chapter}[display]{\bf\filcenter
}{\chaptertitlename\ \thechapter}{11pt}{\bf\filcenter}

\titlespacing*{\chapter}{0pt}{-0.275in}{22pt}
\newgeometry{margin=1in,lmargin=1.25in,footskip=\chapterfootskip,includefoot}

\printbibliography[heading=myheading]
\end{spacing}


%
%


\restoregeometry

\backmatter

\end{document}